\documentclass[12pt]{article}
\usepackage[round]{natbib}
\usepackage{amsmath}
\usepackage{amssymb}
\usepackage{amsthm}
\usepackage{dsfont}
\usepackage[margin=1in]{geometry}
\usepackage{hyperref}
\usepackage{multirow}
\usepackage{color}
\usepackage{tikz}
\usepackage{graphicx}
\usepackage{setspace}
\usepackage{ulem}
\usepackage{algorithm2e}
\usepackage{float}
\usepackage{lscape}
\usepackage{rotating}
\usepackage{longtable}
\usepackage{subcaption}
\usepackage{etoc}
\usepackage{authblk}
\usepackage{mdframed}
\newtheorem{lemma}{Lemma}
\newtheorem*{lemma*}{Lemma}
\newtheorem{theorem}{Theorem}
\newtheorem*{theorem*}{Theorem}

\newtheorem{corollary}{Corollary}
\hypersetup{
    colorlinks=true,
    linkcolor=black,
    filecolor=magenta,
    urlcolor=blue,
    citecolor=black,
}

\newcommand{\indep}{\mathrel{\text{\scalebox{1.07}{$\perp\mkern-10mu\perp$}}}}
\DeclareMathOperator*{\argmin}{arg\,min}
\DeclareMathOperator*{\argmax}{arg\,max}
\newcommand{\E}{\mathbb{E}}
\newcommand{\V}{\mathbb{V}}

\newcommand{\bx}{\mathbf{x}}
\newcommand{\phihat}{\widehat{\phi}}
\newcommand{\bX}{\mathbf{X}}
\newcommand{\MG}{{\tt MG}}

\newcommand{\MGxt}{{\tt MG}(\phihat, \bx, t)}
\newcommand{\MGXt}{{\tt MG}(\phihat, \bX, t)}

\newcommand{\Nxt}{N(\phihat, \bx, t)}
\newcommand{\NXt}{N(\phihat, \bX, t)}

\newcommand{\ind}{\mathbb{I}}

\newcommand{\Wixt}{W_i(\phihat, \bx, t)}
\newcommand{\WiXt}{W_i(\phihat, \bX, t)}

\newcommand{\taux}{\tau(\bx)}

\newcommand{\Dnx}{\mathcal D(\phihat, \bx, t)}
\newcommand{\Rnx}{B(\phihat, \bx, t)}

\newcommand{\muxt}{\mu(\bx, t)}
\newcommand{\muxthat}{\hat{\mu}(\bx, t)}

\newcommand{\pconv}{\overset{p}{\rightarrow}}
\newcommand{\dconv}{\overset{d}{\rightarrow}}

\newcommand{\xiNix}{\xi_{n,i}(\bx)}
\newcommand{\xiNjx}{\xi_{n,j}(\bx)}

\newcommand{\Fnix}{\mathcal{F}_{n,i}(\bx)}
\newcommand{\Fnimox}{\mathcal{F}_{n,i-1}(\bx)}

\newcommand{\muXid}{\mu(\bX_i, t)}

\newcommand{\muxd}{\mu(\bx, t)}
\newcommand{\muXd}{\mu(\bX, t)}
\newcommand{\sigmaXid}{\sigma^2(\bX_i, t)}
\newcommand{\sigmaxd}{\sigma^2(\bx, t)}
\newcommand{\etaxd}{\eta(\bx, t)}
\newcommand{\etaXid}{\eta(\bX_i, t)}

\newcommand{\mudxhat}{\hat{\mu}(\bx, t)}

\newcommand{\mudxtilde}{\tilde{\mu}(\bx, t)}
\newcommand{\mudXtilde}{\tilde{\mu}(\bX, t)}

\newcommand{\bu}{\mathbf{u}}
\newcommand{\bv}{\mathbf{v}}
\newcommand{\bz}{\mathbf{z}}
\newcommand{\br}{\mathbf{r}}

\newcommand{\Rx}{{R_\phi(\phihat, \bx, t)}}

\newcommand{\fbX}{f_{\bX}}

\newcommand{\FbXt}{F_{\bX|T=t}}

\newcommand{\sigmasqxt}{\sigma^2(\bx, t)}
\newcommand{\sigmasqxthat}{\hat{\sigma}^2(\bx, t)}

\newcommand{\nux}{\nu_{\phi}(\bx, t)}

\newcommand{\Gamman}{\Gamma_n}

\newcommand{\muhatDR}{\hat{\mu}^{DR}}

\newcommand{\tauhatDR}{\hat{\tau}^{DR}}
\newcommand{\psihat}{\widehat{\psi}}
\newcommand{\muhat}{\hat{\mu}}
\newcommand{\ehat}{\hat{e}}
\newcommand{\tauhat}{\hat{\tau}}
\newcommand{\tauhatDRATE}{\tauhatDR_{ATE}}
\newcommand{\tauhatDRATT}{\tauhatDR_{ATT}}
\newcommand{\sigmasq}{\sigma^2}
\newcommand{\sigmasqhat}{\hat{\sigma}^2}

\newcommand{\R}{\mathbb{R}}

\newcommand{\Xmax}{\bX_{max}}

\newcommand{\X}{\mathbb{X}}

\newcommand{\bbeta}{\boldsymbol{\beta}}
\newcommand{\blambda}{\boldsymbol{\lambda}}
\newcommand{\bdelta}{\boldsymbol{\delta}}
\newcommand{\Dphiq}{{D_\phi^q}}
\newcommand{\Dphiqhat}{{D_{\phihat}^q}}

\newcommand{\Dnl}{\mathcal{O}_{\setminus \ell}}

\newcommand{\On}{{\mathcal{O}_n}}
\newcommand{\bO}{{\mathbf{O}}}
\newcommand{\bZ}{\mathbf{Z}}

\newcommand{\cP}{\mathcal{P}}

\begin{document}
\title{\bf Matched Machine Learning:  
A Generalized Framework for 
Treatment Effect Inference 
With Learned Metrics
}
\author[1]{Marco Morucci}
\author[2]{Cynthia Rudin}
\author[3]{Alexander Volfovsky}
\affil[1]{Center for Data Science, New York University}
\affil[2]{Department of Computer Science, Duke University}
\affil[3]{Department of Statistical Science, Duke University}
\date{}
\maketitle

\begin{abstract}
\singlespacing
We introduce Matched Machine Learning, a framework that combines the flexibility of machine learning black boxes with the interpretability of matching, a longstanding tool in observational causal inference. Interpretability is paramount in many high-stakes application of causal inference. Current tools for nonparametric estimation of both average and individualized treatment effects are black-boxes that do not allow for human auditing of estimates. Our framework uses machine learning to learn an optimal metric for matching units and estimating outcomes, thus achieving the performance of machine learning black-boxes, while being interpretable. Our general framework encompasses several published works as special cases. We provide asymptotic inference theory for our proposed framework, enabling users to construct approximate confidence intervals around estimates of both individualized and average treatment effects. We show empirically that instances of Matched Machine Learning perform on par with black-box machine learning methods and better than existing matching methods for similar problems. Finally, in our application we show how Matched Machine Learning can be used to perform causal inference even when covariate data are highly complex: we study an image dataset, and produce high quality matches and estimates of treatment effects. 
\end{abstract}
\noindent%

\newpage
\doublespacing

\section{Introduction}
Matching methods have a long history in observational causal inference. Their simplicity makes them interpretable even by non-technical audiences, as well as guarantees fast execution of causal analyses with known statistical properties, all while requiring no parametric assumptions on either outcome or treatment distributions \citep{rosenbaum1983central}. Recently, black-box machine learning (ML) tools for nonparametric estimation have come to supplant matching as the default for treatment effect estimation with contextual covariates \citep[e.g.,][]{hill2011bayesian, wager2018estimation, chernozhukov2018double, hahn2020bayesian}. There are good reasons for this: the complexity of ML black boxes allows them to predict treatment effects with an unprecedented degree of accuracy, which is sometimes not achieved by matching methods. However, the use of black box ML comes at a cost of interpretability in the results. Our main concern is that models that are not interpretable are difficult to \textit{audit}, i.e., it is difficult to assess the robustness and credibility of results from an uninterpretable model using contextual information about the data, problem, or population under study. This in turn is problematic since most causal inference datasets can have myriad forms of hidden noise, as well as unmeasured confounding. In this paper, we propose to bridge the gap between the accuracy of ML and the  auditability of matching by using the first to inform the second. We propose a general framework that first uses flexible ML to \textit{learn} a distance metric for matching units, and then employs a matching algorithm that uses the learned metric to construct high-quality matches. These matches are designed to approximate the predictions of the black-box ML while still being auditable; after the matches are made, the black box is thrown out and the analysis proceeds with the matches only. Analysts can then audit quality of the matched estimate by simply examining the matches themselves. Methods from our framework output estimates for the Conditional Average Treatment Effect (CATE), i.e., the expected effect on any individual unit (also known as individualized treatment effect), the Average Treatment Effect (ATE), and Average Treatment Effect on the Treated (ATT), which are average treatment effects on respectively all or only the treated units.
Our paper makes several key contributions to the literature on matching and treatment effect estimation in general:\\
(i) We introduce a framework for the production of interpretable treatment effect estimates that are still black-box accurate for the CATE.\\ 
(ii) We derive the asymptotic distributions and error bounds for CATE estimates made with any matching method falling under our framework.\\
(iii) We expand our framework into a general, doubly-robust matching framework for valid $\sqrt{n}$-asymptotic inference for the ATE/ATT.\\
We also show empirically that the finite-sample performance of our tools does indeed match that of black box ML, and theoretically, that our matched estimates are asymptotically normal and with known and estimable variance for conditional treatment effects. In addition, we leverage Double Machine Learning (DML) \citep{chernozhukov2018double} methodologies to combine CATEs from matching into a doubly-robust estimator to obtain estimates for average treatment effects that are asymptotically normal at a rate of $\sqrt{n}$. Importantly, this last method allows us to sidestep the issue of the slow-vanishing asymptotic bias of many common matching methods brought up by \cite{abadie2006large}. 
Our framework generalizes many known and widely-employed matching methods, some of which are summarized in Table \ref{tab:generalization}. Importantly, the theoretical guarantees that we establish for our framework can be applied to \textit{any} of the matching methods it generalizes. We expand on how our framework generalizes the matching methods in Table \ref{tab:generalization} in Section \ref{sec:extensions}. 
\begin{table}[!htbp]
    \vspace{-1cm}
    {\centering
    \caption{Some existing matching methods that are special cases of M-ML. }
    \label{tab:generalization}
    \resizebox{\textwidth}{!}{%
    \begin{tabular}{c|c|c}
        \hline\hline
        Method & Choice of $\phi$ & Choice of $q$ \\
        \hline\hline
        Nearest Neighbor &  $\phi(\bx) = \bx,$ & $2$ \\\hline
        Propensity Score Matching & $\phi(\bx) = h(\bx),\; h \in \argmin_{h \in \mathcal{H}} \E_{\bX}[(h(\bX) - \Pr(T_i=t|\bX))^2]$ & $1$ \\
        \citep{rosenbaum1983central} & & \\\hline
        Prognostic Score Matching & $\phi(\bx) = h(\bx),\; h \in \argmin_{h \in \mathcal{H}} \E_{\bX,Y_i(t)}[(h(\bX) - Y_i(t))^2]$ & $1$  \\
        \citep{hansen2008prognostic} & & \\\hline
        Adaptive Hyper-Boxes$^*$ & $\phi(\bx) = [h_t(\bx), h_{t'}(\bx)]$ &   \\
        \citep{morucci2020adaptive}& $h_t \in \argmin_{h \in \mathcal{H}} \E_{\bX,Y_i(t)}[(h(\bX) - Y_i(t))^2]$ & $1$\\
        & $h_{t'} \in \argmin_{h \in \mathcal{H}} \E_{\bX,Y_i(t')}[(h(\bX) - Y_i(t'))^2]$ & \\\hline
        Coarsened Exact Matching & $\phi(\bx) = \tilde{\Gamma}\bx$ & $\infty$ \\
        \citep{iacus2012causal}&$\tilde{\Gamma} \in \R_{diag}^{p\times p}$ is a matrix of dimension-wise calipers & \\\hline
        Genetic & $\phi(\bx) = \mathbf{M}\bx$ & $2$ \\
        \citep{diamond2013genetic} &$\mathbf{M} \in \argmin_{\mathbf{M}\in \R_{diag}^{p\times p}}\sum_{i,j \in \textrm{Tr}}\textrm{Maha}(\bx_i, \bx_j, \mathbf{M})$& \\\hline
        MALTS & $\phi(\bx) = \mathbf{M}\bx,$ & $2$ \\
        \citep{parikh2018malts} &$\mathbf{M} \in \argmin_{\mathbf{M}\in \mathbb{R}_{diag}^{p\times p}} \left|\sum\limits_{i,j \in \textrm{Tr}}\frac{(y_i - y_j)\exp(-\textrm{Maha}(\bx_i, \bx_j, \mathbf{M})\ind(T_i=T_j))}{\sum_{k \in Tr} \exp(-\textrm{Maha}(\bx_i, \bx_k, \mathbf{M})\ind(t_i=t_k))}\right|$& \\\hline
        Fine Balance$^*$ & $\phi(\bx) = h(\bx),\; h \in \argmin_{h \in \mathcal{H}} \E_{\bX}[(h(\bX) - \Pr(T_i=t|\bX))^2]$ & 1 \\
        \citep{rosenbaum2007minimum} & & \\
        \hline\hline
    \end{tabular}}}
    \footnotesize{Note: $^*$ Requires additional constraints on the matching problem. The parameters $\phi$ and $q$ are M-ML hyperparameters that will be defined later; setting them to the values in the table give us the methods in the left column. $\phi$ is a mapping from the original space to a useful feature space. It is estimated using a machine learning model. That machine learning model minimizes the loss function shown in the table. $q$ defines a distance metric in the learned feature space. See Section \ref{sec:extensions} for an in-depth explanation of this table. Treatment represented by the random variable $T$, a specific treatment level is $t$, covariates are $\bX$ (or $\bx$ if not random), potential outcomes are $Y(t)$, $\text{Tr}$ is a training set, $\textrm{Maha}(\bx_i, \bx_j, \mathbf{M})$ is the Mahalanobis distance of the covariate vectors of units $i$ and $j$ weighted by the positive, real-valued diagonal matrix $\mathbf{M}$.}
    \vspace{-0.4cm}
\end{table}
We will demonstrate the power and flexibility of our method by matching on images, where we learn a low-dimensional representation of image covariates with a convolutional neural network. We will show that our method produces interpretable matched groups even when input covariates are complex and high-dimensional, like images. While the representation we match on is uninterpretable, the results are visually auditable: humans will be able to visually inspect the matched groups of images and easily assess their quality and trustworthiness. This will enable us to study whether brand responsiveness to consumers on social media is associated with an increase in consumer interaction with the brand. This is an important problem in marketing and consumer behavior \citep[e.g., ][]{laroche2013or}, but so far it has not been studied at the level of granularity that our method enables. 
Our paper will proceed as follows: In Section \ref{sec:matching}, we introduce matching methods in general. In Section \ref{sec:MML}, we outline the Matched Machine Learning framework for estimation of Conditional Average Treatment Effects by matching units on learned distance metrics. In Section \ref{sec:asymptotics}, we present large-sample theoretical properties of our methodology. In Section \ref{sec:DML}, we propose an algorithm for asymptotically efficient matched estimation of the Average Treatment Effect. Section \ref{sec:simulations}  presents empirical evidence of the performance of our methods on simulated datasets. Finally, in Section \ref{sec:application}, we apply our methods to the problem of studying the impact of brand responsiveness on social media on the amount of consumer interaction with the brand. 

\section{Matching and Observational Causal Inference}\label{sec:matching}
We first introduce notation. We have a sample of $n$ units, $i = 1, \dots, n$, having potential outcomes $Y_i(1),\dots, Y_i(M) \in \mathbb{R}$ for a treatment that can take $M$ possible values, $t \in \{1,\dots, M \}$. Assigned treatments are denoted by the random variable $T_i \in \{1, \dots, M\}$. We never observe the full vector of potential outcomes for each unit, but instead we observe the outcome variable $Y_i = \sum_{i=1}^nY_i(t)\ind[T_i=t]$, where $\ind[E]$ is the indicator function for event $E$. Each unit has is assigned a $p$-dimensional random vector of covariates $\bX_i$ taking values in  $\X$, where $\X \subset \R^p$ is a compact set; observed covariate vectors are $\bx_i \in \X$. For an arbitrary random variable $A$, will use the notation $f_A$ to denote the Probability Mass Function (PMF) or  Probability Distribution Function (PDF) of $A$, and use $F_A$ to denote the Cumulative Distribution Function (CDF) of $A$, and we will also use $f_A$ to denote the distribution of $A$. We use the notation $\E_{A}[A]$  and $\V_{A}[A]$ to denote expectation and variance of $A$ with respect to the distribution of $A$. When we use this notation without subscripts we mean that the expectation operator is with respect to all the random quantities inside of the square brackets.
We make the following classical assumptions, for all $i$:

\noindent\textbf{A1 (Data Distribution)}: \\
(a) The data $\On = \{\bO_i\}_{i=1}^n = \{Y_i, \bX_i, T_i\}_{i=1}^n$ is a set of $n$ i.i.d. copies of $\bO$. \\
(b) The domain of the covariate distribution, $\mathbb{X}$ is a compact subset of $\R^p$. \\
(c) The covariates have marginal distribution  with differentiable CDF (w.r.t. the lebesgue measure) $F_{\bX}(\bx)$, and constants $c_{\fbX}, C_{\fbX}$, such that $0 < c_{\fbX} < \fbX(\bx) < C_{\fbX} < \infty$ everywhere over $\mathbb{X}$.\\
\textbf{A2 (Overlap)}: For all $\bx \in \mathbb{X}$ and $t=1,\dots,M$ we have $0 < \Pr(T=t|\bX=\bx) < 1$.\\
\textbf{A3 (Conditional Ignorability)}: $T \indep (Y(1), \dots, Y(M))|\bX$. \\
\textbf{A4 (Bounded Higher Moments)}: For all $t, t' \in \{1, \dots, M\}$, all $\bx \in \mathbb{X}$ and  for some $\delta > 0$ and a constant $C_\delta$ we have: $\E[|Y(t)|^{2 + \delta}|\bX=\bx, T=t'] \leq C_\delta$.

This paper is largely focused on estimating the Conditional Response Function (CRF) and Conditional Average Treatment Effect for a given covariate vector, $\bx$. These are denoted, respectively, by: $\mu(\bx, t) = \E[Y(t)|\bX=\bx]$, and $\taux = \mu(\bx, t) - \mu(\bx, t')$, for two treatments $t,t'$. Note that Assumption 3 allows us to write: $\E[Y|\bX=\bx, T=t] = \E[Y(t)|\bX = \bx]$, and, therefore, implies that our quantities of interest can be consistently estimated from our observed data. We will also use the notation $\sigmasqxt = \V[Y(t)|\bX=\bx]$ to denote the conditional variance of the potential outcomes. Later in this paper, we will be concerned with estimating averaged versions of the CRF and the CATE, which are defined as follows: Average Response Function (ARF) $\mu(t) = \E[Y(t)]$, Average Treatment Effect (ATE): $\tau(t, t') = \E[Y(t)] - \E[Y(t')]$, and Average Treatment Effect on the Treated (ATT): $\delta(t, t') = \E[Y(t)|T=t] - \E[Y(t')|T=t]$. We will see how consistent estimation of the CRF allows for consistent estimation of all the other quantities. The main idea of matching is to create a Matched Group, i.e., a subset of units: $\MGxt \subset \{1, \dots, n\}$ that contains units whose observed outcomes will be used to estimate the quantities of interest defined above for the desired value $\bx$. It follows that the main problem of a matching procedure is to select which units we should include in $\MGxt$. In the following section we outline our strategy to do so. 
\section{Matched Machine Learning: A General Procedure} \label{sec:MML}
The key idea of this paper is to take advantage of powerful black-box machine learning models to inform how we should make matched groups, and, consequently, estimate CRFs and CATEs. To do this, we propose using machine learning to construct a $d$-dimensional representation of the covariates, and to subsequently match on these representations instead of the raw covariate values: ML will flexibly learn representations that are informative about the relationship between $\bX$, $T$, and $Y$, leading to higher-quality matches.  To accomplish the goal just described, we introduce the function $\phi: \X \mapsto \R^d$, which is a representation function for the observed covariates. We assume that $\phi(\bx)$ exists for all $\bx$ in their respective domains and is continuous. It is also possible for $\phi$ to depend on the treatment level, meaning that one separate $\phi$ will be estimated per treatment group. We omit this from the notation for ease of readership. We then propose that $\phi$ is learned with a ML method on a separate training set. The idea behind this representation is to \textit{map the covariates to a space where units are close if their potential outcomes are close, which is ultimately the goal of matching}. For example, if we considered similarity between three units $i,j,k$ based on $\bX$ alone, we might might match $i$ to $k$ even though $i$ is more similar to $j$ in terms of (unobserved) potential outcomes. Instead, if we considered some transformation $\phi(\bX)$ that is informative as to the relationship between $\bX$ and $Y(t)$, we might be able to correctly conclude that $i$ should be matched to $j$ instead of $k$.
Specifying a useful map $\phi$ can lead to great improvement in match quality over simple matches on raw covariate values. This idea has already been explored in the existing literature on matching, and  popular examples of $\phi$ that have been used previously include the propensity score for covariates $\bx$ and a desired treatment level, $t$, $\bx$: $\phi(\bx) = \Pr(T=t|\bX=\bx)$ \citep{rosenbaum1983central}, the prognostic score, or expected potential outcome under control: $\phi(\bx) = \E[Y(0)|\bX=\bx]$ \citep{hansen2008prognostic} (assuming that $t=0$ represents the control condition), and the Mahalanobis distance: $\phi(\bx) = \mathbf{M}\bx$, with $\mathbf{M}$ being a diagonal $p\times p$ matrix of positive weights \citep{rubin1980bias,diamond2013genetic,parikh2018malts}. 
Once we have formulated a representation of the covariates that we wish to use for matching, we will need a distance metric that encodes similarity of units on the transformed covariates. We employ a general $L_q$ norm for this purpose. For any $(\bu, \bv) \in \X \times \X$, let: $\Dphiq(\bu, \bv)  = \left(\sum_{j=1}^d|\phi(\bu)_j - \phi(\bv)_j|^q\right)^{\frac{1}{q}}, \label{eq:distdef}$
be the $q$-norm distance between $\bu$ and $\bv$. $q$ can be any integer or infinity, which we define as usual as $D_\phi^\infty(\bu, \bv) = \max_{j=1, \dots, d}|\phi(\bu)_j - \phi(\bv)_j|$. In practice, the most popular choices of norm are $q=2$ \citep[e.g.,][]{rubin1980bias, abadie2011bias, diamond2013genetic}, for matching on the covariates themselves on the $L_2$ distance, absolute value distance (which is any $L_q$ in 1D) \citep[e.g.,][]{rosenbaum1983central}, for matching on 1-dimensional representations of the covariates, such as the propensity score, and $q=\infty$ \citep[e.g.,][]{rosenbaum1984reducing, iacus2012causal}, for coarsening-based matching methods. Our main algorithm is as follows:


\begin{mdframed}
    \begin{center} \textbf{Algorithm: Matched Machine Learning (M-ML)}\end{center}
\noindent \textbf{Input:} A dataset of $n$ observations $\mathcal{D} = \{\bx_i, y_i, t_i\}_{i=1}^n$ split into a training set and a matching set. A desired covariate value $\bx$, desired treatment level $t$, a matching set of units, and a separate training set, a positive, real-valued caliper, $\Gamman > 0$.\\
\noindent \textbf{Output:} Estimator of conditional response function for covariate value $\bx$ and treatment level $t$.\\
\noindent\textbf{Stage 1:} Using the separate training set, construct an estimator of the representation $\phi(\cdot)$, denoted $\phihat(\cdot)$. Calculate $\phihat(\bx_i)$ for all units $i$ in the matching set as well as the input values: $\phihat(\bx)$. \\
\textbf{Stage 2:} Form the Matched Group by choosing units in the matching set that have the desired treatment level and are at a distance less than $\Gamman$ from $\bx$:
\begin{align}
        \MGxt = \left\{i = 1, \dots, n: \; \Dphiqhat(\bx, \bx_i) \leq \Gamman,\; T_i = t\right\}. \label{eq:MGxtdef}
\end{align}
\textbf{Stage 3:} Construct the estimator:
\begin{align}
    \muxthat &= \frac{1}{|\MGxt|}\sum_{i \in \MGxt} y_i. \label{eq:muhatdef}
\end{align}
\end{mdframed}
In Stage 1, we split the data into a training and matching set, and learn an estimator the function $\phi$ on the training set, and use it to estimate $\phi(\bx_i)$ for every unit in the matching set. In Stage 2, we construct our matches by maximizing the weighted sum of units with distance less than some generalized caliper, $\Gamman$. 
In Stage 3, we use the matched group constructed at Stage 2 to estimate the CRF for $\bx$. We choose to control the size of the matched group and who gets included with a constraint on the representation distance defined in \eqref{eq:distdef}: units with a distance from our target value less than $\Gamman$ are included in the matched group. The way in which $\Gamman$ is defined results in two variants of the M-ML algorithm:\\
\textbf{Caliper M-ML}: In this case, $\Gamman$ is defined to be some positive, real value chosen by the analyst. This is the technique known as caliper matching \citep[e.g.,][]{rosenbaum1985constructing}. This technique guarantees that the distance between units within a matched group will never be less than a known value. Note that in Caliper M-ML, it could happen that there are no units in $\MGxt$: in this case, the estimator cannot be constructed; we should use a larger value of $\Gamman$ so that matches can be constructed.\\
\textbf{KNN M-ML}: In this case, $\Gamman = \Dphiqhat(\bx, \bx_{(k_n)})$, where $(k_n)$ is the $k_n^{th}$ order statistic of the vector $(\Dphiqhat(\bx, \bx_1), \dots, \Dphiqhat(\bx, \bx_n))$. This technique is known as K-Nearest-Neighbor matching \citep[e.g.,][]{rubin1976multivariate}. This method guarantees that each matched group will contain exactly $k_n$ units (assuming no ties), however, in this case the caliper $\Gamman$ will be a function of $\bx$, and of the data, which implies that users will not be able to directly control its value. \\
We will show in the next section that these two variants of the M-ML algorithm are essentially equivalent asymptotically, under appropriate choices of $\Gamman$ and $k_n$. \\
In Stage 3, a canonical matching estimators for $\muxd$ is constructed with matches made at the prior step. An estimator for the CATE, $\tau(\bx, t, t')$ can be intuitively constructed by running the M-ML algorithm twice: once with $\bx$ and $t$ as inputs, obtaining $\mudxhat$ as output, and once with $\bx$, $t'$ as inputs, obtaining $\hat{\mu}(\bx, t')$ as output; and finally by taking the difference between the two: $\hat{\tau}(\bx, t, t') = \mudxhat - \hat{\mu}(\bx, t').$ We will see in an upcoming section that this estimator shares the desirable asymptotic properties of $\mudxhat$ constructed with M-ML. 

\section{Matched Machine Learning:  Asymptotic Properties} \label{sec:asymptotics}
We now study the statistical behavior of our M-ML estimator. This is important because doing so will allow us not only to give bounds on estimation error as $n$ grows, but also because it will allow us to construct approximate confidence intervals for our CATE and ATE/ATT estimates. Quantifying uncertainty in this way is paramount in virtually all scientific application of matching, and is necessary to improve the trustworthiness of matched estimates. Since M-ML is nonparametric, we focus on establishing results concerning its asymptotic properties, and study finite-sample behavior empirically via simulations. We will show in this section that CRF and CATE estimates for any $\bx$ in the domain of the distribution of our data can be estimated consistently and efficiently at the nonparametrically optimal rate in the sense of \citet{stone1982optimal}. We will see in Section \ref{sec:DML} that achieving this rate allows us to use Double Machine Learning methods with M-ML estimates as inputs to obtain root-$n$ asymptotically normal estimates of ARF, ATE and ATT. It follows from these results that conventional sample variance estimators applied to $\MGxt$ are consistent for  $\V[Y(t)|\bX=\bx]$, and that, therefore, approximate confidence intervals can be constructed for M-ML CRF and CATE estimates. We will see that letting either the caliper $\Gamman$ shrink or the number of matches $k_n$ grow as $n$ grows is fundamental to achieve consistency for M-ML estimates. Additionally, one key intuition behind our results is that the first-stage estimates do not affect the asymptotic behavior of the matching estimators, only its convergence rate, and the matching estimator can be understood asymptotically as if matches were made on the true value of $\phi$, as we will show. All proofs for the results below are available in the supplement. They key assumption is that the following condition holds with respect to $\Dphiq$: \\
\noindent\textbf{A5 (Lipschitz Condition)}: For all $\bx, \bz \in \X$ and $t \in \{1,\dots, M\}$ there exists a constant $C_L$ such that: (a) $|\mu(\bx, t) - \mu(\bz, t)| \leq C_L\Dphiq(\bx, \bz)$, and (b) $|\sigmasq(\bx, t) - \sigmasq(\bz, t)| \leq C_L\Dphiq(\bx, \bz)$.\\
This (or a similar) smoothness condition on the outcome function is a common assumption in virtually all nonparametric estimation frameworks similar to matching \citep[e.g.,][]{kallus2020generalized, farrell2021deep,wager2018estimation}, but the key difference here is that we would like it to hold for our \textit{transformed} covariates, but not necessarily on the \textit{raw} covariates. Assuming smoothness on the raw covariates, as is commonly done in matching and nonparametric methods, is a much stronger assumption: it would directly imply that the condition is also respected for the transformed covariates, as long as $\phi$ is Lipschitz-continuous in the covariate values, which is a simple and widely-satisfied requirement for many choices of $\phi$.

The final component of our framework is a flexible ML method to estimate $\phi$ from the data. To this end, we introduce a separate \textit{training} set, of size $\rho n$, for some fraction $\rho \in (0, 1)$. This training set can be obtained by randomly subsetting the whole data into two sets: a training set, and a matching set. For notational simplicity, we will assume that the total number of units is $n + \rho n$. We then assume that a ML method will be applied to the training data, to construct an estimator of $\phi$ denoted by $\phihat$. In practice $\phi$ can be modeled as the minimizer of some population loss function over a space of functions, and $\phihat$ as its empirical counterpart, but this need not always be the case. The requirement that we will need on $\phihat$ for our theoretical results to hold is that $\phihat$ is a consistent estimator of $\phi$, as well as both being differentiable functions. 
In order to establish the asymptotic properties of M-ML estimates, we need to make one additional but reasonable assumption for the first-stage distance metric estimates:\\
\noindent\textbf{A6 (Representation function)}: There exists a function $\phihat(\bx, \On): \mathbb{X} \times \Omega \mapsto \R^d$ such that, for real-valued $r_{ML} > 0$:  (a) The functions $\phihat(\bx, \On)$ and $\phi(\bx)$ are $f_\bO$-almost surely continuous with respect to $\bx$ at all $\bx \in \mathbb{X}$, (b) $\|\phihat(\bx, \On) - \phi(\bx) \|_{\cP, q} = o(n^{-r_{ML}})$ almost surely over $\fbX$, (c) $\|\phihat(\bX, \On) - \phi(\bX) \|_{\cP, q} = o(n^{-r_{ML}})$.\\
Part a) of this assumption limits potential representation to continuously differentiable functions of $\bx$, which is a common property of most ML algorithms. Part b) of this assumption states that the process used for learning $\phihat$ must lead to a quantity with a proper distribution for all possible inputs. Part c) is satisfied as long as $\phihat$ is learned from an independent training sample. Finally, part d) of this assumption states that first-stage ML estimates of $\phi$ must converge to the true value of $\phi$ both point-wise and in mean square. This assumption is relatively standard in nonparametric two-stage estimation settings \citep{chernozhukov2018double}, and has been verified for a number of different ML methods such as LASSO \citep{belloni2014pivotal}, Random Forests \citep{wager2018estimation}, Support Vector Machines \citep{devroye2013probabilistic}, and Deep Neural Networks \citep{farrell2021deep}, which are all also shown to converge at a rate $r \geq 1/4$ under relatively mild assumptions. 


Consider a unit outside of the data, which has observed covariates $\bx$. The probability that any of the units in our data are a match for $\bx$ is given in the following lemma.
\begin{lemma}\label{thm:iinmg}
Let A1-A6 hold. For $\bx \in \X$, let matches be made with Caliper M-ML for a fixed $\Gamman \geq 0$, i.e.: $\MGxt = \{i = 1, \dots, n:\, \Dphiqhat(\bX_i, \bx) \leq \Gamman, T_i=t\}$. Then, if $\Gamman \rightarrow 0$ as $n \rightarrow \infty$ we have, for arbitrary $i \in 1,\dots,n$:
\begin{enumerate}
    \item $\Gamman^{-d}\Pr_{\On, \bX_i, T_i}(i \in \MGxt) \rightarrow V_de(t)f_{\phi(\bX)|T=t}(\phi(\bx))$ for all $\bx \in \X$,
    \item $\Gamman^{-d}\Pr_{\bX, \On, \bX_i, T_i}(i \in \MGXt) \rightarrow V_de(t)\E_\bX[f_{\phi(\bX)|T=t}(\phi(\bX))],$
\end{enumerate}
where $e(t) = \Pr(T=t)$, $V_d = \frac{2Ga(\frac{2}{q} + 1)^d}{Ga(\frac{d}{q} + 1)}$, where $Ga$ is the Gamma function, and $f_{\phi(\bX)|T=t}(\phi(\bx))$ is the pdf of $\phi(\bX)$ conditional on $T=t$.
\end{lemma}
Note that the first statement concerns almost sure convergence over $f_{\bX}$, while the second concerns convergence in expectation over the same distribution. The above result is both intuitive and interesting: its proof does not rely on the ML convergence rate $r$ at all, but instead takes advantage of continuity of $\phihat$, together with a generalized change of variables to establish the result. The theorem shows that $\Pr(i \in \MGxt)$ is asyptotically proportional to $\Gamman^d$: this has the important consequence that, asymptotically, the rate at which our caliper $\Gamman$ contracts is the only relevant one for the probability of matching any unit $i$, and that the dimensionality of $\phi$, $d$, will influence this rate regardless of the original number of covariates. 

After this result is established, it can be used to prove the main theoretical result for CRF estimates obtained with M-ML, using the constants defined in the assumptions A4 and A5:
\begin{theorem}{(Asymptotic Behavior of Caliper M-ML CRF Estimates)}\label{thm:CRFasym}\\
Let A1-A6 hold, with A5 holding for some $\delta > 0$. Let $e(t)$, $V_d$, and $f_{\phi(\bX)|T=t}(\phi(\bx))$ be defined as in Lemma \ref{thm:iinmg}. Let $r = \min\left(\frac{1}{2 + d}, r_{ML}\right)$. 
For matches made with Caliper M-ML we have:\\
(i) For caliper $\Gamman = Kn^{\frac{2r-1}{d}}$: $n^r(\mudxtilde - \muxd) \dconv \mathcal{N}\left(0, \frac{\sigmasqxt}{K^dV_de(t)f_{\phi(\bX)|T=t}(\phi(\bx))}\right).$\\
(ii) For caliper $\Gamman = n^{-\frac{1}{2 + d}}$ and $s = 2 + \delta$: $\|\mudXtilde - \muXd\|_{\cP, s} = O(n^{-\frac{1}{2 + d}}) + o(n^{-r_{ML}})$,
and this bound is minimal over all possible values of $\Gamman$.
\end{theorem}
The same result holds for KNN M-ML: the following theorem establishes that, under suitable conditions on $k_n$, M-ML CRF estimates made with this methodology are asymptotically equivalent to estimates made by controlling $\Gamman$ directly. 
\begin{theorem} (Asymptotic Behavior of KNN M-ML CRF estimates) \label{thm:knnasym}\\
Let A1-A6 hold. Let $\MGxt = \{i = 1, \dots, n:\, \Dphiqhat \leq \Gamman,\, T_i=t\}$, with  $\Gamman$ equal to the $k_n^{th}$ order statistic of the vector $(\Dphiqhat(\bx, \bX_i), \dots, \Dphiqhat(\bx, \bX_n))$ (note that in this case $\Gamman$ is a function of $\bx$ and $\bX_1, \dotsm \bX_n$) with $k_n$ being a positive integer. Let $r = \min\left(\frac{1}{2 + d}, r_{ML}\right)$. We have:\\
(i) For $k_n = \lfloor K n^{2r}\rfloor$, for a positive constant $K >0$:
$n^r(\muhat(\bx, t) - \muxd) \dconv \mathcal{N}\left(0, \frac{\sigmasqxt}{K}\right).$\\
(ii) For $k_n = \lfloor n^{\frac{2}{2 + d}}\rfloor $ and integer $s > 2$: 
$\|\muhat(\bx, t) - \mu(\bx, t)\|_{\cP, s} = O(n^{-\frac{1}{2 + d}}) + o(n^{-r_{ML}})$,
and this bound is minimal over all possible values of $k_n$. 
\end{theorem}
Lemma \ref{thm:iinmg}, which gives us a way to establish the asymptotic order of the size of the matched group, $|\MGxt|$, in Theorem \ref{thm:CRFasym} and of $\Gamman$ in Theorem \ref{thm:knnasym}, is of fundamental importance in the proof of both theorems. An important implication of both theorems is that the choice of $\Gamman$ and $k_n$ directly impacts the convergence rate and asymptotic variance of M-ML estimates. In the case of Caliper M-ML, using the same reasoning as in the proof of Theorem \ref{thm:CRFasym}, the asymptotic variance can be made equal to 1 by setting $\Gamman = \left(n^{r-1}\frac{K^dV_de(t)f_{\phi(\bX)|T=t}(\phi(\bx))}{\sigmasqxt}\right)^{\frac{1}{p}}$, and it can be made arbitrarily small by multiplying the above by a positive constant. Obviously, this decrease in variance is paid for by an increase in bias due to the larger matched group, making it impractical to achieve an asymptotic variance lower than 1 in most cases. An analogous relationship is true for  KNN M-ML:  here one option to choose $k_n$ is to set it to the integer closest to $n^rK$. In this case, when we choose $K = 1$, the asymptotic variance is exactly $\sigmaxd$, however, depending on the data, there might not be $n^r$ high quality matches for $\bx$, and $K < 1$ might have to be chosen, leading to larger asymptotic variance. We note that our results on asymptotic normality of KNN regression generalize 
those of \citet{stute1984asymptotic}
by incorporating a distance metric learning step.

The bounds  given in Theorems \ref{thm:CRFasym} and \ref{thm:knnasym} have two important consequences. First, that the optimal convergence rates of KNN matching and caliper matching are the same, determining the asymptotic equivalence of these two long-standing matching procedures. 
Second, our bounds are directly related to the results on the convergence rates for KNN classification and regression on matches made on the $L_q$ distance of the raw covariates established in various works and summarized by \cite{gyorfi1981rate, gyorfi2002distribution}. Notably, the main difference between our bound and other bounds on non-transformed matching have a difference of a factor of $o(n^{-r_{ML}})$, which is due to the covariate transformation having to be learned in our case. This fact has an important consequence for our setting: gains in performance due to learning a distance metric, rather than matching on raw values of the covariates, can only be made in finite samples, rather than asymptotically. This conclusion is supported by recent work of \cite{rimanic2020convergence}, who derive a similar bound for KNN regression but under different assumptions on the representation function $\phi$. We will show in our simulations section that these finite-sample gains are substantial. Even more importantly, the bound implies that \textit{greatly improving the predictive accuracy of matching by adding a distance-learning step via ML comes at almost no cost in terms of convergence rate}, since the matching portion of the error bound decreases at its nonparametric rate.


Let us move on to discuss the asymptotic variance of the potential outcomes, $\sigmaxd$, which is of importance for asymptotic inference. This quantity can also be estimated via M-ML by applying the traditional variance estimator to the matched groups constructed with M-ML. This is stated formally in the following theorem:
\begin{theorem} (Consistency of Sample Variance Estimator)\label{thm:varasym}
Let the sample variance estimator for $\sigmasqxt$ be defined as: $\sigmasqxthat = \frac{1}{\Nxt}\sum_{i\in \MGxt} (Y_i - \muxthat)^2.$
Let A1-A6 hold, and let $\MGxt$ be constructed either with Caliper M-ML or KNN M-ML. Then we have: $\sigmasqxthat \pconv \sigmasqxt$ for all $t$ as $n \rightarrow \infty$. 
\end{theorem}
Note that the result holds independently of whether the number of units to match to $\bx$ is chosen or whether $\Gamman$ is chosen to control the radius of $\MGxt$. 

Finally, the theorems just introduced have the following direct consequence as a corollary, which permits us to construct asymptotic confidence intervals for the CATE. 
\begin{corollary} (Asymptotic Normality of CATE  estimates) \label{thm:CATEasym}
Let A1-A6 hold, and let let $r = \min\left(\frac{1}{2 + d}, r_{ML}\right)$. For two treatment levels $t, t' \in \{1, \dots, M\}$ and a real $K > 0$: as $n \rightarrow \infty$:

\noindent (i) If matches are made with Caliper M-ML and $\Gamman = Kn^{\frac{2r-1}{p}}$:
    $n^{r}(\tauhat(\bx, t, t') - \tau(\bx, t, t')) \dconv \mathcal{N}\left(0, \frac{\sigmasqxt}{K^dV_de(t)f_{\phi(\bX)|T=t}(\phi(\bx))} + \frac{\sigma^2(\bx, t')}{K^dV_de(t')f_{\phi(\bX)|T=t'}(\phi(\bx))}\right).$

\noindent (ii) If matches are made with KNN M-ML and $k_n = \lfloor n^{2r} K \rfloor $:  $n^{r}(\tauhat(\bx, t, t') - \tau(\bx, t, t')) \dconv \mathcal{N}\left(0, \frac{\sigmasqxt}{K} + \frac{\sigma^2(\bx, t')}{K}\right).$
\end{corollary}
This corollary is possible because only units with observed treatment $T_i=t$ are used to construct $\muxthat$, and only units with $T_i=t'$ are used to construct $\muhat(\bx, t')$: this renders the two estimators independent of each other. With the two estimators independent of one another, the continuous mapping theorem applied to the vector $(\mudxhat, \muhat(\bx, t'))$ allows the conclusion in the corollary to be reached. The result in the corollary implies that an approximate $1-\alpha$ confidence interval can be constructed for a fixed $\bx$ with:$CI(\bx, t, t') = \left[\tauhat(\bx, t, t') \pm \Phi^{-1}\left(1-\alpha/2\right)\sqrt{\frac{\sigmasqxthat}{c(\bx, t)} + \frac{\hat{\sigma}^2(\bx, t')}{c(\bx, t')}}\right]$,
where $c(\bx, t) = n^rK^dV_de(t)f_{\phi(\bX)|T=t}(\phi(\bx))$ for caliper matches (Thm. \ref{thm:CATEasym}), or $c(\bx, t) = k_n$ for KNN matches (Thm. \ref{thm:knnasym}). This will permit analysts to quantify the uncertainty around their CATE estimates in a way that is widely accepted, and has known guarantees. 

\section{Matched Double Machine Learning For Average Estimands}\label{sec:DML}
The M-ML framework can naturally be extended to nonparametric estimation of ATE and ATT in a way that maintains the auditability of matching, as well as the ability to construct asymptotically valid confidence intervals. This is possible because M-ML estimates can be used as input for first-stage estimates in Augmented Inverted Propensity Weighted (AIPW) estimators \citep{robins1994estimation}. These estimators have been found to behave normally asymptotically and with known variance, provided that first-stage estimates of CRF and propensity converge sufficiently fast \citep{chernozhukov2018double, van2006targeted}, which is a property that our methods have, as shown in the analysis above. 

Thanks to this property, we can formulate a version of the M-ML algorithm for ATE and ATT estimation modeled after the DML algorithm of \cite{chernozhukov2018double}. The algorithm we propose is called Matched Double Machine Learning (M-DML), and is defined as follows:
\begin{mdframed}
    \begin{center} \textbf{ATE or ATT Estimation: Matched Double Machine Learning (M-DML)}\end{center}

\noindent\textbf{Stage 1:} Randomly split the data $\{\bx_i, y_i, t_i\}_{i=1}^n$ into $L$ folds each of size $n/L$. Let $S_\ell$ denote all the indices of units in fold $\ell$, and $S_{\setminus \ell}$ all the indices of units not in that fold. The units in fold $\ell$ are used for estimating the ATE and ATT.

\noindent Repeat the following steps for $\ell=1,\dots, L$.\\
\textbf{Stage 2: } Further split the units in $S_{\setminus \ell}$ into a training set, denoted by $\mathcal{TS}_\ell$ and a matching set, denoted by $\mathcal{MS}_\ell$. Using only the units in $\mathcal{TS}_\ell$, learn the representation function $\phihat(\bx)$. Using all units in $S_{\setminus \ell}$, construct a consistent estimator of the propensity for each treatment level $t$, denoted by $\ehat(\bx, t)$, and of the marginal propensity for receiving treatment $t$: $\Pr(T=t)$, the latter denoted by $\hat{e}(t)$.\\
\textbf{Stage 3:} For each unit $i \in S_\ell$, and for treatment levels $t,t'$, run the M-ML algorithm for each treatment level, with all the units in $\mathcal{MS}_\ell$ as candidates for matching to obtain $\muhat(\bx_i, t), \muhat(\bx_i, t')$. Additionally predict the propensity score of $i$ for each treatment level, $\ehat(\bx_i, t)$, using the propensity models learned in Stage 2.\\
\textbf{Stage 3:} For each unit $i \in S_\ell$, using the outputs of the previous stage, construct the doubly robust score function:
$\psihat(\bx_i, t) = \muhat(\bx_i, t) + \frac{\ind[t_i=t](y_i - \muhat(\bx_i, t))}{\ehat(\bx_i, t)} \label{eq:psiate}$, and compute $\psihat(\bx_i, t')$ analogously, then compute:
$\psihat(\bx_i, t, t') =  \frac{\ind[t_i=t](y_i - \muhat(\bx_i, t'))}{\ehat(t)} - \frac{\ind[t_i=t']\ehat(\bx_i, t)(y_i - \muhat(\bx_i, t'))}{\ehat(t)\ehat(\bx_i, t')}.\label{eq:psiatt}$\\
\textbf{Stage 4:} Construct the estimators: $\muhatDR_\ell(t) = \frac{L}{n}\sum_{i \in S_\ell} \psihat(\bx_i, t)$, for both t,t',  $\tauhatDRATE(t,t')_\ell =  \muhatDR_\ell(t) - \muhatDR_\ell(t')\label{eq:atehat}$, $\tauhatDRATT(t, t')_\ell = \frac{L}{n}\sum_{i \in S_\ell} \psihat(\bx_i, t, t').\label{eq:atthat}$\\
\textbf{Stage 5:} Average across folds: 
$\hat{\mu}^{\textit{DR}}(t) = \frac{1}{L} \sum_{\ell}\muhatDR_\ell(t)$,  $\hat{\tau}^{\textit{DR}}_{\textit{ATE}}(t,t') = \frac{1}{L} \sum_{\ell}\tauhatDRATE(t,t')_\ell$, and $\hat{\tau}^{\textit{DR}}_{\textit{ATT}}(t,t') = \frac{1}{L} \sum_{\ell}\tauhatDRATT(t, t')_\ell$. These three quantities are the output of M-DML.
\end{mdframed}
The algorithm works by splitting the data into $L$ folds (Stage 1), which are then further split into training and matching set. Note that this is a 3-way data split, which is intuitively required by the matching step added on top of the ML prediction step, a similar splitting procedure is also used in \citep{wang2021flame}. The M-ML algorithm is then fit separately to each fold (Stage 2) and results are averaged to obtain a final estimate of the parameter of interest (Stage 5). Note that if one has enough data, then one could use only a single held out training split and eliminate the need to further split units outside fold $\ell$ into training and matching sets. 
This avoids increasing the complexity of cross-fitting, which could be computationally expensive for large datasets.


Given two treatment levels of interest to the user, $t$ and $t'$, the M-ML algorithm is run twice on each fold, once for treatment level $t$ and once for $t'$. The estimators constructed using $\muhat(\bx, t)$ output by M-ML are given in Equations \eqref{eq:psiate} and \eqref{eq:psiatt} and they are the doubly robust score functions, used in the AIPW estimators of \cite{robins1994estimation}. The idea of these estimators is to use the observed data to correct the first-stage predictions, and thus ensure asymptotic normality of estimates. As shown in the algorithm, one needs a consistent estimate of the propensity score, $e(\bx_i, t)$ for all units, $i$, and treatment levels of interest. Many good estimators exist for this quantity, and the use of any estimator based on a ML method that satisfies the requirements on convergence rates given in Theorem \ref{thm:DML} will result in the asymptotic guarantees on M-DML given in the theorem. 

The M-DML algorithm is a special case of the DML2 algorithm given in Definition 3.2 of \cite{chernozhukov2018double}. To match M-DML to that definition, M-ML is taken to be the first stage estimator in the definition. The properties of the AIPW estimators, together with the convergence rates of M-ML give us guarantees on the asymptotic normality and convergence rate of M-DML. This is stated in the following theorem. 
\begin{theorem}\label{thm:DML}
Let the observed outcome and treatment be defined as follows: $Y = \mu(\bX, D) + U,\quad \E[U|\bX, D] = 0$, and $D = e(\bX, D) + V,\quad \E[V|\bX] = 0$. Let A1-A6 hold, and assume further that the user has chosen a propensity score estimator that satisfies, for a positive, real $r_e$: (i) $\|\ehat(\bX, t) - e(\bX, t)\|_{\cP, 2} = O(n^{-r_e})$, (ii) $\|\ehat(\bX, t) - 1/2\|_{\cP, \infty} \leq 1/2 - \epsilon$, for some $\epsilon > 0$, (iii) $0 < \ehat(\bx, t) < 1$ for all $\bx$ and $t$. Let $r = \min\left(\frac{1}{2 + d}, r_{ML}\right)$ and assume that $r + r_e \geq 1/2$. Let $\MGXt$ be constructed either with Caliper M-ML or with KNN M-ML with either $\Gamman = Kn^{\frac{2r -1}{d}}$, or $k_n = K n^{2r}$, for a fixed integer $K > 0$. Then the following holds for M-DML estimates:\\
1) Asymptotic Normality: $\sqrt{n}(\muhatDR(t) - \mu(t)) \dconv \mathcal{N}(0, \sigma^2(t))$, with $\sigma^2(t) = \E_\bX[\psi(\bX, t)^2]$, $\sqrt{n}(\tauhatDRATE(t, t') - \tau(t, t')) \dconv \mathcal{N}(0, \sigmasq_{ATE}(t, t'))$, with $\sigmasq_{ATE}(t, t') = \E_\bX[(\psi(\bX, t) - \psi(\bX, t'))^2]$, and $\sqrt{n}(\tauhatDRATT(t, t') - \tau(t, t')) \dconv \mathcal{N}(0, \sigmasq_{ATT}(t, t'))$, with $\sigmasq_{ATT}(t, t') = \E_\bX[\psi(\bX, t, t')^2]$.\\
2) Consistency of the sample variance estimators applied to the second-stage estimators, i.e.: $\hat{\sigma}^2(t) = \frac{1}{n}\sum_{i=1}^n (\psihat(\bX_i, t) - \muhatDR(t))^2 \pconv \sigmasq(t)$, $\hat{\sigma}^2_{ATE}(t,t') = \frac{1}{n}\sum_{i=1}^n (\psihat(\bX_i, t) - \psihat(\bX_i, t') - \tauhatDRATE(t, t'))^2 \pconv \sigmasq_{ATE}(t, t')$, and, $\hat{\sigma}^2_{ATT}(t,t') = \frac{1}{n}\sum_{i=1}^n (\psihat(\bX_i, t, t') - \tauhatDRATT(t, t'))^2 \pconv \sigmasq_{ATT}(t, t').$
3) Approximate confidence intervals:\\
Let $\widehat{\Delta}$ be a M-DML estimator from Stage 5 of the M-DML algorithm, $\Delta$ its corresponding estimand, and $\sigmasqhat$ its respective asymptotic variance. An approximate $1 - \alpha$ confidence interval for the parameter of interest is: $CI(\delta) = \left[\widehat{\Delta} \pm \Phi^{-1}\left(1 - \frac{\alpha}{2}\right)\sqrt{\frac{\sigmasqhat}{n}}\right]$,
where $\Phi^{-1}(a)$ is the $a^{th}$ quantile of the standard normal distribution. 
\end{theorem}
The statement follows almost directly from our Theorems \ref{thm:CATEasym}, \ref{thm:knnasym} and Theorem 5.1 in \cite{chernozhukov2018double}. This result establishes asymptotic normality at a $\sqrt{n}$ rate for ATE and ATT estimates obtained with M-DML. This is of primary importance because it enables us to approximate confidence intervals on our average parameters of interest with the asymptotic distribution of our estimators, thus providing the uncertainty quantification that is needed for causal inference. Note that, if the same ML estimator is used for both $\phihat$ and $\ehat$, then the requirement on its rate becomes: $r_{ML} \geq \frac{1}{4}$, which is the same rough requirement given in \cite{chernozhukov2018double}. Note that this rate can be achieved by M-ML and first stage methods under the condition that $\beta \geq \frac{p}{2}$, i.e., if outcomes are smooth enough as a function of the covariates, which is a common requirement in nonparametric estimation frameworks. In order to achieve the rate needed it is also important to choose the dimensionality of the representation $\phi$ in such a way that the condition $\frac{1}{2 + d} + r_{e} \geq \frac{1}{2}$ holds. This can be achieved by choosing $d \leq \frac{2r_{e}}{\frac{1}{2} - r_{e}}$. If $r_e$ is the optimal nonparametric convergence rate with a $\beta$-smooth propensity score and $p$ covariates \citep{stone1982optimal}, then the condition reduces to $d \leq \frac{4\beta}{p}$. 

\section{M-ML and M-DML: Examples and Extensions} \label{sec:extensions}
In this section, we present some practical examples of how M-ML and M-DML might be used, and give some case-specific considerations that apply to the algorithms in these settings. 
In addition, we present an extension to the M-ML algorithm that allows the user to add arbitrary constraints to the matching optimization problem at Stage 2 of the M-ML algorithm. 
\subsection{Matching to explain ML predictions} The most straightforward application of M-ML is to match on the estimated potential outcomes from a black-box ML model in order to audit the model's predictions with case-based reasoning. In the causal inference literature, matching on an estimate of $\muxt$ is known as prognostic score matching \citep{hansen2008prognostic}, which displays many of the properties of propensity score matching. In this case, we would define $\phi$ as $\min_h\E[\ell(Y(t), \bX, h)]$, where $\ell$ is a loss function. For regression problems, for example, one could use the loss $\ell(Y(t), \bX, h) = (Y(t) - h(\bX))^2$. It is easy to show that in this case $\phi(\bx) = \E[Y(t)|\bX=\bx] = \muxt$. Conventional ML methods will estimate $\phi$ by minimizing the empirical risk $\frac{1}{n}\sum_{i=1}^n \ell(Y_i(t), \bX_i, h)$ over a space of hypothesis functions, $\mathcal{H}$, which is potentially very large, complex, uninterpretable, and therefore almost impossible to audit. As argued before, matching will remedy this lack of interpretability by replacing the output of black box $h(\bx, t)$ with the average of the observed outcomes of nearby units; if we construct the distance metric well, those units will have similar predicted $Y(t)$ values. Importantly, in this case $d=1$, implying that the convergence rate for M-ML will be $\min(n^{\frac{1}{3}}, n^{r_{ML}})$, which will almost always equal $r_{ML}$, since it is very unlikely that any ML method may achieve a convergence rate greater than $1/3$, especially when $p>1$. This consequently implies that adding matching on top of ML for auditability comes at virtually no cost in terms of convergence rate of the ML predictions. Notably, in this case the dimensionality of $\phi$ will be exactly $d=1$. This implies that the rate of convergence for the matching portion of our estimators will be exactly equal to the nonparametrically optimal one (see \cite{stone1982optimal} as well as Sec. 6.3 of \cite{gyorfi2002distribution}). This implies that, in this case, the rate of convergence of the CATE M-ML estimator will be $r = r_{ML}$, i.e., the M-ML estimator will converge as fast as any backend ML estimator of $\phi$ can. This has the important implication that \textit{adding a layer of interpretability on top of ML predictions with matching comes at no cost in terms of convergence rates}.  

\subsection{Matching on the Mahalanobis distance with learned weights}
One potential use of M-ML is to construct covariate weights that describe the importance of each feature for outcome generation, and to then match on a weighted $L_2$ distance with those weights. This is already accomplished by several existing matching methods \citep{wang2021flame,parikh2018malts, diamond2013genetic}, each targeting a different set of weights that ensures different desirable properties of the matches. This setting can be expressed in the M-ML framework by setting $\phi(\bx) = \mathbf{M}\bx$, where $\mathbf{M}$ is a $p\times p$ diagonal matrix of weights that are either known or learned from the data. In this case, $\phi(\bx)$ is invertible, and the asymptotic probability of a match, $n^rK^dV_de(t)f_{\phi(\bX)|T=t}(\phi(\bx))$, used in our framework to estimate the asymptotic variance of caliper matches, becomes: $n^rK^dV_de(t)f_{\bX|T=t}(\bx)\prod_{j=1}^pm_j$, where $m_j$ is the true value of the weight on covariate $j$.

\subsection{M-ML as a General Matching Framework}
As the previous examples suggest, the M-ML algorithm can also be seen as a generalization of several other popular matching methods when matching is done with replacement, either with a caliper on the pair-wise distance of units to be matched, or with a fixed number of matches. We give a description of methods that are special cases of M-ML here, and a summary in Table \ref{tab:generalization}. Specifically, the following methods are special cases of the M-ML framework: Nearest Neighbors, Propensity Score Matching, Prognostic Score Matching, Prognostic Score Matching, Adaptive Hyperboxes, Coarsened Exact Matching, Genetic Matching, Genetic Matching, Matching After Learning to Stretch (MALTS), and Fine Balance. 

Let us describe in more detail how these are special cases of M-ML, starting with Mahalanobis distance matching, which includes MALTS and Genetic Matching.
Mahalanobis distance matching \citep{rubin1980bias} matches units that are close in terms of Mahalanobis distance, which is defined for two vectors $\bu, \bv \in \R^p$ and a square matrix $\mathbf{M} \in \R^{p\times p}$ as: $\textrm{Maha}(\bu, \bv,\mathbf{M}) = \sqrt{(\bu - \bv)^T\mathbf{M}(\bu - \bv)}$. $\mathbf{M}$ is usually chosen to be diagonal, and we will use diagonal $\mathbf{M}$. Mahalanobis matching can be implemented as M-ML by choosing $\phi(\bx) = \mathbf{M}\bx$ and then matching on the $L_2$ distance. Setting $\mathbf{M} = \mathbf{I}$, where $\mathbf{I}$ is the identity matrix in the same scenario will result in simple Nearest-Neighbor Matching, used, for example, by \citet{abadie2006large}. Methods like Genetic Matching \citep{diamond2013genetic} and MALTS \citep{parikh2018malts} perform Mahalanobis matching just as described, but they add a first step for learning an optimal $\mathbf{M}$, thus also taking advantage of the learning component of M-ML. For a version of Genetic Matching that learns its distance on a separate training set, this first step can be expressed as finding $\mathbf{M}$ that solves: $\mathbf{M} \in \argmin\limits_{\mathbf{M}\in \R_{diag}^{p\times p}}\sum\limits_{i,j \in \textrm{Tr}}\textrm{Maha}(\bx_i, \bx_j, \mathbf{M})$, where $\textrm{Tr}$ is a set of indices indicating which of the units belong to the training set. In the case of MALTS, $\mathbf{M}$ is found by optimizing: $\mathbf{M} \in \argmin\limits_{\mathbf{M}\in \mathbb{R}_{diag}^{p\times p}} \left|\sum\limits_{i,j \in \textrm{Tr}}\frac{(y_i - y_j)\ind(t_i=t_j)\exp(-\textrm{Maha}(\bx_i, \bx_j, \mathbf{M}))}{\sum_{k \in \textrm{Tr}} \ind(t_i=t_k)\exp(-\textrm{Maha}(\bx_i, \bx_k, \mathbf{M}))}\right|$. 

Methods that use predictors of treatment and outcomes and match on those can also be implemented as special cases of M-ML. For propensity score matching \citep{rosenbaum1985bias}, $\phi(\bx) = h(\bx)$ and $h$ is chosen to be the best predictor of treatment assignment within a class of functions, $\mathcal{H}$, i.e: $h \in \argmin_{h \in \mathcal{H}} \E_\bX[(h(\bX) - \Pr(T_i=t|\bX))^2]$. For prognostic score matching \citep{hansen2008prognostic}, the same is done, but with $h$ being the best predictor of the control outcome, denoted here by $t'$: $h \in \argmin_{h \in \mathcal{H}} \E_{\bX, Y(t')}[(h(\bX) - Y(t'))^2]$. A pre-trained version of the Adaptive Hyperboxes (AHB) matching algorithm \citep{morucci2020adaptive} can also be cast as a version of M-ML, by adopting the same form for $h$ as in prognostic score matching, but for both treatment levels of interest, $t, t'$, and constructing the vector: $\phi(\bx) = (h_t(\bx), h_{t'}(\bx))$, where $h_t \in \argmin_{h \in \mathcal{H}} \E_{\bX, Y(t)}[(h(\bX) - Y(t))^2]$, and $h_{t'}$ is defined in an analogous manner. In addition to this, AHB matches units in a hyperrectangular region of the covariate space that is learned from the data: while this is not directly achievable within M-ML, it is possible to pre-specify a rectangular region of the covariate space, $\mathbf{H}(\bx)$, and to add a constraint to the M-ML matched group that requires matched units to be within that region: $\MGxt = \{i = 1, \dots, n:\, t_i=t,\, \Dphiqhat(\bx_i, \bx) \leq \Gamman,\, \bx_i \in \mathbf{H}(\bx)\}$. After estimating $\phi$, all of these matching methods match using the $L_1$ distance, which could be chosen for M-ML as well.  
Coarsened Exact Matching \citep{iacus2012causal} can be implemented as M-ML by creating a diagonal matrix of dimension-wise calipers, $\tilde{\Gamma}$, where the diagonal is given by the vector $\left[\frac{1}{\gamma_1}, \dots, \frac{1}{\gamma_p}\right]$, and each value of $\gamma_j$ is the maximum allowed distance for two units on the $j^{th}$ dimension. To fully emulate CEM, matches should then be made on the sup norm, i.e., by setting $q=\infty$. In a similar vein, matching with near-fine balance on the $j^{th}$ covariate \citep{rosenbaum2007minimum} can be implemented as M-ML by adding a constraint to the matched group that takes the form $|x_{ij} - x_j|\leq \gamma_j$, where $\gamma_j$ is a small caliper on the $j^{th}$ dimension; fine balance can be achieved by setting $\gamma_j = 0$. Fine balance matches can be made with any choice of $\phi$ and $q$, but to emulate the implementation of \citet{rosenbaum2007minimum}, one would choose $\phi$ to be the propensity score, and set $q=1$.

Finally, analysts could be interested in creating optimal matched groups by solving a weighted version of an optimization problem, where the number of units to include in $\MGxt$ is chosen as the optimum of a weighted combination of cumulative distance and number of units. This is a type of bias-variance trade-off, as larger groups have lower variance, but higher bias since they include points that are farther away. The following lemma establishes that the matched group formulation defined in Eq. \eqref{eq:MGxtdef} is also an optimal solution to such a weighted problem:
\begin{lemma}{(M-ML is a solution to the weighted matching problem.)}\label{thm:mixproblem}
Let $\Wixt = \ind[i \in \MGxt]$ represent whether unit $i$ is included in $\MGxt$, then the matched group $\MGxt$, defined in \eqref{eq:MGxtdef}, is also a solution to the problem: 
 $\min_{W_1(\bx,t), \dots, W_n(\bx, t): \in \{0, 1\}^n} \sum_{i=1}^n \Dphiqhat(\bx, \bX_i) - \Gamman \sum_{i=1}^n \Wixt.$
\end{lemma}
The above formulation is used, for example, in \cite{morucci2020adaptive}, or in some of the optimization problems of \citet{zubizarreta2012using}, together with additional constraints on the matching problem. This lemma importantly shows that the M-ML framework incorporates matching algorithms that target combined optimization problems, implying that the asymptotic and empirical results we obtain for M-ML can also be extended to such methods. 

In conclusion, we have shown that our methodology and theoretical results apply generally to many existing matching algorithms, and this in turn enables easy computation of asymptotic confidence intervals for these algorithms. 

\subsection{Related Work}
The idea of matching on a learned function of the covariates has been previously explored in the literature on matching in various specialized and restricted settings. Of these settings, the first to emerge and to be extensively studied was Propensity Score Matching (PSM) \citep{rosenbaum1983central}. Of the various analyses of propensity score matching, the two closest to our setting are that of \cite{rubin1992affinely}, and that of \cite{abadie2016matching}. The former shows that theoretical guarantees on finite-sample bias and variance can be derived for matching methods when covariate transformations are affine, and outcomes have ellipsoidal distributions. The latter shows that propensity score matching does indeed exhibit efficient asymptotic behavior when propensity scores are linear and for average effects only. Other recent work that considers matching on a transformation of the covariates includes \cite{luo2020matching}, who consider matching on linear transformations of the covariates, and \cite{kallus2020generalized}, who proposes a method for estimation of the ATT that is $\sqrt{n}$-consistent without requiring additional nonparametric adjustments, and relying on a kernel mapping of the covariates. Our paper is more general in that our matching framework works with \textit{any} transformation of the covariates and require minimal distributional assumptions on the outcome data, as well as encompassing estimation of  both ATE/ATT and CATE.
Aside from tools for statistical inference, the literature on matching for treatment effect estimation has seen a proliferation of methods to make matches on the raw, untransformed values of the covariates \citep[e.g.,][]{iacus2012causal, diamond2013genetic, zubizarreta2012using}, but most of these methods focus on matching to optimize some aggregate metric of quality across units, and therefore perform poorly when estimating CATEs, unlike our proposed approach, and for the ATE/ATT they are prone to the issues of convergence outlined by \cite{abadie2006large}. That work shows that nearest-neighbor matching methods fail to attain the nominal, $\sqrt{n}$, convergence rate for the ATE/ATT, a problem important for our setting. Recent work by \cite{savje2022inconsistency}  shows that matching methods that match without replacement fail to attain this rate as well. We introduce a methodology based on recent results for efficient two-stage estimation \citep{chernozhukov2018double, van2006targeted} that allows our average estimates to be consistent at the nominal rate. Other methods to address this issue include work by \cite{abadie2011bias} and \citet{otsu2017bootstrap}, but both of those methods require combining matching with independent nonparametric estimates of outcome and propensity functions that do not involve matching and therefore render final estimates hard to audit. Another existing method that addresses this issue is that of \cite{wang2019Large}, who show that matching methods that target average balance, rather than nearest-neighbor balance per unit, can achieve the nominal rate. Our results and proposed framework differ in that it uses nearest-neighbor matches, which is both computationally faster than optimizing the full set of matches simultaneously (which is done by mixed-integer program), and enables direct estimation of unit-level treatment effects, unlike their method. There also exists a literature on matching methods for individualized treatment effect estimation that combines machine learning of distance functions with matching \citep{dieng2019interpretable, parikh2018malts, morucci2020adaptive, wang2021flame}: our paper aims to generalize all these methods under a single framework, and to provide users of these methods with a way to perform inference for their output estimates.  Finally, the literature on nonparametric CATE estimation is related to our work. This literature has mainly focused on powerful black-box methods \citep{chipman2010bart, wager2018estimation, farrell2021deep} whose predictions are not auditable by analysts and decision-makers, leading to substantially less trustworthy and potentially wrong results in many settings. Our method explicitly addresses this problem with matching.

\section{Simulations}\label{sec:simulations}

\begin{figure}[!htbp]
    \centering
    \caption{Estimation error}
    \label{fig:CATEerr}
    \includegraphics[width=\textwidth]{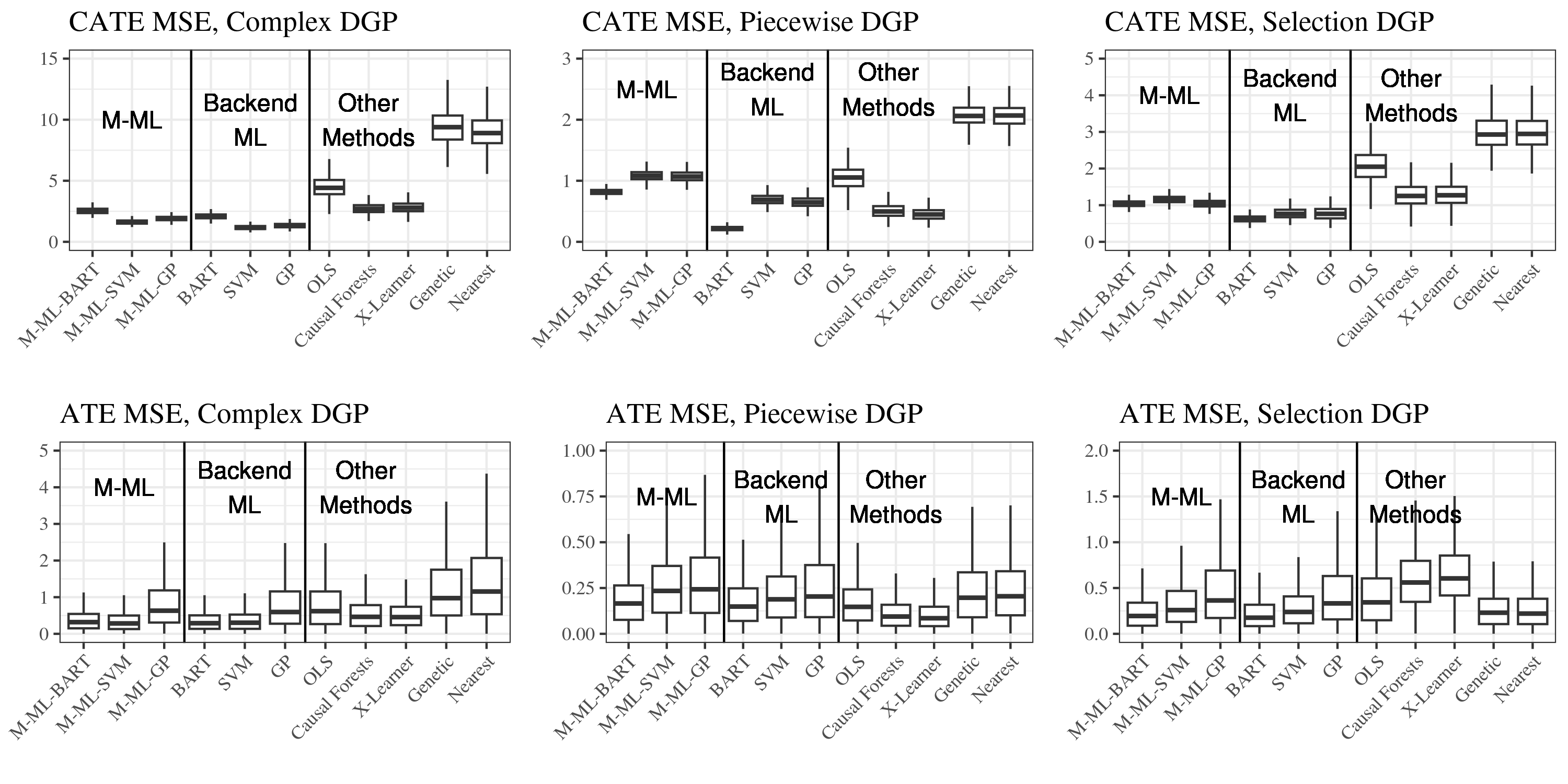}\\
    \footnotesize{Note: Top Row: CATE, Bottom Row: ATE. Different methods compared are on the horizontal axis, and the vertical axis is the mean absolute estimation error at each each iteration. Acronyms are described in Table 
    \ref{tab:simmethods}.}
\end{figure}

\begin{figure}[!htbp]
    \centering
    \caption{95\% Asymptotic Confidence Interval Coverage}
    \label{fig:coverage}
    \includegraphics[width=\textwidth]{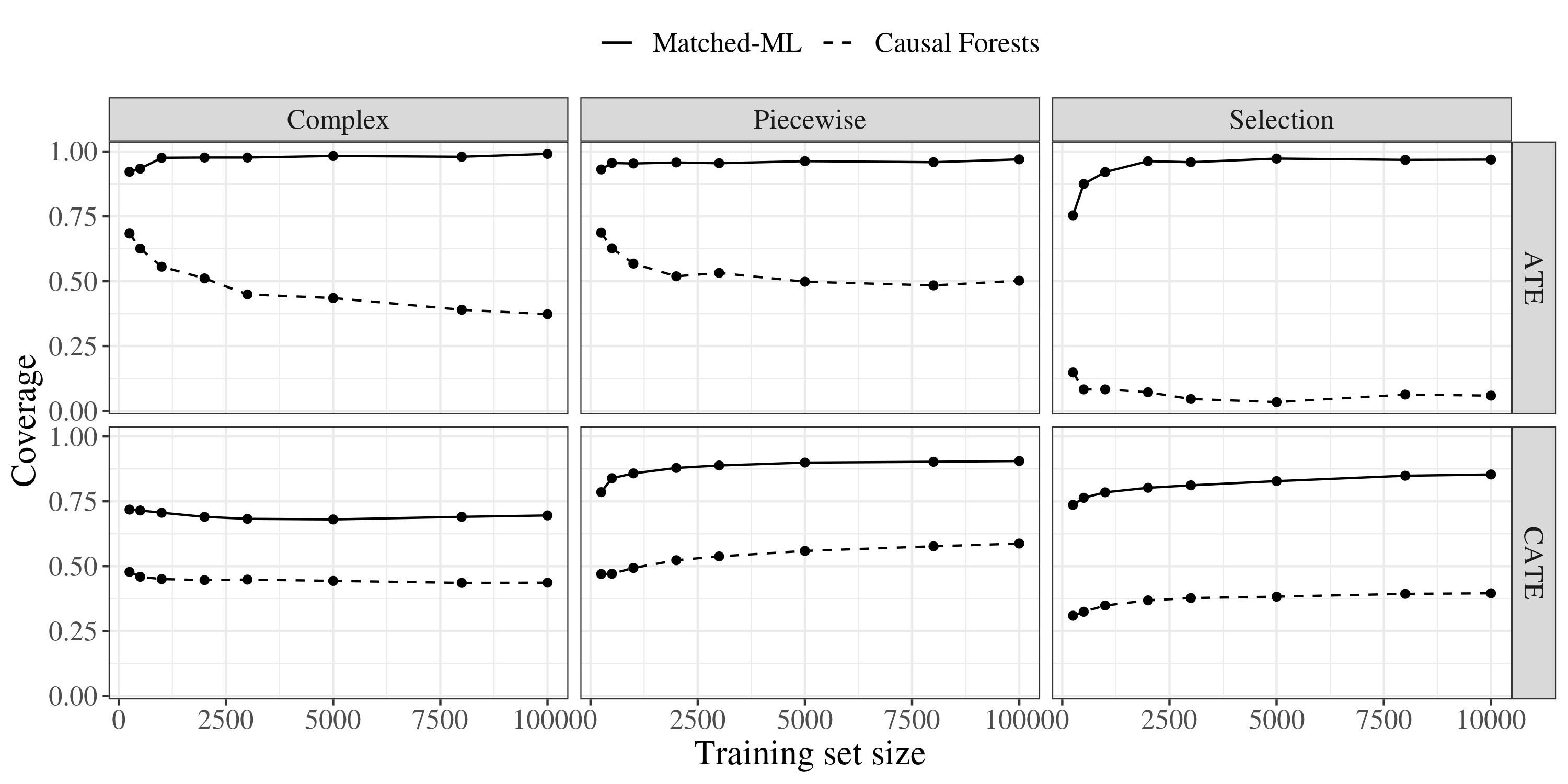}
\end{figure}

We present results from an empirical evaluation of the performance of M-ML for CATE and ATE estimation on several simulated datasets for which we know ground truth causal effects. As a setting, we focus on the application of M-ML as an auditing tool for ML by matching on outcome predictions made by black-box algorithms, as this is one of the most natural uses of M-ML. We show that, on average, M-ML performs comparably to black-box methods that it is based on, and in some settings even improves on their performance. 
Global to all simulations, we generate data for $n=20000$ units and $p=20$ covariates, where 5000 units are used for training and the remaining for matching/estimation. For $i=1,\dots,n$, we generate:
$\bX_{i} \sim \mathcal{N}_p(1,1),\; \sigma_i \sim \textrm{Uniform}(1, 2),\; \blambda \sim \textrm{Uniform}_p(-4, 4),\; \bbeta^{lin} = \blambda,\; \bbeta^{qua} \sim \textrm{Uniform}_p( 0, 1) + \blambda,\; \bbeta^{cos} \sim \textrm{Uniform}_p(0, 1) + \blambda,\; \bdelta \sim \textrm{Uniform}_p(-1, 1),\; \bdelta^{int} \sim \textrm{Uniform}_{p\times p}(-0.5, 0.5),\; \tau = 5,\; \alpha = 5,\; \epsilon_{i} \sim \mathcal{N}(0, \sigma_i$
where, for some vector $\bu \in \R^p$,  $\bu \sim f_p$ denotes a vector made up of $p$ draws from the same distribution, $f$. 
We then generate outcomes according to the following DGP, for two treatment levels $t=0, 1$: \\
\text{\textbf{Nonlinear:}} $Y_i(t) = \alpha + t\tau + \bX_i\bbeta^{lin} + \bX_i^2\bbeta^{qua} + \cos(\bX_i)\bbeta^{cos} + t\bX_i\bdelta + t\sum_{j=1}^p\sum_{k=1}^p X_{ij}X_{ik}\delta_{jk}^{int} + \epsilon_i$\\
\text{\textbf{Piecewise: }}$Y_i(t) = \alpha + t\tau + \sum_{j=1}^p\ind(X_{ij} > 0)\beta^{lin}_{j} + t\sum_{j=1}^p \ind(X_{ij} > 0)\delta_j + \epsilon_i$\\
\text{\textbf{Selection: }} $Y_i(t) = \alpha + t\tau + \sum_{j=1}^{10} X_j\beta^{lin}_j +t\sum_{j=1}^{10}X_j\delta_j + \epsilon_i.$\\
We choose the DGPs above because they simulate three settings that are complicated to deal with nonparametrically, but that may occur in applied scenarios. The nonlinear setting is one in which the outcome is a complex function of the covariates that may vary in unexpected ways, the piecewise setting is simpler, but each covariate is considered as a simple threshold, which adds a stepwise component to the function. Finally, the selection setting is also linear, but involves a variable selection component, as only 10 of the 20 simulated covariates are used to generate outcomes, and estimation methods need to be flexible enough to exclude or downweight the unused covariates to attain optimal results. 
Finally, we also generate propensity scores and treatment indicators with:
\begin{align*}
    u_i \sim \textrm{Uniform} (0.1, 1),\; e(\bX_i, 1) &= \frac{\exp(u_i(Y_i(1) + Y_i(0))/2 - \epsilon_i)}{1+\exp(u_i(Y_i(1) + Y_i(0))/2 - \epsilon_i)},\;  T_i \sim \textrm{Bernoulli}(e(\bX_i, 1)),
\end{align*}
note that the correlation between potential outcomes and treatment assignment is controlled by the random variable $u_i$, which we introduce to avoid fully correlated treatment and outcomes and potential overlap violations.  Our simulations will include BART, Gaussian Process, SVM as baseline predictors for M-ML as well as several other comparison methods including Causal Forests. A complete list of methods used can be found in Table \ref{tab:simmethods} of the Appendix.For all the M-ML methods employed in our simulations, we match on the unweighted L2 distance, where $\phihat$ is either the propensity score or the difference in potential outcomes estimated with one of three ML methods. We use KNN M-ML matching with fixed $k_n$ and $k_n$ set to $\lfloor n^{1/2}\rfloor$, so that $\sqrt{k_n}$ is a lower bound on the first stage convergence rates of most ML methods, under sufficient regularity assumptions on the data \citep{chernozhukov2018double}. 
We first present results for CATE estimation in the top row of Figure \ref{fig:CATEerr}. CATEs in this setting are estimated as simple differences of predicted potential outcomes fitted by each method under consideration. Our results for this setting show that, while M-ML does not generally outperform the non-parametric black box methods that we compare it to, it also does not generally underperform them. This lends evidence to the idea that \textit{adding matching on top of ML methods can boost interpretability, auditability and enable uncertainty quantification, while leading to minimal or no loss in performance}. Results also show that the DGP does have an influence on performance: specifically, M-ML seems to perform better under the Nonlinear and Selection DGPs, and outperforms nonparametric causal methods such as causal forests and X-learner in these settings. 
The bottom row of Figure \ref{fig:CATEerr} presents results for a similar set of simulations, but for ATE estimation. We use the M-DML algorithm with the KNN M-ML algorithm to construct first-stage predictions of conditional response functions. Causal forests, and bias-corrected matching methods, are used with the estimators provided in the original papers and packages that implement them. We compare to matching methods intended for ATE estimation, such as GenMatch, as well as the other nonparametric ML methods. Results are presented for 500 simulation rounds, at each of which a dataset of $n=5000$ units was generated with the same DGP as before, and 3000 units were separated as a training set. Parameters other than $\epsilon, \bX, $, and $Y$ were generated first and kept the same for all 500 rounds, while other variables were generated at each round. Again we see that M-ML performs comparably with other ML methods, and can outperform some of them in certain settings. Again, choice of baseline algorithm and DGP seem to have an influence on performance. 
Finally, we conduct a set of simulations to study the coverage of 95\% asymptotic confidence intervals obtained with M-ML and a BART ML backend. We choose to compare the coverage of M-ML against Causal Forests \citep{wager2018estimation}, as this is the only other method we know of that produces asymptotically valid confidence intervals for CATE estimation. We run the same set of simulations as before, but vary the size of the training set each time. At each train set size, we randomly draw 250 CATEs to estimate from the distribution of $\bX$, and compute the proportion covered by their respective estimated CI. This procedure is repeated and averaged over 1000 simulations for each setting. Results are reported in the bottom row of Figure \ref{fig:coverage} . We see that M-ML performs much better than Causal Forests in Figure \ref{fig:coverage}, \textit{having larger coverage in all our simulation settings}. Additionally we can see that M-ML still does not reach the nominal coverage level: this is expected as existing methods for CATE estimation also rarely do \citep{kunzel2019metalearners}, given the hardness of the problem. 
Turning to the ATE, we compare coverage of M-DML to coverage obtained by causal forests with the same simulation setup as before. Results are shown in the top row of Figure \ref{fig:coverage}. Clearly, M-DML achieves nominal (95\%) coverage in all settings, while Causal Forests greatly underperforms. Additional results comparing interval size for both CF and M-DML are presented in Figures \ref{fig:intsize} and \ref{fig:atesize} available in the Appendix. Generally, these results show that M-ML and M-DML output marginally larger intervals than Causal Forests does, however this increase in size is largely justified by the far superior coverage achieved by our methods. Thus, our simulation results have shown that M-ML can add the ability to audit high-accuracy black-box machine learning methods, without leading to loss of performance. This applies to both CATE and ATE estimation.

\section{Application: Matching with Image Data}\label{sec:application}
We apply our method to the study of the returns of brand responsiveness to social media followers. This is a well-studied issue in online marketing and consumer behavior: there is a cyclical relationship between brand relationships and engagement on social media. Engaging with a brand on social media can strengthen the consumer-brand relationship \citep{laroche2013or, labrecque2014fostering}. Social media engagement strengthens the consumer-brand relationship when the consumer feels like the brand is responsive to them \citep{labrecque2014fostering}. However, considerable evidence also shows that consumers who already have strong relationships with a brand are more likely to engage with that brand on social media \citep{john2017does, simon2018does}. Because of this, understanding the effect of brand responsiveness to consumers on social media presents a clear causal inference challenge that we try to address here.  Specifically, in order to control for potential confounders of the relationship in question, we match posts on metadata, such as date/time and number of comments, but \textit{also on the image that was posted}. Matching on images is important because the content of an image in a post likely has an influence on the likelihood of interaction with that post. This application also demonstrates how M-ML allows matching of units on complex covariates such as images, and how one can audit results by simple inspection of the matches. 

\subsection{Data and Methodology}
We use a dataset of Instagram image posts made by 31 food brands between May 15th 2020 and May 15th 2021. The treatment is 1 if the post had any responses from the brand itself to comments left by the viewers in the same day it was posted, and the outcome variable is the number of comments received by a post a day or more after it was posted. The control covariates are hour, month and weekday that the original post was made, and total number of comments left by users on the post on the day it was posted, as well as the image of the post itself. The latter allows us to control directly for the content of the post.  For this application, we first constructed our ML representation function by training a variational autoencoder \citep{kingma2013auto} on the training set. We used the representation function to construct a 100-dimensional representation of each image within the matching set. Our encoder-decoder architecture  consists of two identical CNNs with 4 layers. We trained the model on 70\% of the post data and used the remaining 30\% for inference. This left us with $n=1677$ posts that we matched and estimated CATEs for. 
We chose the size of the matched groups as follows: for $t=0, 1$ we set $k_{nt}$ set to the integer closest to $n_t^{1/2}$, where $n_t$ is the size of the group with treatment $t$ in the matching set, using this procedure for both the treated and control group, we obtained $k_1 = 7$ to estimate $\mu(\bx, 1)$ and $k_0 = 6$ units to estimate $\mu(\bx, 0)$. 

\subsection{Results}
\begin{figure}[!htbp]
    \centering
    \caption{CATE by brand}
    \label{fig:atebrand}
    \includegraphics[width=\textwidth]{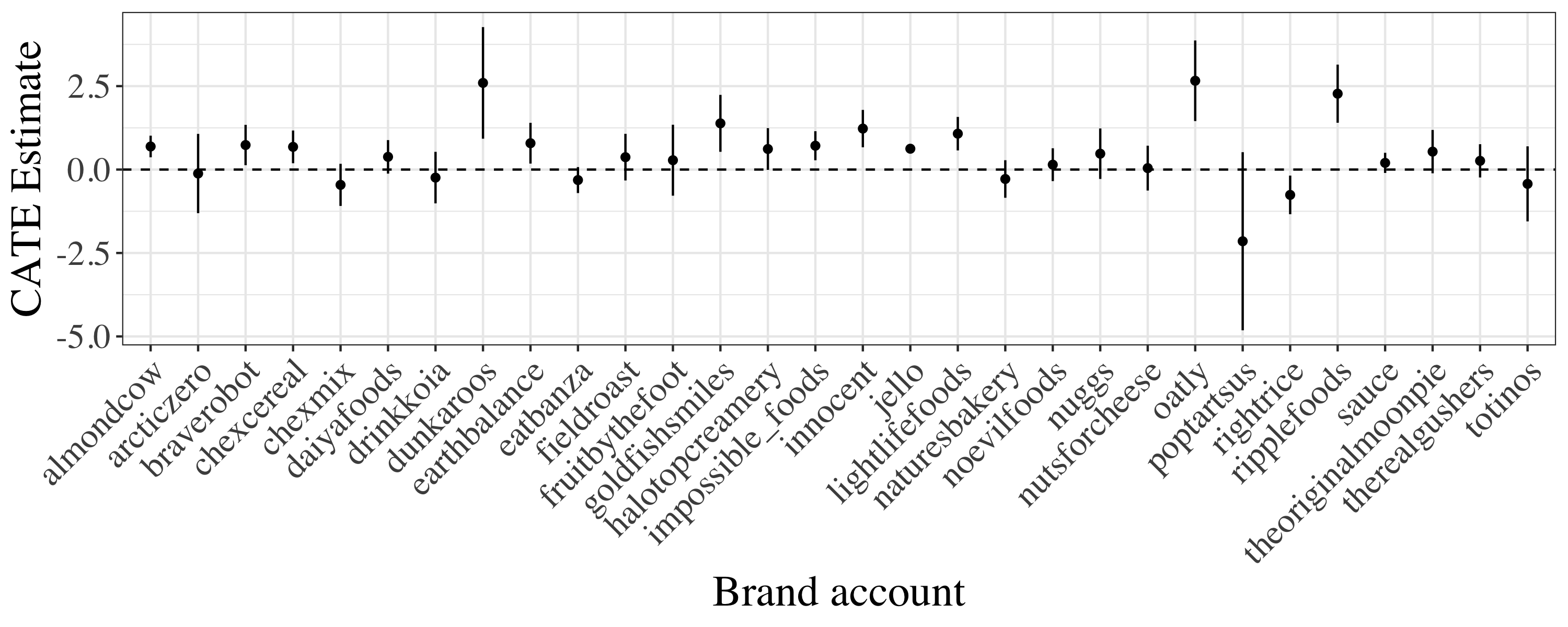}
\end{figure}

\begin{figure}[!htbp]
    \centering
    \caption{Sample matched groups}
    \label{fig:MG}
\fbox{\includegraphics[width=0.23\textwidth]{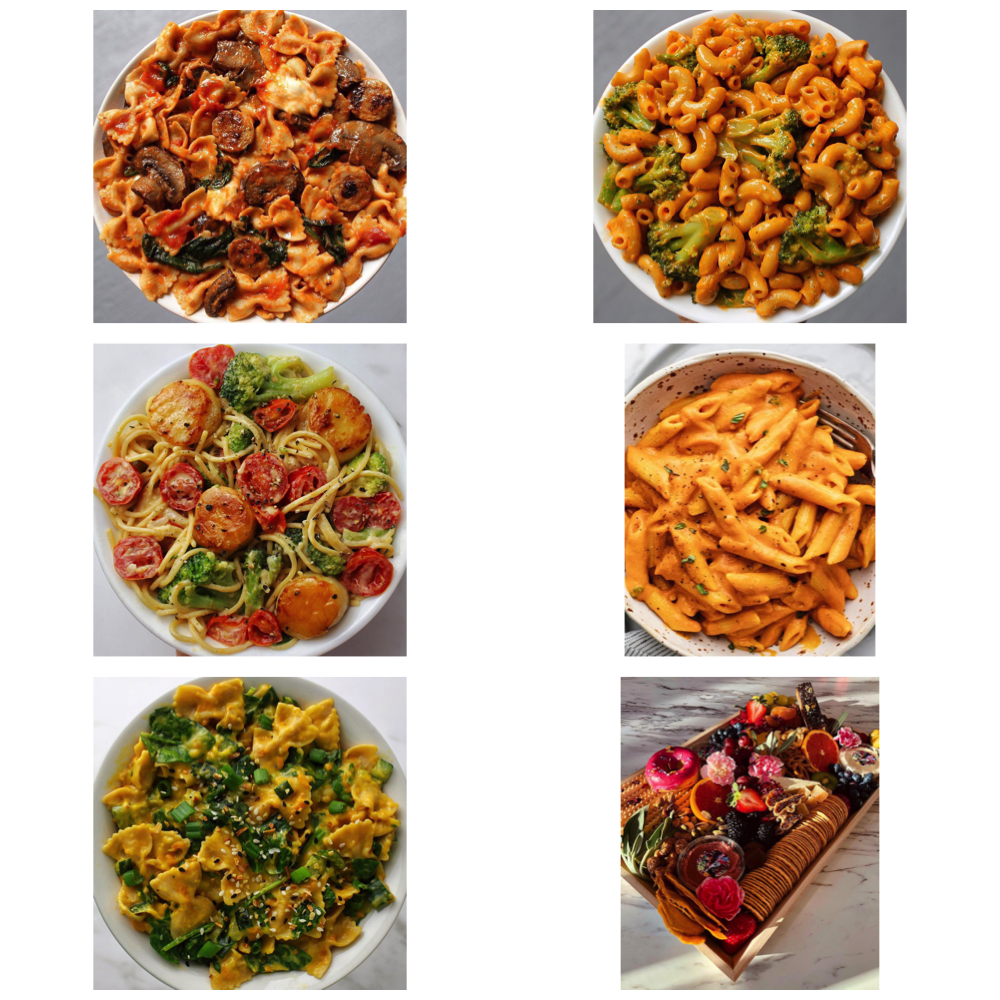}}
\fbox{\includegraphics[width=0.23\textwidth]{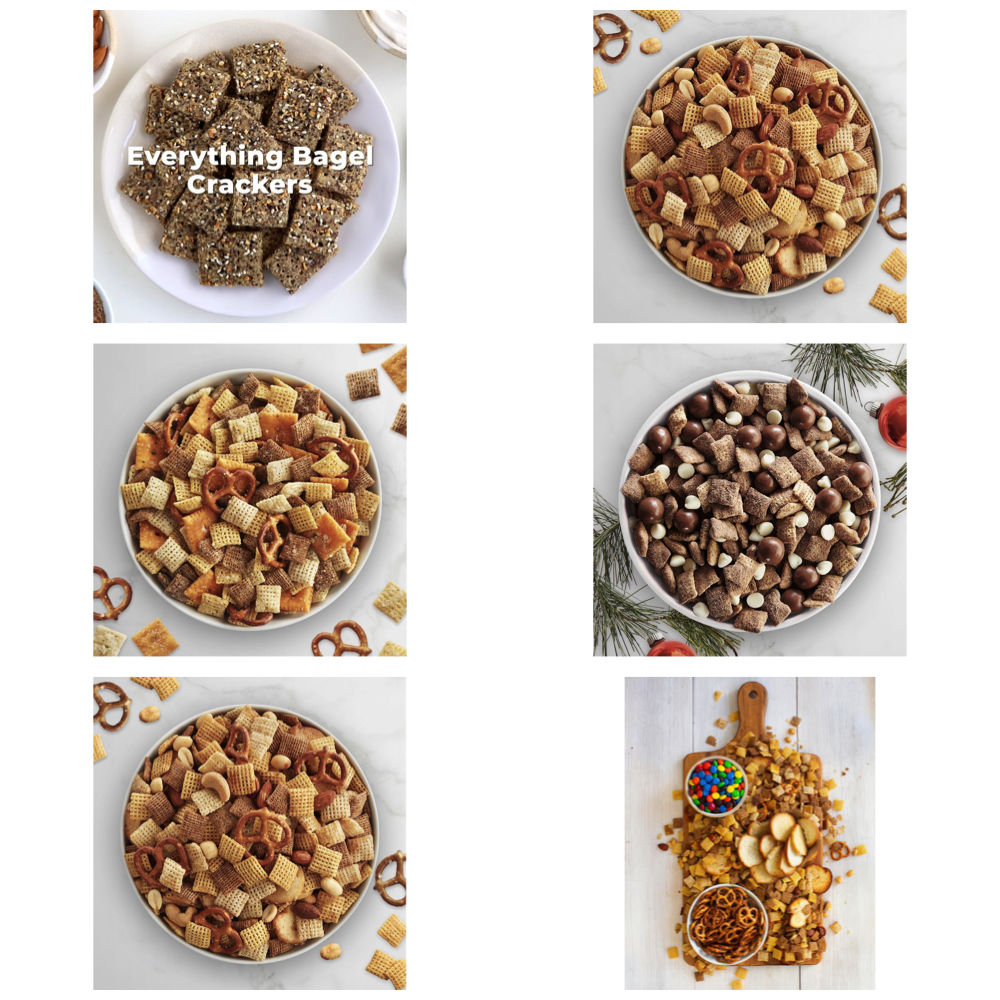}}
\fbox{\includegraphics[width=0.23\textwidth]{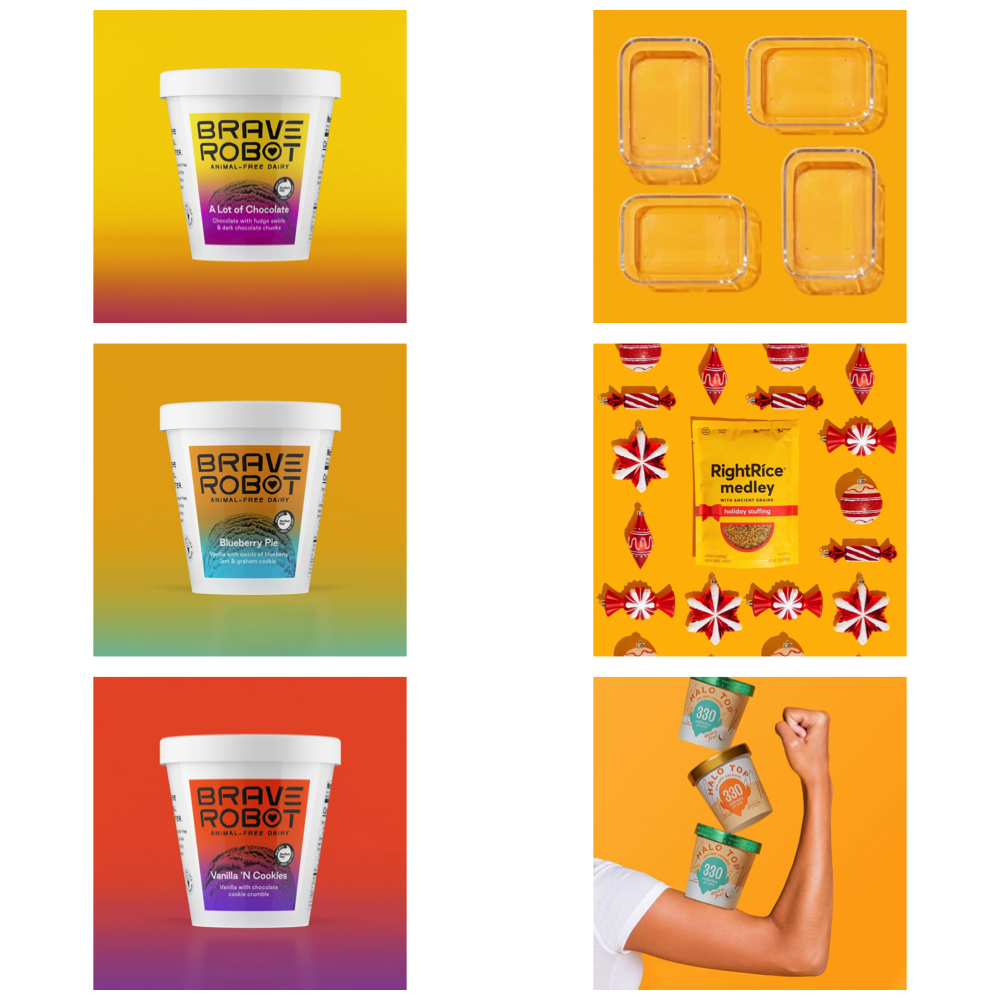}}
\fbox{\includegraphics[width=0.23\textwidth]{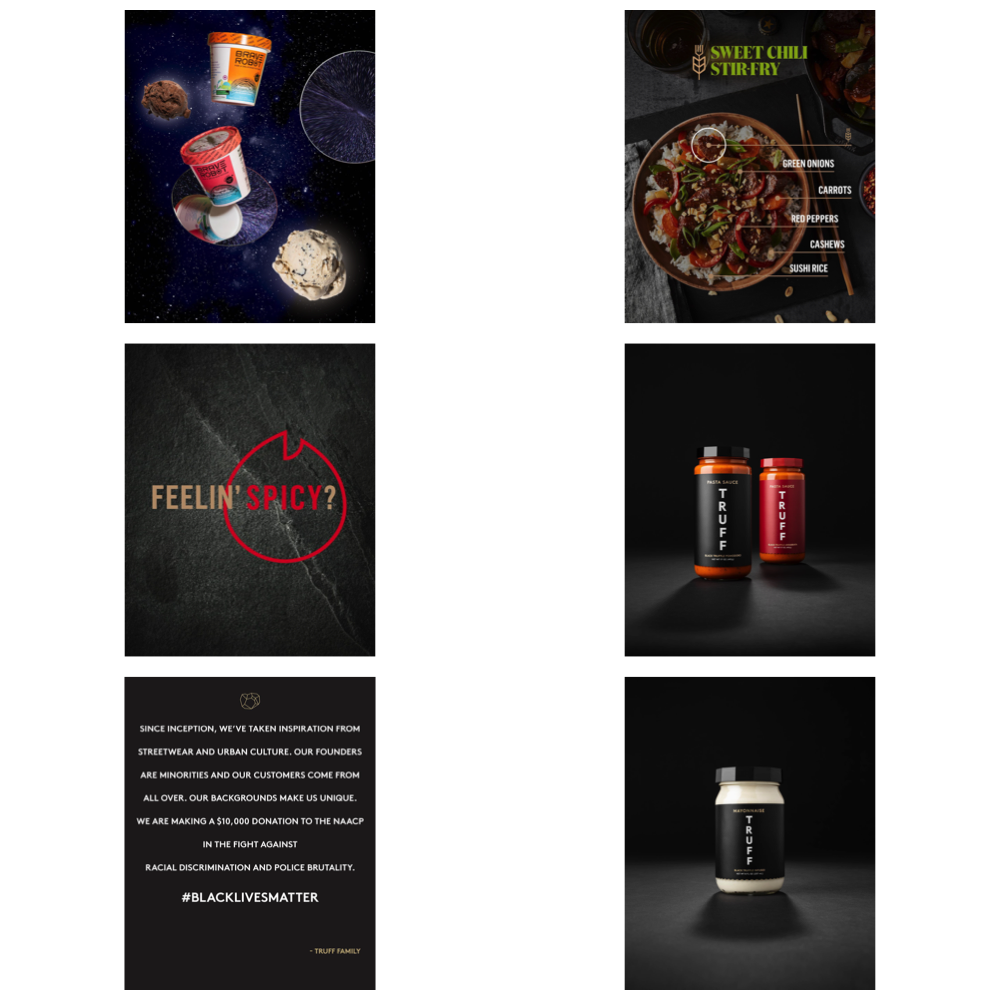}}
\end{figure}

The ATE of having at least one brand interaction during the same day a post was made on the number of comments received by that post in the following days was 0.36 with a 95\% asymptotic confidence interval between 0.27 and 0.63. Since the outcome variable is the natural logarithm of a count, the result can be interpreted as saying that \textit{adding a brand interaction within the day of posting produces approximately a 44\% increase in the number of comments received by that post in the following days}. 
Additional results are presented in Figure \ref{fig:atebrand}, which shows individual ATE estimates for each brand. These estimates were obtained by aggregating CATEs for individual posts for each brand with the M-DML estimator. Notably, most brands seem to exhibit a similar positive treatment effect, while only RightRice has a negative and statistically significant treatment effect. This suggests the presence of heterogeneities in the treatment effect that could motivate further investigation. 
Finally, we present some sample images from the matched groups created by M-ML with the VAE back-end. This is important because it allows us to better audit our estimates by looking at the cases that were used to generate them, i.e., the matched groups: if matched groups do not make intuitive sense, then there is reason to doubt their usefulness and overall trustworthiness for treatment effect estimation. The sample groups shown in Figure \ref{fig:MG} highlight how images with similar elements are matched together: most of the groups contain posts from the same  brand and contain visual representations of similar foods. Additional sample matched groups displaying a similar pattern are available in the supplement. This shows that our results are interpretable on a human level, and based on clear visual cues present in the matching covariates. 
\section{Conclusion}
Interpretability is paramount in the high-stakes decision making settings in which causal inference is used because it enables estimates to be audited based on analysts' contextual knowledge. In this paper we have introduced Matched Machine Learning, a method that aims to combine the predictive capabilities of black-box ML methods with the auditability and user-friendliness of matching. We have presented M-ML algorithms for both CATE and ATE estimation in a general framework. Many different choices of matching metrics, representation functions, and additional constraints can be formulated as M-ML. We have theoretically shown that, under reasonable conditions, M-ML for CATE estimation achieves asymptotic normality and consistency at a rate close to the nonparametrically optimal one. We have also shown that, by using M-ML estimates as inputs to AIPW estimators, ATE and ATT can be estimated consistently at a $\sqrt{n}$ rate. Empirically, we have shown that M-ML does not compromise accuracy for auditability. In our application, we have shown how M-ML can be used to analyze causally non-standard data such as images. Overall, M-ML expands the boundary of research in interpretable but accurate causal inference. Potential extensions of M-ML include matching with continuous treatments that are transformed via ML methods, as well as developing milder theoretical conditions on first-stage ML estimates than those introduced in the present paper.

\bibliography{biblio}

\begin{thebibliography}{50}
\providecommand{\natexlab}[1]{#1}
\providecommand{\url}[1]{\texttt{#1}}
\expandafter\ifx\csname urlstyle\endcsname\relax
  \providecommand{\doi}[1]{doi: #1}\else
  \providecommand{\doi}{doi: \begingroup \urlstyle{rm}\Url}\fi

\bibitem[Abadie and Imbens(2006)]{abadie2006large}
Alberto Abadie and Guido~W Imbens.
\newblock Large sample properties of matching estimators for average treatment
  effects.
\newblock \emph{econometrica}, 74\penalty0 (1):\penalty0 235--267, 2006.

\bibitem[Abadie and Imbens(2011)]{abadie2011bias}
Alberto Abadie and Guido~W Imbens.
\newblock Bias-corrected matching estimators for average treatment effects.
\newblock \emph{Journal of Business \& Economic Statistics}, 29\penalty0
  (1):\penalty0 1--11, 2011.

\bibitem[Abadie and Imbens(2012)]{abadie2012}
Alberto Abadie and Guido~W Imbens.
\newblock A martingale representation for matching estimators.
\newblock \emph{Journal of the American Statistical Association}, 107\penalty0
  (498):\penalty0 833--843, 2012.

\bibitem[Abadie and Imbens(2016)]{abadie2016matching}
Alberto Abadie and Guido~W Imbens.
\newblock Matching on the estimated propensity score.
\newblock \emph{Econometrica}, 84\penalty0 (2):\penalty0 781--807, 2016.

\bibitem[Belloni et~al.(2014)Belloni, Chernozhukov, Wang,
  et~al.]{belloni2014pivotal}
Alexandre Belloni, Victor Chernozhukov, Lie Wang, et~al.
\newblock Pivotal estimation via square-root lasso in nonparametric regression.
\newblock \emph{Annals of Statistics}, 42\penalty0 (2):\penalty0 757--788,
  2014.

\bibitem[Chernozhukov et~al.(2018)Chernozhukov, Chetverikov, Demirer, Duflo,
  Hansen, Newey, and Robins]{chernozhukov2018double}
Victor Chernozhukov, Denis Chetverikov, Mert Demirer, Esther Duflo, Christian
  Hansen, Whitney Newey, and James Robins.
\newblock Double/debiased machine learning for treatment and structural
  parameters, 2018.

\bibitem[Chipman et~al.(2010)Chipman, George, McCulloch,
  et~al.]{chipman2010bart}
Hugh~A Chipman, Edward~I George, Robert~E McCulloch, et~al.
\newblock Bart: Bayesian additive regression trees.
\newblock \emph{The Annals of Applied Statistics}, 4\penalty0 (1):\penalty0
  266--298, 2010.

\bibitem[Devroye et~al.(2013)Devroye, Gy{\"o}rfi, and
  Lugosi]{devroye2013probabilistic}
Luc Devroye, L{\'a}szl{\'o} Gy{\"o}rfi, and G{\'a}bor Lugosi.
\newblock \emph{A probabilistic theory of pattern recognition}, volume~31.
\newblock Springer Science \& Business Media, 2013.

\bibitem[Diamond and Sekhon(2013)]{diamond2013genetic}
Alexis Diamond and Jasjeet~S Sekhon.
\newblock Genetic matching for estimating causal effects: A general
  multivariate matching method for achieving balance in observational studies.
\newblock \emph{Review of Economics and Statistics}, 95\penalty0 (3):\penalty0
  932--945, 2013.

\bibitem[Dieng et~al.(2019)Dieng, Liu, Roy, Rudin, and
  Volfovsky]{dieng2019interpretable}
Awa Dieng, Yameng Liu, Sudeepa Roy, Cynthia Rudin, and Alexander Volfovsky.
\newblock Interpretable almost-exact matching for causal inference.
\newblock In \emph{The 22nd International Conference on Artificial Intelligence
  and Statistics}, pages 2445--2453. PMLR, 2019.

\bibitem[Drucker et~al.(1997)Drucker, Burges, Kaufman, Smola, Vapnik,
  et~al.]{drucker1997support}
Harris Drucker, Chris~JC Burges, Linda Kaufman, Alex Smola, Vladimir Vapnik,
  et~al.
\newblock Support vector regression machines.
\newblock \emph{Advances in neural information processing systems}, 9:\penalty0
  155--161, 1997.

\bibitem[Farrell et~al.(2021)Farrell, Liang, and Misra]{farrell2021deep}
Max~H Farrell, Tengyuan Liang, and Sanjog Misra.
\newblock Deep neural networks for estimation and inference.
\newblock \emph{Econometrica}, 89\penalty0 (1):\penalty0 181--213, 2021.

\bibitem[Gy\"{o}rfi(1981)]{gyorfi1981rate}
L~Gy\"{o}rfi.
\newblock The rate of convergence of k\_n-nn regression estimates and
  classification rules (corresp.).
\newblock \emph{IEEE Transactions on Information Theory}, 27\penalty0
  (3):\penalty0 362--364, 1981.

\bibitem[Gy{\"o}rfi et~al.(2002)Gy{\"o}rfi, Kohler, Krzy{\.z}ak, and
  Walk]{gyorfi2002distribution}
L{\'a}szl{\'o} Gy{\"o}rfi, Michael Kohler, Adam Krzy{\.z}ak, and Harro Walk.
\newblock \emph{A distribution-free theory of nonparametric regression},
  volume~1.
\newblock Springer, 2002.

\bibitem[Hahn et~al.(2020)Hahn, Murray, and Carvalho]{hahn2020bayesian}
P~Richard Hahn, Jared~S Murray, and Carlos~M Carvalho.
\newblock Bayesian regression tree models for causal inference: Regularization,
  confounding, and heterogeneous effects (with discussion).
\newblock \emph{Bayesian Analysis}, 15\penalty0 (3):\penalty0 965--1056, 2020.

\bibitem[Hansen(2008)]{hansen2008prognostic}
Ben~B Hansen.
\newblock The prognostic analogue of the propensity score.
\newblock \emph{Biometrika}, 95\penalty0 (2):\penalty0 481--488, 2008.

\bibitem[Hill(2011)]{hill2011bayesian}
Jennifer~L Hill.
\newblock Bayesian nonparametric modeling for causal inference.
\newblock \emph{Journal of Computational and Graphical Statistics}, 20\penalty0
  (1):\penalty0 217--240, 2011.

\bibitem[Iacus et~al.(2012)Iacus, King, and Porro]{iacus2012causal}
Stefano~M Iacus, Gary King, and Giuseppe Porro.
\newblock Causal inference without balance checking: Coarsened exact matching.
\newblock \emph{Political analysis}, pages 1--24, 2012.

\bibitem[John et~al.(2017)John, Emrich, Gupta, and Norton]{john2017does}
Leslie~K John, Oliver Emrich, Sunil Gupta, and Michael~I Norton.
\newblock Does “liking” lead to loving? the impact of joining a brand's
  social network on marketing outcomes.
\newblock \emph{Journal of Marketing Research}, 54\penalty0 (1):\penalty0
  144--155, 2017.

\bibitem[Kallus(2020)]{kallus2020generalized}
Nathan Kallus.
\newblock Generalized optimal matching methods for causal inference.
\newblock \emph{Journal of Machine Learning Research}, 21\penalty0
  (62):\penalty0 1--54, 2020.

\bibitem[Kingma and Welling(2013)]{kingma2013auto}
Diederik~P Kingma and Max Welling.
\newblock Auto-encoding variational bayes.
\newblock \emph{arXiv preprint arXiv:1312.6114}, 2013.

\bibitem[K{\"u}nzel et~al.(2019)K{\"u}nzel, Sekhon, Bickel, and
  Yu]{kunzel2019metalearners}
S{\"o}ren~R K{\"u}nzel, Jasjeet~S Sekhon, Peter~J Bickel, and Bin Yu.
\newblock Metalearners for estimating heterogeneous treatment effects using
  machine learning.
\newblock \emph{Proceedings of the national academy of sciences}, 116\penalty0
  (10):\penalty0 4156--4165, 2019.

\bibitem[Labrecque(2014)]{labrecque2014fostering}
Lauren~I Labrecque.
\newblock Fostering consumer--brand relationships in social media environments:
  The role of parasocial interaction.
\newblock \emph{Journal of interactive marketing}, 28\penalty0 (2):\penalty0
  134--148, 2014.

\bibitem[Laroche et~al.(2013)Laroche, Habibi, and Richard]{laroche2013or}
Michel Laroche, Mohammad~Reza Habibi, and Marie-Odile Richard.
\newblock To be or not to be in social media: How brand loyalty is affected by
  social media?
\newblock \emph{International journal of information management}, 33\penalty0
  (1):\penalty0 76--82, 2013.

\bibitem[Li and Racine(2007)]{li2007nonparametric}
Qi~Li and Jeffrey~Scott Racine.
\newblock \emph{Nonparametric econometrics: theory and practice}.
\newblock Princeton University Press, 2007.

\bibitem[Luo and Zhu(2020)]{luo2020matching}
Wei Luo and Yeying Zhu.
\newblock Matching using sufficient dimension reduction for causal inference.
\newblock \emph{Journal of Business \& Economic Statistics}, 38\penalty0
  (4):\penalty0 888--900, 2020.

\bibitem[Morucci et~al.(2020)Morucci, Orlandi, Roy, Rudin, and
  Volfovsky]{morucci2020adaptive}
Marco Morucci, Vittorio Orlandi, Sudeepa Roy, Cynthia Rudin, and Alexander
  Volfovsky.
\newblock Adaptive hyper-box matching for interpretable individualized
  treatment effect estimation.
\newblock In \emph{Conference on Uncertainty in Artificial Intelligence}, pages
  1089--1098. PMLR, 2020.

\bibitem[Otsu and Rai(2017)]{otsu2017bootstrap}
Taisuke Otsu and Yoshiyasu Rai.
\newblock Bootstrap inference of matching estimators for average treatment
  effects.
\newblock \emph{Journal of the American Statistical Association}, 112\penalty0
  (520):\penalty0 1720--1732, 2017.

\bibitem[Parikh et~al.(2022)Parikh, Rudin, and Volfovsky]{parikh2018malts}
Harsh Parikh, Cynthia Rudin, and Alexander Volfovsky.
\newblock Malts: Matching after learning to stretch.
\newblock \emph{Journal of Machine Learning Research}, 2022.

\bibitem[Rasmussen(2003)]{rasmussen2003gaussian}
Carl~Edward Rasmussen.
\newblock Gaussian processes in machine learning.
\newblock In \emph{Summer school on machine learning}, pages 63--71. Springer,
  2003.

\bibitem[Rimanic et~al.(2020)Rimanic, Renggli, Li, and
  Zhang]{rimanic2020convergence}
Luka Rimanic, Cedric Renggli, Bo~Li, and Ce~Zhang.
\newblock On convergence of nearest neighbor classifiers over feature
  transformations.
\newblock \emph{arXiv preprint arXiv:2010.07765}, 2020.

\bibitem[Robins et~al.(1994)Robins, Rotnitzky, and Zhao]{robins1994estimation}
James~M Robins, Andrea Rotnitzky, and Lue~Ping Zhao.
\newblock Estimation of regression coefficients when some regressors are not
  always observed.
\newblock \emph{Journal of the American statistical Association}, 89\penalty0
  (427):\penalty0 846--866, 1994.

\bibitem[Rosenbaum and Rubin(1983)]{rosenbaum1983central}
Paul~R Rosenbaum and Donald~B Rubin.
\newblock The central role of the propensity score in observational studies for
  causal effects.
\newblock \emph{Biometrika}, 70\penalty0 (1):\penalty0 41--55, 1983.

\bibitem[Rosenbaum and Rubin(1984)]{rosenbaum1984reducing}
Paul~R Rosenbaum and Donald~B Rubin.
\newblock Reducing bias in observational studies using subclassification on the
  propensity score.
\newblock \emph{Journal of the American statistical Association}, 79\penalty0
  (387):\penalty0 516--524, 1984.

\bibitem[Rosenbaum and Rubin(1985{\natexlab{a}})]{rosenbaum1985bias}
Paul~R Rosenbaum and Donald~B Rubin.
\newblock The bias due to incomplete matching.
\newblock \emph{Biometrics}, pages 103--116, 1985{\natexlab{a}}.

\bibitem[Rosenbaum and Rubin(1985{\natexlab{b}})]{rosenbaum1985constructing}
Paul~R Rosenbaum and Donald~B Rubin.
\newblock Constructing a control group using multivariate matched sampling
  methods that incorporate the propensity score.
\newblock \emph{The American Statistician}, 39\penalty0 (1):\penalty0 33--38,
  1985{\natexlab{b}}.

\bibitem[Rosenbaum et~al.(2007)Rosenbaum, Ross, and
  Silber]{rosenbaum2007minimum}
Paul~R Rosenbaum, Richard~N Ross, and Jeffrey~H Silber.
\newblock Minimum distance matched sampling with fine balance in an
  observational study of treatment for ovarian cancer.
\newblock \emph{Journal of the American Statistical Association}, 102\penalty0
  (477):\penalty0 75--83, 2007.

\bibitem[Rubin(1976)]{rubin1976multivariate}
Donald~B Rubin.
\newblock Multivariate matching methods that are equal percent bias reducing,
  i: Some examples.
\newblock \emph{Biometrics}, pages 109--120, 1976.

\bibitem[Rubin(1980)]{rubin1980bias}
Donald~B Rubin.
\newblock Bias reduction using mahalanobis-metric matching.
\newblock \emph{Biometrics}, pages 293--298, 1980.

\bibitem[Rubin and Thomas(1992)]{rubin1992affinely}
Donald~B Rubin and Neal Thomas.
\newblock Affinely invariant matching methods with ellipsoidal distributions.
\newblock \emph{The Annals of Statistics}, pages 1079--1093, 1992.

\bibitem[S{\"a}vje(2022)]{savje2022inconsistency}
Fredrik S{\"a}vje.
\newblock On the inconsistency of matching without replacement.
\newblock \emph{Biometrika}, 109\penalty0 (2):\penalty0 551--558, 2022.

\bibitem[Simon and Tossan(2018)]{simon2018does}
Fran{\c{c}}oise Simon and Vesselina Tossan.
\newblock Does brand-consumer social sharing matter? a relational framework of
  customer engagement to brand-hosted social media.
\newblock \emph{Journal of Business Research}, 85:\penalty0 175--184, 2018.

\bibitem[Stone(1982)]{stone1982optimal}
Charles~J Stone.
\newblock Optimal global rates of convergence for nonparametric regression.
\newblock \emph{The annals of statistics}, pages 1040--1053, 1982.

\bibitem[Stute et~al.(1984)]{stute1984asymptotic}
Winfried Stute et~al.
\newblock Asymptotic normality of nearest neighbor regression function
  estimates.
\newblock \emph{The Annals of Statistics}, 12\penalty0 (3):\penalty0 917--926,
  1984.

\bibitem[Van Der~Laan and Rubin(2006)]{van2006targeted}
Mark~J Van Der~Laan and Daniel Rubin.
\newblock Targeted maximum likelihood learning.
\newblock \emph{The international journal of biostatistics}, 2\penalty0 (1),
  2006.

\bibitem[Wager and Athey(2018)]{wager2018estimation}
Stefan Wager and Susan Athey.
\newblock Estimation and inference of heterogeneous treatment effects using
  random forests.
\newblock \emph{Journal of the American Statistical Association}, 113\penalty0
  (523):\penalty0 1228--1242, 2018.

\bibitem[Wang et~al.(2021)Wang, Morucci, Awan, Liu, Roy, Rudin, and
  Volfovsky]{wang2021flame}
Tianyu Wang, Marco Morucci, M~Usaid Awan, Yameng Liu, Sudeepa Roy, Cynthia
  Rudin, and Alexander Volfovsky.
\newblock Flame: A fast large-scale almost matching exactly approach to causal
  inference.
\newblock \emph{Journal of Machine Learning Research}, 22\penalty0
  (31):\penalty0 1--41, 2021.

\bibitem[Wang(2005)]{wang2005volumes}
Xianfu Wang.
\newblock Volumes of generalized unit balls.
\newblock \emph{Mathematics Magazine}, 78\penalty0 (5):\penalty0 390--395,
  2005.

\bibitem[Wang and Zubizarreta(Forthcoming)]{wang2019Large}
Yixin Wang and Jos{\'e}~R Zubizarreta.
\newblock Large sample properties of matching for balance.
\newblock \emph{Statistica Sinica}, Forthcoming.

\bibitem[Zubizarreta(2012)]{zubizarreta2012using}
Jos{\'e}~R Zubizarreta.
\newblock Using mixed integer programming for matching in an observational
  study of kidney failure after surgery.
\newblock \emph{Journal of the American Statistical Association}, 107\penalty0
  (500):\penalty0 1360--1371, 2012.

\end{thebibliography}

\pagebreak

\singlespacing
\setcounter{section}{0}
\renewcommand{\thesection}{\Alph{section}}
\setcounter{page}{1}

\part*{Supplement}
\section{Preliminaries}
\subsection{Assumptions and Main Notation}
Here we restate the main notation and assumptions of the paper. Let $(\Omega, \mathcal{F}, \cP)$ be a probability space, with $\Omega = \mathbb{R} \times \mathbb{X} \times \{1, \dots, M\}$, and let $\bO = (Y, \bX, T)$ be a set of random variables on this space, with $Y = \sum_{j=1}^M Y(j)\ind[T=j]$. Note that $Y(t)$ has domain in $\R$, $\bX$ in $\X$ and $T$ in $\{1,\dots, M\}$. Denote the joint distribution of $\bO$ by $\cP$. For a random variable $A$, we use $F_A$ to denote its CDF and $f_A$ to denote its PDF, as well as $\E_A$ and $\V_A$ to denote expectation and variance wrt $A$. When the notation $\E[\cdot]$ or $\V[\cdot]$ is used without any indices it is taken to be with respect to all the random variates within the brackets.  For a function $g: \R^p \mapsto \R^d$, define the distance function:  $D_{g}^q(\bu, \bv) := \|g(\bu) - g(\bv)\|_q$, where $\|\cdot\|_q$ is the standard $q$-norm. Let $A$ be a random variable over $\R^p$, and let $f(A): \R^p \mapsto \R^d$. We will use the notation $\|f(A)\|_{\cP, q} = \sqrt[q]{\int \max_{j=1,\dots,d}|f(a)_j|^qd\cP(a)}$ to denote the $L_q$ norm wrt measure $\cP$. We use the notation: $\muxd := \E[Y|\bX=\bx, T=t]$, $\sigma^2(\bx, t) = \V[Y|\bX=\bx, T=t]$, and $\tau(\bx, t, t') = \mu(\bx, t) - \mu(\bx, t')$ to refer to quantities of interest. \\

\noindent We assume the following:\\
\noindent\textbf{A1 (Data Distribution)}: \\
(a) The data $\On = \{\bO_i\}_{i=1}^n = \{Y_i, \bX_i, T_i\}_{i=1}^n$ is a set of $n$ i.i.d. copies of $\bO$. \\
(b) The domain of the covariate distribution, $\mathbb{X}$ is a compact subset of $\R^p$. \\
(c) The covariates have marginal distribution  with differentiable CDF (w.r.t. the lebesgue measure) $F_{\bX}(\bx)$, and constants $c_{\fbX}, C_{\fbX}$, such that $0 < c_{\fbX} < \fbX(\bx) < C_{\fbX} < \infty$ everywhere over $\mathbb{X}$.\\
\textbf{A2 (Overlap)}: For all $\bx \in \mathbb{X}$ and $t=1,\dots,M$ we have $0 < \Pr(T=t|\bX=\bx) < 1$.\\
\textbf{A3 (Conditional Ignorability)}: $T \indep (Y(1), \dots, Y(M))|\bX$. \\
\textbf{A4 (Bounded Higher Moments)}: For all $t, t' \in \{1, \dots, M\}$, all $\bx \in \mathbb{X}$ and  for some $\delta > 0$ and a constant $C_\delta$ we have: $\E[|Y(t)|^{2 + \delta}|\bX=\bx, T=t'] \leq C_\delta$.\\
\textbf{A5 (Lipschitz Condition)}: For all $\bx, \bz \in \X$ and $t \in \{1,\dots, M\}$ there exists a constant $C_L$ such that:\\
(a) $|\mu(\bx, t) - \mu(\bz, t)| \leq C_L\Dphiq(\bx, \bz)$\\
(b) $|\sigmasq(\bx, t) - \sigmasq(\bz, t)| \leq C_L\Dphiq(\bx, \bz)$\\
\textbf{A6 (Representation function)}: There exists a function $\phihat(\bx, \On): \mathbb{X} \times \Omega \mapsto \R^d$ such that, for real-valued $r_{ML} > 0$, and $q > 0$: \\
(a) The functions $\phihat(\bx, \On)$ and $\phi(\bx)$ are $f_\bO$-almost surely continuous with respect to $\bx$ at all $\bx \in \mathbb{X}$. \\
(b) $\|\phihat(\bx, \On) - \phi(\bx) \|_{\cP, q} = o(n^{-r_{ML}})$ almost surely over $\fbX$. \\
(c) $\|\phihat(\bX, \On) - \phi(\bX) \|_{\cP, q} = o(n^{-r_{ML}})$.\\

Throughout this appendix we will also make use of some specialized notation to refer to matching operations. We will use $\MGxt \subset \{1, \dots, n\}$ to denote the matched group made around covariate value $\bx$, treatment value $t$, and with representation function $\phihat$. Note that the definition of $\MGxt$ will vary depending on whether Caliper M-ML or KNN-M-ML is used, and what definition is used will be specified in each theorem. We will also use the notation $\Wixt = \ind[i \in \MGxt]$ to denote membership of units $i$ in $\MGxt$, and $\Nxt = \sum_{i=1}^n \Wixt$ to count the number of units in $\MGxt$. 

\section{Proofs}
\subsection{Proof of Lemma \ref{thm:iinmg}}

\begin{proof}
Before proving the result we establish some important facts. Consider the quantity $\phihat(\bx, \On) =: \phihat(\bx)$, where we remove the explicit dependence of $\phihat$ on $\On$ for notational simplicity. By A6 (b) and (c), we know that it must be a random variable (i.e., measurable function) over some subset $A \subset \R^d$. Denote the CDF of this random variable by $F_{\phihat(\bx)}$. By A6 (b), we have that, for any $\bu \in A$: 
$$\lim_{n \rightarrow \infty} F_{\phihat(\bx)}(\bu) = F_{\phi(\bx)}(\bu) = \begin{cases}1 &\mbox{ if } u_j \geq \phi(\bx)_j,\; j=1,\dots,d \\ 0 &\mbox{ otherwise.}\end{cases},$$
due to the fact that $\phi(\bx)$ is a constant with respect to $\bu$. This also implies that $dF_{\phi(\bx)}(\bu) = \delta(\phi(\bx) - \bu)d\bu$, where $\delta(\bu)$ is Dirac's delta function that puts density 1 at 0 and 0 everywhere else. By the above, we can also conclude that the joint CDF of the pair $(\phihat(\bx), \phihat(\bz))$, denoted by $F_{(\phihat(\bx), \phihat(\bz))}(\bu, \bv)$ will converge to $F_{\phi(\bx)}(\bu)F_{\phi(\bz)}(\bv)$.

For arbitrary $\bx, \bz \in \X$ and $\Gamman \geq 0$, we can write the quantity: $\Pr_{\On}(\|\phihat(\bx) - \phihat(\bz)\|_q \leq \Gamma_n)$ (where the randomness is over the training data) as a function of the quantities just studied: 
\begin{align*}
    \Pr_{\On}(\|\phihat(\bx) - \phihat(\bz)\|_q \leq \Gamman) &= \int_{\R^d} \int_{\R^d}\ind[\|\bu - \bv\|_q \leq \Gamman]dF_{(\phihat(\bx), \phihat(\bz))}(\bu, \bv).
\end{align*}
Let us change variables from $\bu, \bv$ to $\bu,\br$, where $\br = \frac{\bu - \bv}{\Gamman}$ with Jacobian determinant equal to $\Gamman^d$. We have:
\begin{align*}
    \Pr_{\On}(\|\phihat(\bx) - \phihat(\bz)\|_q \leq \Gamman) &= \Gamman^d\int_{\R^d} \int_{\R^d}\ind[\|\br\|_q \leq 1]dF_{(\phihat(\bx), \phihat(\bz))}(\bu, \bu - \Gamman \br).
\end{align*}
Consider now the limiting behavior of $\Gamman^{-d}\Pr_{\On}(\|\phihat(\bx) - \phihat(\bz)\|_q \leq \Gamman)$, we have:
\begin{align*}
    \lim_{n\rightarrow\infty}\Gamman^{-d}\Pr_{\On}(\|\phihat(\bx) - \phihat(\bz)\|_q \leq \Gamman) &= \lim_{n\rightarrow \infty}\int_{\R^d} \int_{\R^d}\ind[\|\br\|_q \leq 1]dF_{(\phihat(\bx), \phihat(\bz))}(\bu, \bu - \Gamman \br)\\
    &=  \int_{\R^d} \int_{\R^d}\ind[\|\br\|_q \leq 1] \lim_{n\rightarrow \infty}dF_{(\phihat(\bx), \phihat(\bz))}(\bu, \bu - \Gamman \br)\\
    &= \int_{\R^d} \int_{\R^d}\ind[\|\br\|_q \leq 1] dF_{\phi(\bx)}(\bu)F_{\phi(\bz)}(\bu) \\
    &= \int_{\R^d} \int_{\R^d}\ind[\|\br\|_q \leq 1] \delta(\phi(\bx) - \bu)\delta(\phi(\bz) - \bu)d\bu d\br\\
    &= \int_{\R^d}\ind[\|\br\|_q \leq 1]  d\br\int_{\R^d}\delta(\phi(\bx) - \bu)\delta(\phi(\bz) - \bu)d\bu \\
    &= \frac{2Ga(\frac{2}{q} + 1)^d}{Ga(\frac{d}{q} + 1)}\delta(\phi(\bx) - \phi(\bz)) = V_d \delta(\phi(\bx) - \phi(\bz)). 
\end{align*}
where the second equality follows from the dominated convergence theorem (DCT), which we can apply because the indicator function is upper bounded by 1. The fourth equality follows from the fact that $F_{\phi(\bx)}$  is absolutely continuous wrt. the Lebesgue measure, and therefore has density equal to $\delta(\phi(\bx))$. The final equality follows from the definition of the volume under the $d$-dimensional unit ball around 0 defined by the $q$-norm, i.e., $V_d = \int_{\R^d}\ind[\|\br\|_q \leq 1]d\br = \frac{2Ga(\frac{2}{q} + 1)^d}{Ga(\frac{d}{q} + 1)}$,  where $Ga$ is the Gamma function \citep[see e.g., ][]{wang2005volumes}, and from known properties of Dirac's $\delta$ function. 

We can now study the limiting behavior of $\Pr_{\On}(\bZ \in \MGxt)$, for $(\bZ, T) \sim f_{\bX, T}$, using the result just obtained and applying the DCT once again. Define first the density function $f_{\phi(\bX)}(\bu) = \int_{\X} \fbX(\bz)\delta(\phi(\bz) - \bu)d\bz$. Fix now an arbitrary unit $i$ in the matching data with covariates $\bX_i = \bZ$ and treatment indicator $T_i = T$. We have: 
\begin{align*}
   \lim_{n\rightarrow \infty}\Gamman^{-d}\Pr_{\On, \bZ, T}(\bZ \in \MGxt) & = \lim_{n\rightarrow\infty}\Gamman^{-d}\Pr_{\On, \bZ, T}(\Dphiqhat(\bZ, \bx) \leq \Gamman, T = t) \\
    &= \lim_{n\rightarrow\infty}\Gamman^{-d}\Pr(T=t)\E_{\bZ}[\Pr_{ \On }(\Dphiqhat(\bZ, \bx) \leq \Gamman | T = t)|\bZ]\\
    &= \lim_{n\rightarrow\infty}\Gamman^{-d}e(t)\int_{\X} \Pr_{\On}(\Dphiqhat(\bz, \bx) \leq \Gamman|\bZ=\bz, T=t)d\FbXt(\bz)\\
    &= \lim_{n\rightarrow\infty}\Gamman^{-d}e(t)\int_{\X} \Pr_{\On}(\Dphiqhat(\bz, \bx) \leq \Gamman)d\FbXt(\bz)\\
    &= e(t)\int_{\X} \lim_{n\rightarrow\infty}\Gamman^{-d}\Pr_{\On}(\|\phihat(\bx) - \phihat(\bz)\|_q \leq \Gamman)d\FbXt(\bz)\\
    &= e(t)\int_{\X} V_d \delta(\phi(\bx) - \phi(\bz))d\FbXt(\bz)\\
    &= V_de(t)\int_{\X}\fbX(\bz)\int_{\R^d}\delta(\phi(\bx) - \bu)\delta(\phi(\bz) - \bu)d\bu d\bz\\
    &= V_de(t)\int_{\R^d}\delta(\phi(\bx) - \bu)\int_{\X}\fbX(\bz)\delta(\phi(\bz) - \bu)d\bz d\bu \\
    &= V_de(t)\int_{\R^d}\delta(\phi(\bx) - \bu)f_{\phi(\bX)|T=t}(\bu)d\bu\\
    &= V_de(t)f_{\phi(\bX)|T=t}(\phi(\bx)). 
\end{align*}
Note that the independence of the probabilities in the second and fourth equalities come from the fact that $\bZ$  and $T$ are drawn independently of $\On$, and that the fifth equality comes from an application of the DCT, made possible by the fact that the probability inside the integral is upper bounded by 1. This proves point (i), we now move to point (ii). 

Since $\Pr_{\On, \bZ, T}(\bZ \in \MGxt)$ is upper bounded by 1, another application of the DCT to the integral $\E_{\bX}[\Pr_{\On, \bZ, T}(\bZ \in \MGXt)] = \int_{\X} \fbX(\bx) \Gamman^d\Pr_{\On, \bZ, T}(\bZ \in \MGxt)d\bx$ is sufficient to verify the second statement. 
\end{proof}

\subsection{Proof of Theorem \ref{thm:CRFasym}}
Throughout the rest of this section we will be referring to the corrected estimator: $$\mudxtilde = \frac{1}{\Nxt + 1}\sum_{i\in \MGxt} Y_i.$$
This estimator has asymptotic behavior identical to $\muhat(\bx, t)$, but is easier to work with as it is still defined even if the matched group for $\bx$, $t$ is empty. We will be proving the results for $\mudxtilde$, but they will apply in the same way to $\muhat(\bx, t)$. 

\begin{proof}
We will first prove the result in (i), and after the result in (ii). 
\paragraph{Claim (i)} Recall that $n^r := \min(n^{\frac{1}{2 + d}}, n^{r_{ML}})$. By Lemma \ref{thm:martingaleCLT} (proved below), we know that we need to verify 3 conditions  in order to prove asymptotic normality for $\mudxtilde$, namely that: 1) $\Rx = o_p(n^{-r})$, 2) $\frac{n^{2r}\Nxt}{(\Nxt+1)^2} \pconv \frac{1}{K}$ for some constant, $K$, and  3)  $\E\left[\left(\frac{n^{2r}}{\Nxt + 1}\right)^{2}\right] = O(1)$. 

Starting from the first condition, let $\Xmax = X_{i^*}$, s.t: $i^* \in \argmax_{i \in \MGxt} \Dphiq(\bX_i, \bx)$, and recall that, by definition: $\Rx = \Dphiq(\Xmax, \bx)$.  By assumption that matches are made with Caliper M-ML, we have $\Dphiqhat(\Xmax, \bx) \leq \Gamman$, and by assumption that $\Gamman  \asymp  n^{\frac{2r - 1}{d}}$ it follows that:
\begin{align}
n^r\Dphiqhat(\Xmax, \bx)&\leq n^r\Gamman = O(n^{r}n^{\frac{2r-1}{d}}) \label{eq:CaliperOrder}\\
&= O(n^{r(1 + \frac{2}{d}) - \frac{1}{d}}) \rightarrow 0, \nonumber
\end{align}
for any $r < \frac{1}{2 + d}$. Using this fact, along with A6c, we can show $n^r\Rx \pconv 0$ and verify Condition 1 as follows:
\begin{align*}
    n^r\Rx &= n^r\Dphiq(\Xmax, \bx)\\
     &= n^r\|\phi(\Xmax) - \phi(\bx) + \phihat(\Xmax) - \phihat(\Xmax) + \phihat(\bx) - \phihat(\bx)\|_q\\
     &\leq n^r\|\phi(\Xmax) - \phihat(\Xmax)\|_q + n^r\|\phihat(\bx) - \phi(\bx)\|_q \\
     &+ n^r\|\phihat(\Xmax) - \phihat(\bx)\|_q\\
     &= o_p(1) + o_p(1) + n^r\Dphiqhat(\Xmax, \bx) = o_p(1). 
\end{align*}
The inequality follows from the triangle inequality, and the last line follows by Assumption A6 and Eq. \eqref{eq:CaliperOrder}. 

Second, we will show that Condition 2 holds by showing that $\frac{n^{2r}\Nxt}{(\Nxt + 1)^2} \pconv \frac{1}{\nux}$, where: $\nux = V_de(t)f_{\phi(\bX)|T=t}(\phi(\bx))$ as defined in Lemma \ref{thm:iinmg}. Note that $\Nxt$ is a binomial random variable with size $n$ and probability $\Pr(i \in \MGxt)$. We know by Lemma \ref{thm:iinmg} that: 
\begin{equation}
    E_{\Nxt}\left[\frac{\Nxt}{n^{2r}}\right] = n^{1-2r}\Pr(i \in \MGxt) \pconv \nux \label{eq:NxtRate}
\end{equation}
whenever $\Gamman \asymp n^{\frac{2r-1}{d}}$ as we have assumed in this theorem. By Markov's inequality, this implies that $\frac{\Nxt}{n^{2r}} \pconv \nux$. Consider now the quantity: $(\Nxt + 1)^2$: we know by Eq. \eqref{eq:NxtRate} that: $\Nxt = O_p(n^{2r})$, which implies that $\Nxt^2 = O_p(n^{4r})$ and therefore:
$$ n^{-2r}(\Nxt + 1)^2 \geq n^{-2r}\Nxt^2 = n^{-2r}O_p(n^{4r}) \pconv \infty.$$ Then we can apply the continuous mapping theorem to $g(\Nxt +1)$, where: $g(a) = \frac{1}{n^{-2r}a} - \frac{1}{n^{-2r}a^2}$, because $a = \Nxt + 1 >0$ in our context, and $g(a)$ as defined is continuous over that domain. Then Condition 2 is verified as follows:
\begin{align*}
    \frac{n^{2r}\Nxt}{(\Nxt + 1)^2} &= \frac{\Nxt + 1 - 1}{n^{-{2r}}(\Nxt + 1)^2} \\
    &=  \frac{1}{n^{-{2r}}(\Nxt + 1)} - \frac{1}{n^{-{2r}}(\Nxt + 1)^2}\pconv \nux^{-1}.
\end{align*}
To verify the last condition needed for Lemma \ref{thm:martingaleCLT} we first examine two related quantities. 
First, we know that $\Nxt$ is binomial with size $n$ and probability $\Pr(i \in \MGxt)$, and, therefore, its second moment is $\E[\Nxt^2] = n(n-1)\gamma^2 + n\gamma$, where $\gamma = \Pr(i \in \MGxt) = O(\Gamman^d) = O(n^{2r-1})$, and $(n^{1-2r}\gamma)^2 \rightarrow \nux^{2}$, by Lemma \ref{thm:iinmg} and the Continuous Mapping Theorem. Therefore we have:
\begin{align*}
    \E\left[\left(\frac{\Nxt}{n^{2r}}\right)^2\right] &= \frac{n(n-1)}{n^{4r}}\gamma^2 + \frac{n}{n^{4r}}\gamma\\
    &= n^{2-4r}\gamma^2 - n^{1-4r}\gamma^2 + n^{1-4r}\gamma\\
    &= (n^{1-2r}\gamma)^2 - n^{1-4r}O(n^{4r-2}) + n^{1-4r}O(n^{2r-1})\\
    &= (n^{1-2r}\gamma)^2 - O(n^{-1}) + O(n^{-2r}) \rightarrow \nux^2.
\end{align*}
Another application of the results just stated gives us  the following: 
\begin{align*}
    \E\left[\left(\frac{\Nxt + 1}{n^{2r}}\right)^2\right] 
    &= \E\left[\frac{\Nxt^2 + 2\Nxt + 1}{n^{4r}}\right]\\
    &=  \E\left[\left(\frac{\Nxt}{n^{2r}}\right)^2\right] + \frac{2}{n^{4r}}\E[\Nxt] + n^{-4r}\\
    &= \E\left[\left(\frac{\Nxt}{n^{2r}}\right)^2\right] + o(1) + o(1) = O(1) \rightarrow \nux^{2}.
\end{align*}
The above display implies, by Markov's inequality, that $\left(\frac{\Nxt + 1}{n^{2r}}\right)^2 \pconv \nux^2$. Additionally, since the function $g(a) = 1/a$ is continuous over the positive reals, and $\Nxt + 1$ is within this domain, we can apply the continuous mapping theorem to conclude that: $\left(\frac{n^{2r}}{\Nxt + 1}\right)^2 \pconv \nux^{-2}$. Since $g(a)$ is also bounded above by 1 over the same domain, the dominated convergence theorem can be applied to conclude that $\E\left[\left(\frac{n^{2r}}{\Nxt + 1}\right)^2\right] \pconv \nux^{-2}$ as well,  which directly implies the condition we needed to verify. Since all three conditions are verified, result (i) in the theorem follows from applying Lemma \ref{thm:martingaleCLT}. 

\paragraph{Claim (ii)} 
Let $\bar \mu(\bx, t) = \frac{1}{\Nxt + 1}\sum_{i=1}^n \Wixt \mu(\bX_i, t)$: we apply the triangle inequality to break up the error into two components featuring this term:
\begin{align}
    \|\tilde{\mu}(\bx, t) - \mu(\bx, t) \|_{\cP, s} &=  \left(\E[|\tilde{\mu}(\bx, t) - \bar \mu(\bx, t) + \bar \mu(\bx, t) - \mu(\bx, t) |^s]\right)^{1/s} \nonumber\\
    &\leq 2^{1-1/s}\left((\underbrace{\E[|\tilde{\mu}(\bx, t) - \bar \mu(\bx, t)|^s]}_{I_1})^{1/s} +(\underbrace{\E[|\bar \mu(\bx, t) - \mu(\bx, t) |^s]}_{I_2})^{1/s}\right).\label{eq:I2decomp}
\end{align}
We proceed by upper bounding $I_1$ and $I_2$ separately. Starting with component $I_2$, recall that $\Nxt = \sum_{i=1}^n \Wixt$ and that $\Nxt$ follows a binomial distribution. Now by Lemma \ref{thm:iinmg} and continuous mapping, we have $(n\Gamman^{d})^{-s}\Nxt^s  \pconv \nux^s$, therefore: $\frac{(n\Gamman^d)^{s}}{(\Nxt + 1)^{s}} \pconv \nux^{-s}$ also by continuous mapping theorem. Finally, by dominated convergence we have:
\begin{equation}
\E\left[\frac{1}{(\Nxt + 1)^{s}}\right] = O((n\Gamman^d)^{-s}). \label{eq:nxtconv}
\end{equation}
\begin{align}
    I_2 &= \E[|\bar \mu(\bx, t) - \mu(\bx, t) |^s]= \E\left[\left|\frac{1}{\Nxt + 1}\sum_{i=1}^n \Wixt \mu(\bX_i,t) -  \mu(\bx, t) \right|^s\right]\nonumber \\
    &= \E\left[\left|\frac{1}{\Nxt + 1}\sum_{i=1}^n \left(\Wixt \mu(\bX_i,t) -  \mu(\bx, t)\right) + \frac{\mu(\bx, t)}{\Nxt + 1} \right|^s\right]\nonumber \\
    &\leq 2^{s-1}\E\left[\left|\frac{1}{\Nxt + 1}\sum_{i=1}^n \left(\Wixt \mu(\bX_i,t) -  \mu(\bx, t)\right)\right|^s + \left|\frac{\mu(\bx, t)}{\Nxt + 1} \right|^s\right]\nonumber \\
    &\leq 2^{s-1}\E\left[\left|\frac{1}{\Nxt + 1}\sum_{i=1}^n \Wixt C_L\Dphiq(\bX_{i}, \bx) \right|^s+ \left|\frac{\mu(\bx, t)}{\Nxt + 1} \right|^s\right]\nonumber\\
    &\leq 2^{s-1}\E\left[\left|\frac{C_L}{\Nxt + 1}\sum_{i=1}^n \Wixt \Rx \right|^s \right] +2^{s-1}\E\left[\left|\frac{\mu(\bx, t)}{\Nxt + 1} \right|^s\right] \nonumber\\
    &= 2^{s-1}C_L^s\E\left[\left(\frac{\Nxt}{\Nxt + 1}\right)^s\Rx^s\right] + 2^{s-1}|\mu(\bx, t)|^s\E\left[\frac{1}{(\Nxt + 1)^s}\right] \nonumber \\
    &\leq 2^{s-1}C_L^s\E\left[\| \phi(\Xmax) - \phihat(\Xmax) + \phihat(\Xmax) - \phihat(\bx) + \phihat(\bx) - \phi(\bx)\|_q^s\right] + O((n\Gamman^{d})^{-s})\nonumber\\
    &\leq 2^{s-1}C_L^s  \E[\|\phihat(\Xmax) - \phi(\Xmax)\|_q^s] + 2^{s-1}C_L^s  \E[\|\phihat(\bx) - \phi(\bx)\|_q^s] \nonumber\\ 
    &\quad + 2^{s-1}C_L^s  \E[\|\phihat(\Xmax) - \phihat(\bx)\|_q^s] + O((n\Gamman^{d})^{-s})\nonumber\\
    &= o(n^{-s r_{ML}}) + o(n^{-s r_{ML}}) + O(\Gamman^{s }) + O((n\Gamman^{d})^{-s}).\label{eq:I2boundcal}
\end{align}
The first inequality follows from A5, the second by definition of $\Rx$, and the fourth from the triangle inequality. Note also that the fraction at the fourth line is always less than 1. The statement in the last line follows from A6, and from the fact that $\|\phihat(\Xmax) - \phihat(\bx)\|_q^s \leq \Gamman$ under caliper M-ML. 

We use Lemma \ref{thm:mombound} in order to upper bound $I_1$. Letting $\tilde{Y} = (Y_1, \dots Y_n)$, $\tilde{\bX} = (\bX_1, \dots, \bX_n)$, and $\tilde{T} = (T_1, \dots, T_n)$, we have:
\begin{align}
    I_1 &= \E\left[\left|\frac{1}{\NXt +1}\sum_{i=1}^n(Y_i - \muXd)\WiXt\right|^s\right]\nonumber\\
    &= \E\left[\left|\frac{1}{\NXt +1}\right|^s\E_{\tilde Y | \tilde \bX, \tilde T}\left[\left|\sum_{i=1}^n(Y_i - \muXd)\WiXt\right|^s\right]\right]\nonumber\\ 
   \mbox{(By Lemma \ref{thm:mombound})} &\leq \E\left[\frac{B_sC_s\Nxt^{s/2}}{(\NXt +1)^s}\right] \nonumber \\
    &\leq B_sC_s\E\left[\frac{\Nxt^{s/2}}{(\NXt )^s}\right] =  B_sC_s\E\left[\frac{1}{(\NXt )^{s/2}}\right]. 
\end{align}
Therefore we have: 
\begin{equation}
    I_1 \leq B_sC_s\E\left[\frac{1}{(\NXt )^{s/2}}\right] = O((n\Gamman^d)^{-s/2}). \label{eq:I1boundcal}
\end{equation}
Putting together \eqref{eq:I2decomp}, \eqref{eq:I1boundcal}, and \eqref{eq:I2boundcal}, and applying Jensen's inequality, we obtain:
\begin{align}
    \|\tilde \mu(\bx, t) - \mu(\bx, t) \|_{\cP, s} &\leq 2^{1 - 1/s}(I_1 + I_2)^{1/s} = O((n\Gamman^d)^{-1/2}) + O(\Gamman) + o(n^{-r_{ML}})\label{eq:calgenbound}.
\end{align}
The bound in the theorem can be obtained by setting $\Gamman = n^{\frac{2r - 1}{d}}$ and $r = \frac{1}{2 + d}$ and plugging into \eqref{eq:calgenbound}: 
\begin{align}
    \|\tilde \mu(\bx, t) - \mu(\bx, t) \|_{\cP, s} &= O(n^{-\frac{1}{2 + d}}) +  o(n^{-r_{ML}}).
\end{align}
Since the domain of $\bX$ is bounded, we can apply the DCT to $\|\mudXtilde - \muXd\|_{\cP, s} = \E_{\bX}[\|\mudXtilde - \muXd\|_{\cP}^s]^{1/s}$ to see that the bound holds in expectation over $\bX$ as well. This concludes the proof.
\end{proof}


\subsection{Proof of Theorem \ref{thm:knnasym}}

\begin{proof}

\textbf{Claim (ii)} We will start by bounding the error: $\|\muhat(\bx, t) - \mu(\bx, t) \|_{\cP, s}$ for arbitrary $\bx \in \X$, as many of the steps of the proof of Claim (i) become simple after introducing this result. The result in Claim (ii) will follow by application of the DCT to this first result. Recall that we have defined $\bar \mu(\bx, t) = \frac{1}{k_n}\sum_{i=1}^n \Wixt \mu(\bX_i, t)$, we apply the triangle inequality to break up the error into two components featuring this term:
\begin{align*}
    \|\muhat(\bx, t) - \mu(\bx, t) \|_{\cP, s} &=  \E[\|\muhat(\bx, t) - \bar \mu(\bx, t) + \bar \mu(\bx, t) - \mu(\bx, t) \|^s]^{1/s} \\
    &\leq 2^{1-1/s}\left[(\underbrace{\E[\|\muhat(\bx, t) - \bar \mu(\bx, t)\|^s]}_{I_1})^{1/s} + (\underbrace{\E[\|\bar \mu(\bx, t) - \mu(\bx, t) \|^s]}_{I_2})^{1/s}\right].
\end{align*}
We use Lemma \ref{thm:mombound} in order to upper bound $I_1$. Letting $\tilde{Y} = (Y_1, \dots Y_n)$, $\tilde{\bX} = (\bX_1, \dots, \bX_n)$, and $\tilde{T} = (T_1, \dots, T_n)$, we have:
\begin{align}
    I_1 = \E\left[\left|\frac{1}{k_n}\sum_{i=1}^n(Y_i - \muXd)\WiXt\right|^s\right] &= \E\left[\left|\frac{1}{k_n}\right|^s\E_{\tilde Y | \tilde \bX, \tilde T}\left[\left|\sum_{i=1}^n(Y_i - \muXd)\WiXt\right|^s\right]\right]\nonumber\\ 
   \mbox{(By Lemma \ref{thm:mombound})} &\leq \E\left[\frac{B_sC_sk_n^{s/2}}{k_n^s}\right] \nonumber  =  B_sC_s\frac{1}{k_n^{s/2}}. 
\end{align}

Before directly upper-bounding bias term $I_2$, we establish a bound on the quantity: $\E[\|\phihat(\bX_{(k_n)}) - \phihat(\bx)\|_q^s] = \Dphiqhat(\bX_{(k_n)}, \bx)^s$. Note that the transformed covariates, $\phihat(\bX)$, have continuous and bounded density by continuity and boundedness of $\fbX$ (the density function of the original covariates), and continuity of $\phihat(\bx)$ at all $\bx \in \X$. With these facts, we can apply  Lemma \ref{thm:knndistbound} to $\phihat(\bX_1), \dots, \phihat(\bX_n)$ as inputs, and with $\lambda = -s$ and $\gamma = 0$ ($\lambda$ and $\gamma$ are defined in Lemma \ref{thm:knndistbound}). From this we obtain:
\begin{equation}
\E[\Dphiqhat(\bX_{(k_n)}, \bx)^s] = O\left(\left(\frac{k_n}{n}\right)^{s/d}\right).\label{eq:knndistbound}
\end{equation}


Using this result we can  now switch to upper-bounding the bias term $I_2$.\\
Recall that $\Rx = \max_{i=1,\dots,n} \Wixt \Dphiq(\bx, \bX_i)$, and note that in this case $\Rx$ is the $\phi$-distance between $\bx$ and its $k_n^{th}$ nearest neighbor within the matching sample. We have:
\begin{align*}
    I_2 &= \E[|\bar \mu(\bx, t) - \mu(\bx, t) |^s]= \E\left[\left|\frac{1}{k_n}\sum_{i=1}^n \Wixt \mu(\bX_i,t) -  \mu(\bx, t) \right|^s\right] \\
    &= \E\left[\left|\frac{1}{k_n}\sum_{i=1}^n \Wixt(\mu(\bX_i,t) -  \mu(\bx, t)) \right|^s\right] \leq \E\left[\left|\frac{1}{k_n}\sum_{i=1}^n \Wixt C_L\Dphiq(\bX_{i}, \bx) \right|^s\right]\\
    &\leq \E\left[\left|\frac{C_L}{k_n}\sum_{i=1}^n \Wixt \Rx \right|^s\right] = C_L\E[\Rx^s]
\end{align*}
where the first inequality follows from Assumption 5 and the second from the definition of $\Rx$. Let $\bX_{(k_n)}$ be the covariates of $\bx$'s $k_n^{th}$ nearest neighbor in terms of $\Dphiqhat$, and note that, using \eqref{eq:knndistbound}, we have:
\begin{align}
    \E[\Rx^s] &= \E[\|\phi(\bX_{(k_n)}) - \phi(\bx)\|_q^s]\nonumber\\
    &\leq 2^{s-1}(\E[\|\phi(\bX_{(k_n)}) - \phihat(\bX_{(k_n)})\|_q^{s}] + \E\|\phihat(\bx) - \phi(\bx)\|_q^{s}] + \E[\|\phihat(\bX_{(k_n)}) - \phihat(\bx)\|_q^s])\nonumber\\
    &= o(n^{-r_{ML}s}) + o(n^{-r_{ML}s}) + 2^{s-1}\E[\Dphiqhat(\bX_{(k_n)}, \bx)^s]\nonumber\\
    &= o(n^{-r_{ML}s}) + O\left(\left(\frac{k_n}{n}\right)^{s/d}\right)\label{eq:Rnkbound}
\end{align}
where the fact that terms like $\E[\|\phihat(\bx) - \phi(\bx)\|_q]$ are $o(n^{-r_{ML}})$ follows from Assumption 6. 

Finally, we can put together the bounds obtained so far to establish the result in the theorem: 
\begin{align*}
    \|\muhat(\bx, t) - \mu(\bx, t) \|_{\cP, s} &\leq 2^{1-1/s}(I_1 + I_2)^{1/s} \\
    &\leq 2^{1-1/s}\left[(B_sC_sk_n^{-s/2})^{\frac{1}{s}} + \left(O\left(\left(\frac{k_n}{n}\right)^{s/d}\right) + (o(n^{-sr_{ML}}))\right)^{\frac{1}{s}}\right]\\
    &\leq 2^{1-1/s}\left[C_sB_s^{1/s}k_n^{-1/2} + O\left(\left(\frac{k_n}{n}\right)^{1/d}\right) + o(n^{-r_{ML}})\right]\\
    &= O(k_n^{-1/2}) + O\left(\left(\frac{k_n}{n}\right)^{1/d}\right) + o(n^{-r_{ML}}).
 \end{align*}
This result can be easily extended to $\|\muhat(\bX, t) - \mu(\bX, t) \|_{\cP, s}$ by application of the DCT to $\|\muhat(\bX, t) - \mu(\bX, t) \|_{\cP, s} = (\E_{\bx \sim \fbX}[\|\muhat(\bx, t) - \mu(\bx, t)\|_{\cP, s}^s])^{1/s}$. The final lower bound in the Theorem can be obtained by plugging $k_n = n^{\frac{2}{2 + d}}$ into the bound above and simplifying. 

\paragraph{Claim (i)} We finish by proving the asymptotic normality result given in (i). We will appeal to Lemma \ref{thm:martingaleCLT} to verify this claim, and therefore need to verify that the three conditions required in the lemma hold in this case. Condition 1 holds since by \eqref{eq:Rnkbound} and Markov's inequality we have: $\Rx = o_p(n^{-r})$ almost surely over $f_\bX$ whenever we set $k_n = Kn^{2r}$, where $r = \min(\frac{1}{2+d}, r_{ML})$. Second, for Conditions 2 and 3, we know that $\Nxt = k_n$, and therefore: $\frac{n^{2r}}{\Nxt} = \frac{n^{2r}}{k_n} \rightarrow \frac{1}{K}$ by assumption. This also implies that $\E\left[\left(\frac{n^{2r}}{\Nxt}\right)^2\right] \rightarrow \frac{1}{K^2} = O(1)$. Therefore Conditions 2 and 3 of Lemma \ref{thm:martingaleCLT} are verified, and the lemma directly implies the result. 
\end{proof}

\subsection{Proof of Theorem \ref{thm:varasym}}

\begin{proof}We will essentially use all the same arguments employed before to show this fact. Let $\etaxd = \E[Y(t)^2|\bX=\bx]$, and recall that we have previously defined $\muxd = \E[Y(t)|\bX=\bx]$, and $\Wixt$ to be a binary variable denoting membership in $\MGxt$. We will first concentrate on showing that: $\frac{1}{\Nxt}\sum_{i=1}^n \Wixt Y_i^2 \pconv \etaxd$. To see that this is indeed the case, notice that:
\begin{align*}
    &\frac{1}{\Nxt}\sum_{i=1}^n \Wixt Y_i^2 - \etaxd \\
    &= \frac{1}{\Nxt}\sum_{i=1}^n \Wixt (Y_i^2 - \etaXid) +\frac{1}{\Nxt}\sum_{i=1}^n \Wixt (\etaXid - \etaxd).
\end{align*}
Starting with the first term, we have, for any $i$: $\E[Y_i^2 - \etaXid|T_i, \bX_i] = 0$ by definition of $\etaXid$, which implies that $\E[\Wixt(Y_i^2 - \etaXid)] = \E_{T_i, \bX_i}[\Wixt\E_{Y_i|\bX_i, T_i}[(Y_i^2 - \etaXid)]] = 0$, and therefore, by the weak law of large numbers: $\frac{1}{\Nxt}\sum_{i=1}^n  \Wixt(Y_i^2 - \etaXid) \pconv 0$. 
Moving on to the second term, note first that Assumption A5 implies, for all $\bu, \bv \in \mathcal{X}$: 
\begin{align*}
    |\mu(\bu, t)^2 - \mu(\bv, t)^2| &= |(\mu(\bu, t) - \mu(\bv, t))(\mu(\bu, t) + \mu(\bv, t))|\\
    &\leq |(\mu(\bu, t) - \mu(\bv, t))||(\mu(\bu, t) + \mu(\bv, t))|\\
    &\leq C_L \Dphiq(\bu, \bv)|(\mu(\bu, t) + \mu(\bv, t))| \\
    &\leq C_L \Dphiq(\bu, \bv)2C_\delta ,
\end{align*}
where the first inequality follows by the Cauchy-Schwartz inequality, the second by the Lipschitz condition of Assumption A5, and the third by Assumption A4, which implies that $|\mu(\bu, t)|$ is bounded by some constant $C_\delta$. Applying the above to the second term of the previous expression we see that:
\begin{align*}
    &\frac{1}{\Nxt}\sum_{i=1}^n \Wixt (\etaXid - \etaxd) \\
    &= \frac{1}{\Nxt}\sum_{i\in \MGxt} (\sigmaXid + \muXid^2 - \sigmaxd - \muxd^2)\\
    &= \frac{1}{\Nxt}\sum_{i\in\MGxt} (\sigmaXid- \sigmaxd) + (\muXid^2 - \muxd^2)\\
    &\leq \frac{1}{\Nxt}\sum_{i\in \MGxt} C_L \Dphiq(\bX_i, \bx) + 2C_\delta C_L \Dphiq(\bX_i, \bx)\\
    &\leq (C_L + 2C_\delta C_L)\max_{i \in \MGxt}\Dphiq(\bX_i, \bx) = O_p(\Rx), 
\end{align*}
where the first equality follows by definition of variance, the first inequality by Assumption A5, and the second inequality by the definition of max. By the same argument as the proofs of Thms \ref{thm:CRFasym} and \ref{thm:knnasym}, we know that $\Rx \pconv 0$, both when matches are made with a caliper (by Eq. \eqref{eq:I2boundcal}), and with fixed $k_n$ (by Eq. \eqref{eq:knndistbound}). Therefore, $\frac{1}{\Nxt}\sum_{i=1}^n \Wixt (\etaXid - \etaxd) \pconv 0$. This establishes convergence of $\frac{1}{\Nxt}\sum_{i\in \MGxt} Y_i^2$ to $\etaxd$. Using this result, we can show that the simple variance estimator is indeed consistent for the CATE variance: 
\begin{align*}
    &\frac{1}{\Nxt}\sum_{i \in \MGxt} (Y_i - \mudxhat)^2 = \frac{1}{\Nxt}\sum_{i \in \MGxt}(Y_i^2 - 2Y_i\mudxhat + \mudxhat^2)\\
    &= \left(\frac{1}{\Nxt}\sum_{i \in \MGxt} Y_i^2\right) - \left(\mudxhat\frac{2}{\Nxt}\sum_{i \in \MGxt} Y_i\right) + \mudxhat^2\\
    &= \frac{1}{\Nxt}\sum_{i \in \MGxt} Y_i^2 - \mudxhat^2\\
    & \pconv \etaxd - \muxd^2 = \sigmaxd,
\end{align*}
where converge of $\mudxhat^2$ to $\muxd^2$ follows from the continuous mapping theorem applied to $\mudxhat \pconv \muxd $.
\end{proof}

\subsection{Proof of Theorem \ref{thm:DML}}

\begin{proof} The proof of this statement follows from applying Theorem 5.1 in \citep{chernozhukov2018double} to the estimators $\psihat(\bX_i, t)$. The theorem can be applied because M-DML is a case of the DML2 algorithm in Definition 3.2 of the same paper, which is covered by Theorem 5.1. This theorem requires us to check that our estimators satisfy Assumption 5.1 in the same paper, which is comprised of several conditions. We first examine the primitive conditions of Assumption 5.1 of \cite{chernozhukov2018double}. All of these conditions are satisfied by our main assumptions, specifically:  Condition (i) is satisfied by assumption in the theorem, Condition (ii) is satisfied by A4, Condition (iii)  is satisfied by A2, Condition (iv) is satisfied by A1, Condition (v) is satisfied by A4. 

We now need to verify three conditions on the estimators of $\muhat$ and $\ehat$. First, let $\eta(\bx) = (\mu(\bx, 1), \mu(\bx, 0), e(\bx,1), e(\bx, 0))$, where we use $\eta(\bx)_j$ to refer to the $j^{th}$ component of this vector, for $j=1,\dots, 4$. Additionally, let the respective estimator for this quantity be $\hat{\eta}_{\Dnl}(\bx) = (\muhat(\bx, 1), \mu(\bx, 0), \ehat(\bx, 1), \ehat(\bx, 0))$, where we use the notation $\Dnl$ to emphasize that the estimator $\hat{\eta}_{\Dnl}(\bx)$ depends on data not in fold $\ell$: $\Dnl = \{\bX_i, T_i, Y_i\}_{i \in S_{\setminus \ell}}$. For $s > 2$, define: 
$$\|\hat{\eta}_{\Dnl} - \eta\|_{\cP, s} = \max_{j \in 1,\dots,4}\int_{\Dnl, \bX}\|\hat{\eta}_{D}(\bx)_j - \eta(\bx)_j\|_q f_{\Dnl}(D)\fbX(\bx)d D\bx.$$
In order to show that Assumption 5.1 of \cite{chernozhukov2018double} is satisfied in our setting, we nee to show that $\|\hat{\eta}_{\Dnl} - \eta\|_{\cP, s} \leq C$ for some strictly positive $C$. Note first that $\|\ehat(\bx, t) - e(\bx, t)\|_{\cP, s} < 1$, since $0 < \ehat(\bx, t) < 1$, and $0 < e(\bx, t) < 1$ for all datasets $D$, $\bx$ and $t$ by construction. It remains to show that $\|\muhat - \mu\|_{\cP, s} \leq C$. This follows from our A4 by setting $\delta = s - 2$ therein. Then, for any unit, $i$, $t$, and $\bx$:
\begin{align*}
    \E_{Y_i}[\|Y_i - \mu(\bx, t)\|_s|\bX_i=\bx_i, T_i=t] & \leq \E_Y[\|Y_i\|_s|\bX_i=\bx_i, T_i=t] + \|\mu(\bx, t)\|_s\\
    &= \E_{Y_i}[\sqrt[s]{|Y_i|^s}|\bX_i=\bx_i, T_i=t] + \E_Y[\sqrt[s]{|Y|^s}|\bX=\bx, T=t]\\ 
    &\leq \E_{Y_i}[\|Y_i\|^{s}|\bX_i=\bx_i, T_i=t] + \E_Y[\|Y\|^{s}|\bX=\bx, T=t]\\ 
    &\leq C_s + C_s,
\end{align*}
therefore: 
\begin{align*}
    \|\muhat - \mu\|_{\cP, s} &= \E[\E_{Y_{\setminus \ell}}[\|\muhat(\bx, t) - \mu(\bx, t)\|_q| \bX=\bx, \bX_{\setminus \ell}, T_{\setminus \ell}]]\\
    &= \E\left[\E_{Y_{\setminus \ell}}\left[\biggr\|\frac{1}{\Nxt}\sum_{i \in \MGxt} Y_i - \mu(\bx, t)\biggr\|_q\biggr| \bX=\bx, \bX_{\setminus \ell}, T_{\setminus \ell}\right]\right]\\
    &\leq \E\left[\E_{Y_{\setminus \ell}}\left[\max_{i \in \MGxt}\biggr\|Y_i - \mu(\bx, t)\biggr\|_q\biggr| \bX=\bx, \bX_{\setminus \ell}, T_{\setminus \ell}\right]\right]\\
    &\leq 2C_{s}. 
\end{align*}
Second, we must show that $\|\hat{\eta}_{\Dnl} - \eta\|_{\cP, 2} = o(1)$. This is true for $\|\ehat - e\|_{\cP, 2}$ by assumption, and it holds true for $\|\muhat - \mu\|_{\cP, 2}$ by Theorems \ref{thm:CRFasym}, for Caliper M-ML, and \ref{thm:knnasym} for KNN-MML. Finally, the last requirement for Assumption 5.1 of \cite{chernozhukov2018double} is that:
$$\|\muhat - \mu\|_{\cP, 2} \times \|\ehat - e\|_{\cP, 2} = O(n^{-\frac{1}{2}}),$$
by Theorems \ref{thm:CRFasym}, and \ref{thm:knnasym}, we know that $\|\hat{\mu} - \mu\|_{\cP, 2} = O(n^{-r})$, and by assumption that $\|\ehat - e\|_{\cP, 2} = O(n^{-r_e})$, since we have assumed that $r + r_e = 1/2$, we have:
$$\|\muhat - \mu\|_{\cP, 2} \times \|\ehat - e\|_{\cP, 2} = O(n^{-r}) \times O(n^{-r_e}) \leq O(n^{-1/2}).$$
The same exact argument can be used to show that the same is true for the case of the ATT. 
\end{proof}

\subsection{Proof of Lemma \ref{thm:mixproblem}}
\begin{proof}
    Let $\MG^* = \{i:\, \Dphiqhat(\bX_i, \bx) \leq \Gamman\}$, and define the associated membership indicators: $$W_i^* = \begin{cases}1 &\mbox{ if }\Dphiqhat(\bX_i, \bx) \leq \Gamman\\ 0 & \mbox{ otherwise. }\end{cases}$$ Choose any $W_i, \dots, W_n \in \{0, 1\}^n$ and let $\MG$ be the associated matched group.  We have:
    \begin{align*}
        &\sum_{i \in \MG}^n \Dphiqhat(\bx, \bX_i) - \Gamman \sum_{i=1}^n W_i = \sum_{i=1}^n W_i(\Dphiqhat(\bx,\bX_i) - \Gamman)\\
        &= \underbrace{\sum_{i:\, \Dphiqhat(\bX_i, \bx) > \Gamman} W_i(\Dphiqhat(\bx,\bX_i) - \Gamman)}_{\geq 0} + \underbrace{\sum_{i:\, \Dphiqhat(\bX_i, \bx) \leq \Gamman} W_i(\Dphiqhat(\bx,\bX_i) - \Gamman)}_{\leq 0}\\
        &\geq \sum_{i:\, \Dphiqhat(\bX_i, \bx) \leq \Gamman} W_i(\Dphiqhat(\bx,\bX_i) - \Gamman) \geq  \sum_{i:\, \Dphiqhat(\bX_i, \bx) \leq \Gamman} (\Dphiqhat(\bx,\bX_i) - \Gamman)\\
        &= \sum_{i=1}^n W_i^*(\Dphiqhat(\bx,\bX_i) - \Gamman).
    \end{align*}
    Therefore, $W_1^*, \dots, W_n^*$ and $\MG^*$ are the match indicators and the matched group that minimize the objective. 
\end{proof}

\subsection{Lemma \ref{thm:martingaleCLT}}
\begin{lemma}\label{thm:martingaleCLT}
Let A1-A6 hold. For arbitrary $\bx \in \X$ and $t \in \{1, \dots, M\}$, let the matched group be a subset of the units with treatment $t$: $\MGxt \subset \{1,\dots, n:\, T_i=t\}$. Define the variable $\Wixt = \ind[i \in \MGxt]$ as the indicator denoting membership of unit $i$ in $\MGxt$. Additionally, define the number of units in the matched group as: $\Nxt = |\MGxt|$, and let $\Rx = \max_{i \in \MGxt}\Dphiq(\bX_i, \bx)$ be the radius of the matched group. If $\MGxt$ satisfies the following requirements, for constants $c \geq 0$, and $r > 0$:
    \begin{enumerate}
        \item $\Rx = o_p(n^{-r})$, 
        \item $\frac{n^{2r}\Nxt}{(\Nxt + c)^2} \pconv \frac{1}{K}$ for some constant, $K$, 
        \item  $\E\left[\left(\frac{n^{2r}}{\Nxt + c}\right)^{2}\right] = O(1)$
    \end{enumerate}
Then the estimator: $\mudxhat = \frac{1}{\Nxt + c}\sum_{i = 1}^n \Wixt Y_i$ satisfies: $n^r(\mudxhat - \muxd) \dconv \mathcal{N}(0, \frac{\sigmaxd}{K})$. 
\end{lemma}

\begin{proof}
First, we adapt the martingale representation that \cite{abadie2012} construct for the matching estimator for the Average Treatment effect on the Treated (ATT), to a martingale representation for either matching estimator, $\muxthat$. We begin by writing the estimation error of the matching CATE estimator as the sum of two terms: $\muxthat - \muxt = \Dnx + \Rnx$, as follows:
\begin{align}
    \muxthat - \muxt =& \frac{1}{\Nxt + c}\sum_{i\in \MGxt} Y_i - \muxt\nonumber\\
    &+ \frac{1}{\Nxt + c}\sum_{i\in \MGxt} \muXid  - \frac{1}{\Nxt + c}\sum_{i\in \MGxt} \muXid \nonumber\\
    =& \underbrace{\frac{1}{\Nxt + c}\sum_{i\in \MGxt} Y_i - \muXid }_{\Dnx} + \underbrace{\frac{1}{\Nxt + c}\sum_{i\in \MGxt} \muXid - \muxt}_{\Rnx}.\label{Eq:decomoposition}
\end{align}
We will, for now, disregard the $\Rnx$ term, and come back to it later, as we will see that the conditions needed for asymptotic normality of $\mudxhat$ imply vanishing of this term. Because of this intuition, it will suffice to show asymptotic normality of $n^r\Dnx$ as Slutzky's theorem will imply that the asymptotic distribution of this term is the same as that of $n^r(\mudxhat-\muxd)$. To study the asymptotic distribution of $\Dnx$, we wish to write it as a martingale, which will then enable us to employ central limit theorems for martingale arrays in order to establish its asymptotic normality. 
We can show that $\Dnx$ does indeed constitute a martingale w.r.t. a certain filtration by rewriting it as a sum of martingale differences. Recall that the binary variable $\Wixt = \ind[i \in \MGxt]$ denotes membership of matching set unit $i$ in $\MGxt$. Note that, $\Wixt = 1$ only if $T_i = t$, and define, for any $i \in 1, \dots, n$:
\begin{align*}
    \xiNix &= \frac{n^r}{\Nxt + c}\Wixt(Y_i - \muXid) 
\end{align*}
These quantities will be the martingale differences in the representation we will construct, and by definition: $n^r\Dnx = \sum_{i=1}^n \xiNix$. Finally, define the following $\sigma$-field:
\begin{align}
    \Fnix = \sigma\{\bX_1, \dots, \bX_n, T_1, \dots, T_n, Y_1, \dots, Y_i, \phihat\}. \label{eq:Fnimox}
\end{align}
Note that we include the representation $\phihat$ directly in the filtration defined, but if the representation is learned from a separate training set, then this dataset can be included in the filtration instead of it to obtain the same results. Since the $\xiNix$ have zero mean, and are adapted to the filtration $\Fnix$, then the array:
\begin{align}
    \left\{\sum_{j=1}^i\xiNjx,\; \Fnix,\; 1\leq i\leq n\right\}\label{Eq:martingale}
\end{align}
is a martingale for each $n > 2$ by the same arguments in \cite{abadie2012}. 

We can now apply Lindeberg's central limit theorem for triangular martingale arrays to the martingale array defined in \eqref{Eq:martingale}. The CLT in question states that if:
\begin{enumerate}
    \item (Condition 1) $\E[\xiNix] = 0$
    \item (Condition 2) $\sum_{i=1}^n\E[\xiNix^2|\Fnimox] \pconv \sigma^2$, for some constant, $\sigma^2$, as $n \rightarrow \infty$
    \item (Condition 3) $\forall \epsilon > 0:\; \sum_{i=1}^N\E[\xiNix^2\ind_{[|\xiNix| > \epsilon]}] \rightarrow 0$, as $n \rightarrow \infty$, 
\end{enumerate}
then $n^r\Dnx \dconv \mathcal{N}(0, \sigma^2)$. We will now show that all three conditions hold. 
    
Starting with Condition 1, this condition is easily verified because $\E[\xiNix|\Fnimox] = 0$, since $\xiNix$ is a martingale difference term, and that, therefore, $\E[\xiNix] = \E[\E[\xiNix|\Fnimox]] = \E[0] = 0$.
    
Second, Condition 3, commonly known as Lindeberg's condition, is implied by Lyapunov's condition, which is much easier to check. Lyapunov's condition statest that, for some $\delta > 0$: $$\lim_{n\rightarrow \infty}\sum_{i=1}^N\E[|\xiNix|^{2 + \delta}] = 0.$$ Lyapunov's condition can be seen to hold in our case by the following: fix a $s >0$, then,
\begin{align*}
    \E[|\xiNix|^{2 + \delta}] &= \E\left[\left|\underbrace{\frac{n^r}{\Nxt + c}}_{ \geq 0}
    \underbrace{\Wixt}_{ \in \{0, 1\}}
    (Y_i - \muXid)\right|^{2 + \delta}\right] \\
    &=  \E\left[\frac{n^{2r+r\delta}}{(\Nxt + c)^{2 + \delta}}\Wixt|Y_i - \muXid|^{2 +\delta}\right],
\end{align*}
Recall now that, by Assumption A5, $\E[|Y_i |^{2 + \delta}|\bX_i, T_i] \leq C_\delta$ for some constant $C_\delta < \infty$. Additionally, since $Y_i \indep \phihat$ by A6.b, it follows that: $\E[|Y_i |^{2 + \delta}|\phihat, \bX_i, T_i] \leq C_s$.  Since conditionally on $\phihat$, $\bX_i$, and $T_i$, all of $\muXid$, $\Nxt$, $\Wixt$ are constants, it follows that:
\begin{align*}
    &\E\left[\Wixt\frac{n^{2r+r\delta}}{(\Nxt + c)^{2 + \delta}}|Y_i - \muXid|^{2 + \delta}\right] \\
    &= \E\left[\Wixt\frac{n^{2r+r\delta}}{(\Nxt + c)^{2 + \delta}}\E[|Y_i - \muXid|^{2 + \delta}|\phihat, \bX_i, T_i]\right]\\
    &\leq \E\left[\Wixt\frac{n^{2r+r\delta}}{(\Nxt + c)^{2 + \delta}}C_\delta\right].
\end{align*}
    Putting the bounds together we have, as $n\rightarrow \infty$:
\begin{align*}
    \sum_{i=1}^N\E[|\xiNix|^{2 + \delta}] &\leq  C_\delta  \sum_{i=1}^n \E\left[\Wixt\frac{n^{2r+r\delta}}{(\Nxt + c)^{2 +  \delta}}\right] = C_\delta\E\left[\frac{n^{2r+r\delta}}{(\Nxt+ c)^{1 + \delta}}\right]\\
    &= C_\delta\E\left[\frac{n^{2r(1+\delta)}}{n^{2r(1+ \delta)}}\frac{n^{2r+r\delta}}{(\Nxt+ c)^{1 + \delta}}\right] 
    = C_\delta \frac{1}{n^{r\delta}}\E\left[\left(\frac{n^{2r}}{(\Nxt + c)}\right)^{1 + \delta}\right].
\end{align*}
Finally, by Requirement (3) of this lemma, we have that, for $\delta = 1$:
\begin{align*}
  C_\delta \frac{1}{n^{r\delta}}\E\left[\left(\frac{n^{2\delta}}{\Nxt + c}\right)^{1 + \delta}\right] &= C_\delta \frac{1}{n^{r}}O(1) \rightarrow 0.
\end{align*}
Therefore, Condition 3 holds. 

Moving on to Condition 2, recall first that: $\E[(Y_i - \muXid)^2|\Fnimox] = \sigmaXid$, where $\sigmaXid = \V[Y(t)|\bX = \bX_i]$. In light of this, we can write:
\begin{align}
    \sum_{i=1}^n\E[\xiNix^2|\Fnimox] &= \sum_{i=1}^n\E\left[\left(\frac{n^r}{\Nxt + c}(\Wixt(Y_i - \muXid))\right)^2\biggr|\Fnimox\right]\nonumber\\
    &= \sum_{i=1}^n\frac{n^{2r}}{(\Nxt + c)^2}(\Wixt\E[(Y_i - \muXid)^2|\Fnimox]\nonumber\\
    &= \sum_{i=1}^n\frac{n^{2r}}{(\Nxt + c)^2}\Wixt\sigmaXid\nonumber\\
    &= \sum_{i=1}^n\frac{n^{2r}}{(\Nxt + c)^2}\Wixt (\sigmaXid - \sigmaxd)\nonumber \\
    &+ \sum_{i=1}^n\frac{n^{2r}}{(\Nxt + c)^2}\Wixt\sigmaxd\nonumber\\
    &= \sum_{i=1}^n\frac{n^{2r}}{(\Nxt + c)^2}\Wixt (\sigmaXid - \sigmaxd) \label{eqn:firsterm_var}\\
    &+ \sum_{i=1}^n\frac{n^{2r}}{(\Nxt + c)^2}\sigmaxd.\label{eqn:remainder_var}
\end{align}
Recall that, by Assumption A5: $\sigmaXid - \sigmaxd \leq C_L \Dphiq(\bx, \bX_i) \leq C_L \Rx$ for any $i$, and, therefore: 
\begin{align*}
    &\sum_{i=1}^n\frac{n^{2r}}{(\Nxt + c)^2}\Wixt (\sigmaXid - \sigmaxd)\\
    &\leq C_L\sum_{i=1}^n\frac{n^{2r}}{(\Nxt + c)^2}\Wixt \Rx\\
    &= \frac{n^{2r}\Nxt}{(\Nxt + c)^2}C_L\Rx\\
    &= O_p(1)o_p(1)C_L = o_p(1). 
\end{align*}
where the last equality follows by Requirements (1), and (2) of this lemma.

For the other term (Eq.~\eqref{eqn:remainder_var}) we have: 
\begin{align*}
    \sum_{i=1}^n\frac{n^{2r}}{(\Nxt + c)^2}\sigmaxd &= \frac{n^{2r}\Nxt}{(\Nxt + c)^2}\sigmaxd \pconv \frac{1}{K}\sigmaxd,
\end{align*}
by Requirement (2) of this lemma and Slutzky's theorem. Therefore: 
\begin{align*}
    \sum_{i=1}^n\E[\xiNix^2|\Fnimox] &= o_p(1) + \frac{1}{K}\sigmaxd,
\end{align*}
which proves Condition 2 of Lindenberg's CLT. Since all the conditions are satisfied, the CLT implies that: $n^r\Dnx \dconv \mathcal{N}(0, \frac{1}{K}\sigmaxd)$.

The result  for $\mudxhat - \muxd$ follows immediately by applying the decomposition in Eq. \eqref{Eq:decomoposition} to write: $n^r(\mudxhat - \muxd) = n^r\Dnx + n^r\Rnx$. The first quantity converges to $\mathcal{N}(0, \frac{1}{K}\sigmaxd)$ by Lindenberg's CLT, as just shown, and the second quantity converges to 0 in probability by Requirements (1) and (3) of this lemma:
\begin{align}
    n^r\Rnx &= \frac{n^r}{\Nxt + c}\sum_{i=1}^n \Wixt\muXid - \muxd \nonumber\\
    &= \frac{n^r}{\Nxt + c}\sum_{i=1}^n (\muXid - \muxd)\Wixt + \frac{n^{r}c\muxd}{\Nxt + c} \nonumber\\
    &\leq \frac{n^r}{\Nxt + c}\sum_{i=1}^n C_\mu \Dphiq(\bx, \bX_i)\Wixt+ \frac{n^{r}c\muxd}{\Nxt + c}\nonumber\\
    &\leq C_\mu\frac{n^r}{\Nxt + c}\sum_{i=1}^n \Rx\Wixt+ \frac{n^{r}c\muxd}{\Nxt + c}\nonumber \\
    &\leq \underbrace{C_\mu n^r \Rx}_{o_p(1) \text{ by Req. (1)}} + \underbrace{\frac{n^{r}c\muxd}{\Nxt + c}}_{o_p(1) \text{ by Req. (3)}}. \label{eq:Bnxbound}
\end{align}
Note that Requirement (3) of this lemma implies that $\frac{n^{r}c\muxd}{\Nxt + c} = o_p(1)$ by Markov's inequality. This concludes the proof. 
\end{proof}

\begin{lemma}{(Bound on moments of matched outcomes)}\label{thm:mombound}
Let $\MGxt$ be a collection of units with treatment $t$ matched to $\bx$ and let $\Nxt$ denote the size of this collection. Let A1-A4 hold and let all other queantities be defined as in the rest of the paper. We have, for $s \geq 2$:
\begin{equation}
    \E\left[\left|\sum_{i = 1}^n\Wixt(Y_i - \mu(\bX_i, t))\right|^s\right] \leq B_sC_s\Nxt^{s/2},
\end{equation}
for a constant $B_s$ depending only on $s$, and $C_s$ defined in Assumption 4. 
\end{lemma}
\begin{proof}
    This is a well-known results that holds generally for mean-0 random variables. Here we simply adapt its proof to our setting. Let $\tilde{Y} = (Y_1, \dots, Y_n)$, $\tilde{\bX} = (\bX_1, \dots, \bX_n)$, and $\tilde{T} = (T_1, \dots, T_n)$. Since the random variables $Y_i - \mu(\bX_i, t)$ have mean 0, conditional on $\bX_i$, then by the Marcinkiewicz–Zygmund inequality, there exists a constant $B_s$ such that:
    \begin{align*}
        &\E_{\tilde{Y}|\tilde{\bX}, \tilde{T}}\left[\left|\sum_{i = 1}^n\Wixt(Y_i - \mu(\bX_i, t))\right|^s\right] \leq B_s\E_{\tilde{Y}|\tilde{\bX}, \tilde{T}}\left[\left(\sum_{i=1}^n|\Wixt(Y_i - \mu(\bX_i, t))|^2\right)^{s/2}\right]\\
        &= B_s\Nxt^{s/2}\E_{\tilde{Y}|\tilde{\bX}, \tilde{T}}\left[\left(\frac{1}{\Nxt}\sum_{i=1}^n|\Wixt(Y_i - \mu(\bX_i, t))|^2\right)^{s/2}\right]\\
        \intertext{then by Jensen's inequality we have:}
        &\leq B_s\Nxt^{s/2}\E_{\tilde{Y}|\tilde{\bX}, \tilde{T}}\left[\left(\frac{1}{\Nxt}\sum_{i=1}^n|\Wixt(Y_i - \mu(\bX_i, t))|^s\right)\right]\\
        &= B_s\Nxt^{s/2}\left(\frac{1}{\Nxt}\sum_{i=1}^n\Wixt\E_{Y_i|\bX_i, T_i}[|(Y_i - \mu(\bX_i, t))|^s]\right)\\
        \intertext{and since $\E_{Y_i|\bX_i, T_i}[|(Y_i - \mu(\bX_i, t))|^s] \leq C_s$ by Assumption 4 with $\delta = s-2$:}
        &\leq B_s\Nxt^{s/2}\left(\frac{1}{\Nxt}\sum_{i=1}^n\Wixt C_s\right)\\
        &= B_sC_s\Nxt^{s/2}.
    \end{align*} 
\end{proof}

\begin{lemma}{(Bound on $KNN$ distances, Lemma 14.1 in \citealt{li2007nonparametric})}\label{thm:knndistbound}
Let $\bZ_1, \dots, \bZ_n$ be i.i.d. observations with bounded continuous density $f_{\bZ}$ supported over a subset of $\R^d$. For a point $\bz$ in the support of $f_\bZ$,  let $B(\bz, r)$ be a ball of radius $r$ centered at $\bz$, and define $G(r) = \Pr_{\bZ}(\bZ \in B(\bz, r))$. Additionally let $R_{k}(\bz)$ be the Euclidean distance between $\bz$ and its $k^{th}$ nearest neighbor  among the $\bZ_1, \dots, \bZ_n$. Finally, let $\lambda$ and $\gamma$ be integers such that the function $\Phi(R_{k}(\bz)) := \frac{1}{R_{k}(\bz)^\lambda G(R_{k}(\bz))^\gamma}$ exists. Then:
\begin{equation}
    \E_{\bZ_1, \dots, \bZ_n}[\Phi(R_{k}(\bz))] = O\left(\left(\frac{k}{n}\right)^{-\frac{\lambda}{d}}\right).
\end{equation}
\end{lemma}
This lemma is a restatement of Lemma 14.1 of \cite{li2007nonparametric} and is proven therein.

\section{Methods used in the simulation}
\begin{table}[!htbp]
    \centering
    \caption{Methods used in simulated experiments}
    \label{tab:simmethods}
    \begin{tabular}{l|c|r}
        \hline\hline
        Acronym & Method & Citation\\
        \hline
        Linear Reg. & Linear regression & \\
        BART & Bayesian Additive Regression Trees & \cite{chipman2010bart}\\
        CF & Causal Forests & \cite{wager2018estimation}\\
        GP & Gaussian Process Regression & \cite[see, e.g.,][]{rasmussen2003gaussian}\\
        SVM & Support Vector Machine & \cite{drucker1997support}\\
        XL-RF & X-Learner with Random Forest & \cite{kunzel2019metalearners}\\
        BC GenMatch & Genetic matching with bias correction & \cite{diamond2013genetic}\\
        BC l2Match & L2 distance matching with bias correction & \cite{abadie2011bias}\\
        \hline
        MML-BART-Y & M-ML on BART-estimated outcomes\\
        MML-SVM-Y & M-ML on SVM-estimated outcomes\\
        MML-GP-Y & M-ML on GP-estimated outcomes\\   
        \hline\hline
    \end{tabular}
\end{table}
\section{Additional simulation results}
\begin{figure}[!htbp]
    \centering
    \caption{95\% Asymptotic Confidence Interval Size for the CATE}
    \label{fig:intsize}
    \includegraphics[width=\textwidth]{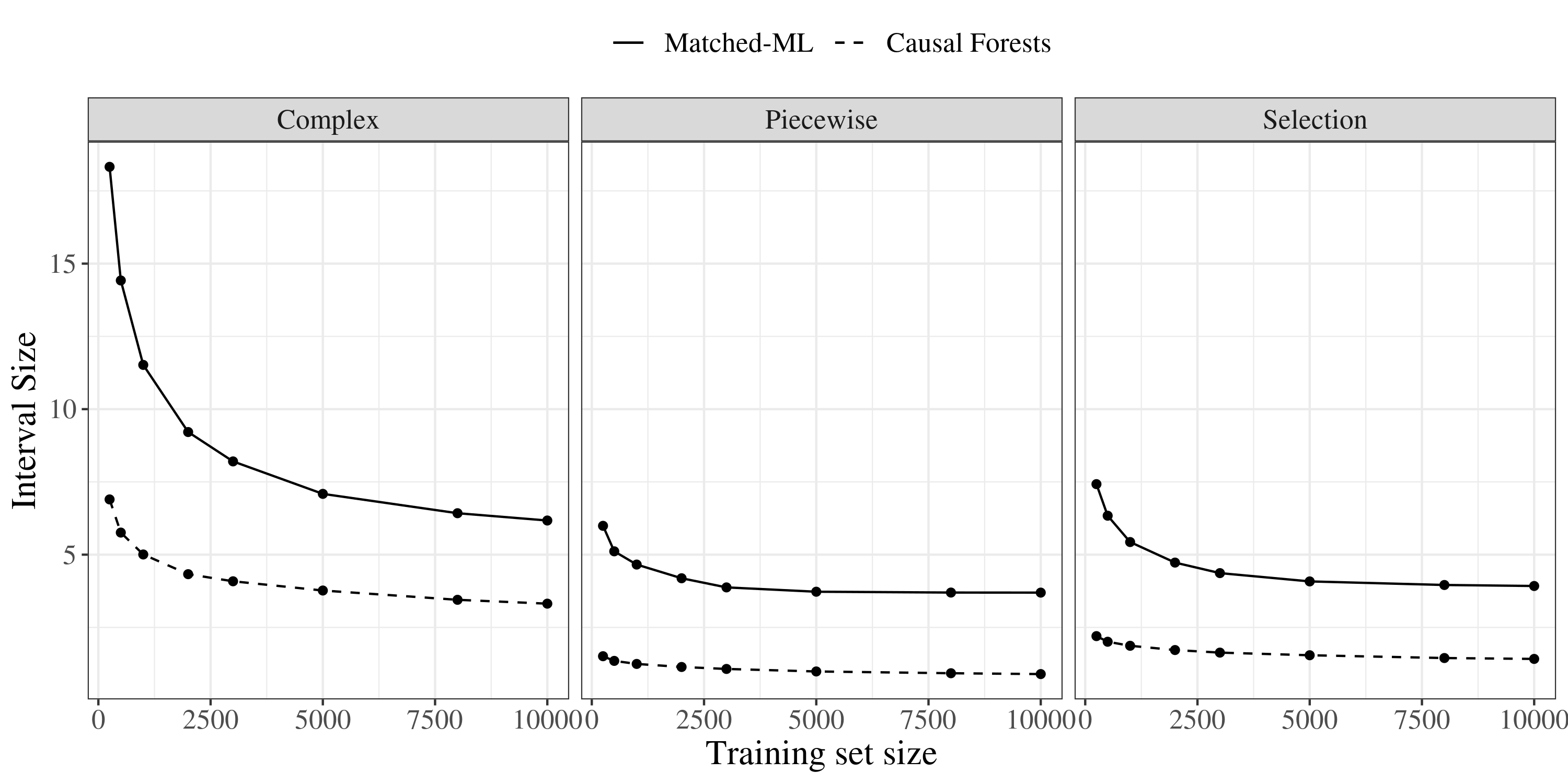}
\end{figure}

\begin{figure}[!htbp]
    \centering
    \caption{95\% Asymptotic Confidence Interval Size for the ATE}
    \label{fig:atesize}
    \includegraphics[width=\textwidth]{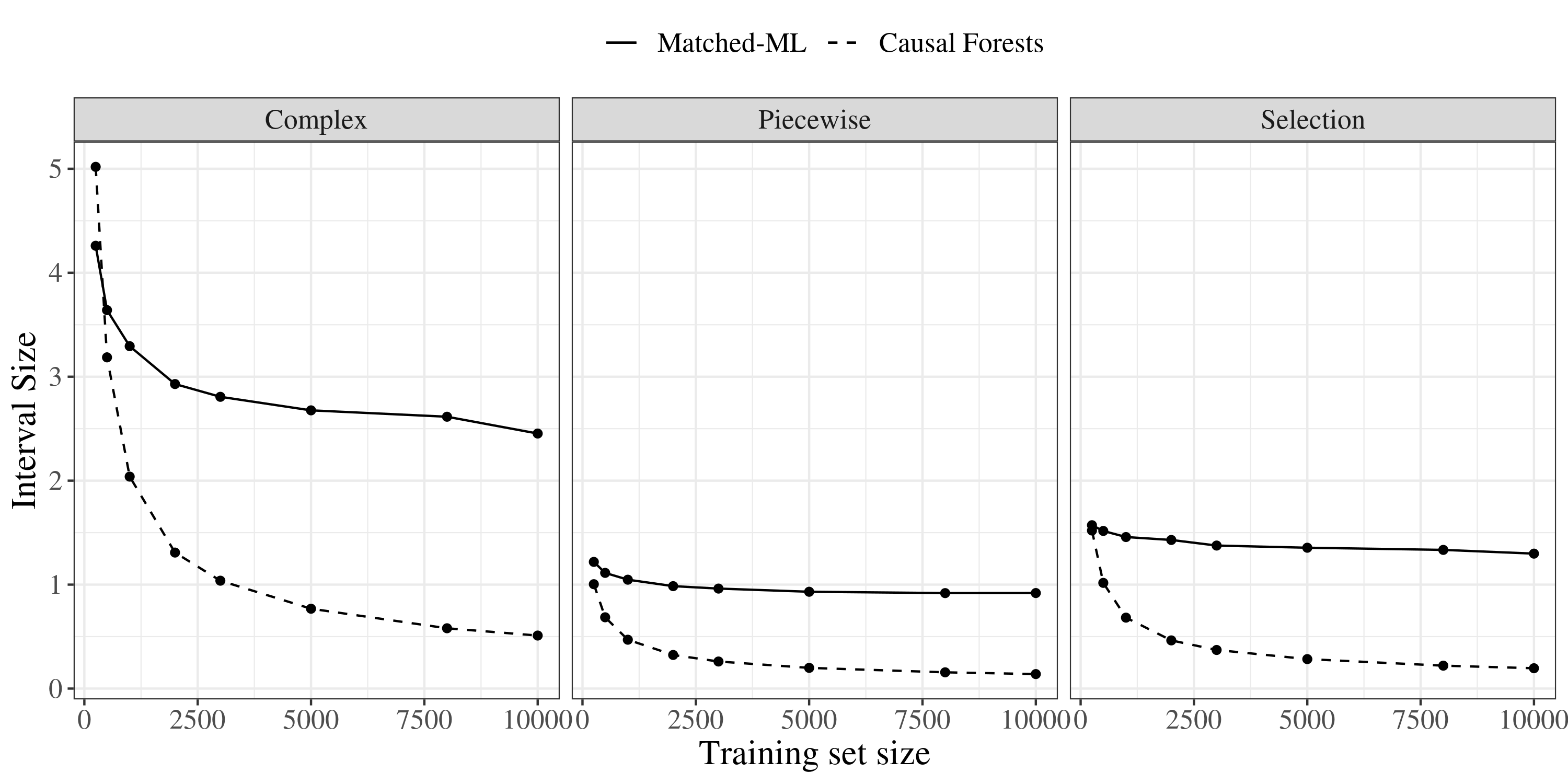}
\end{figure}

\begin{figure}[!htbp]
    \centering
    \caption{Proportion of Confidence Intervals not Containing zero, CATE}
    \label{fig:catezero}
    \includegraphics[width=\textwidth]{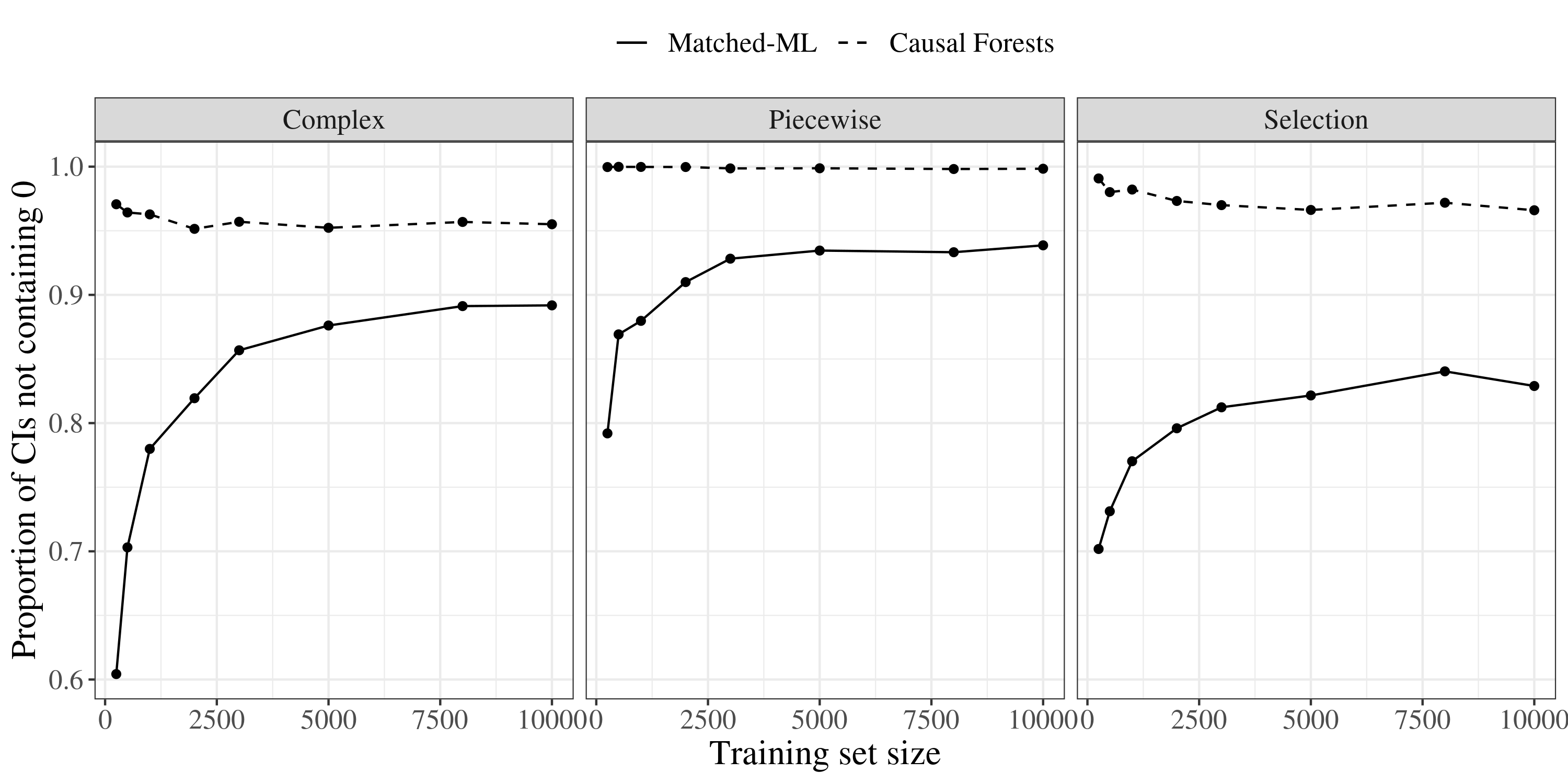}
\end{figure}

\begin{figure}[!htbp]
    \centering
    \caption{Proportion of Confidence Intervals not Containing zero, ATE}
    \label{fig:atezero}
    \includegraphics[width=\textwidth]{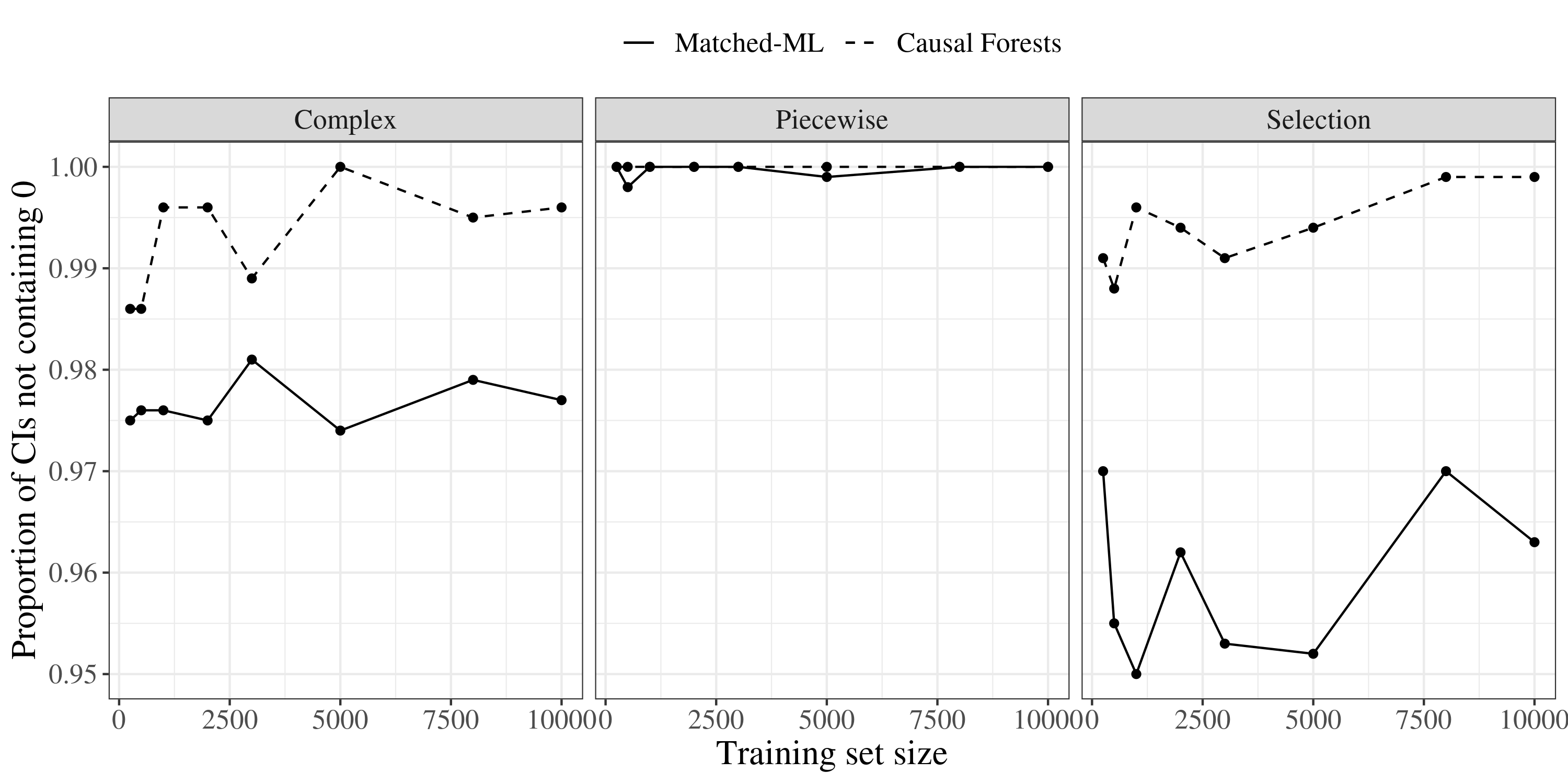}
\end{figure}

\section{Additional sample matched groups on the brand data}

\begin{figure}
    \fbox{\includegraphics[width=0.5\textwidth]{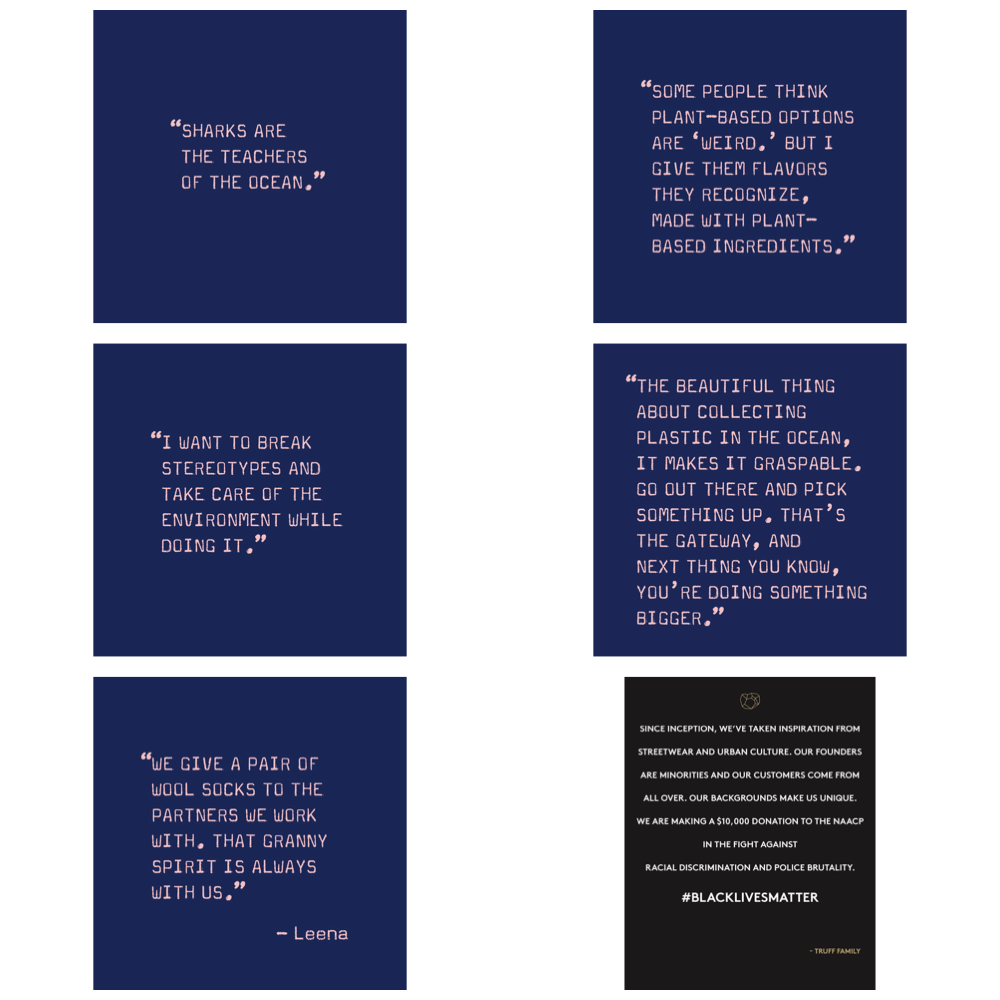}}
    \fbox{\includegraphics[width=0.5\textwidth]{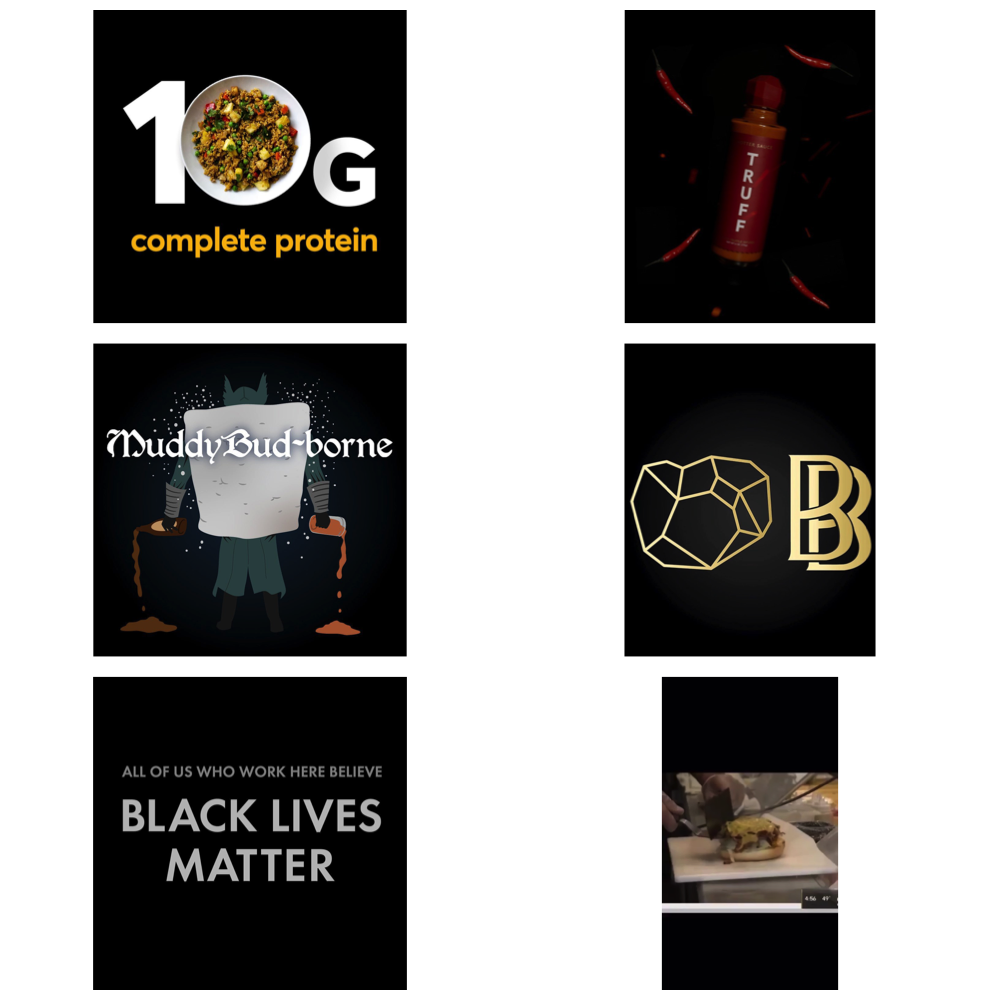}}
    \fbox{\includegraphics[width=0.5\textwidth]{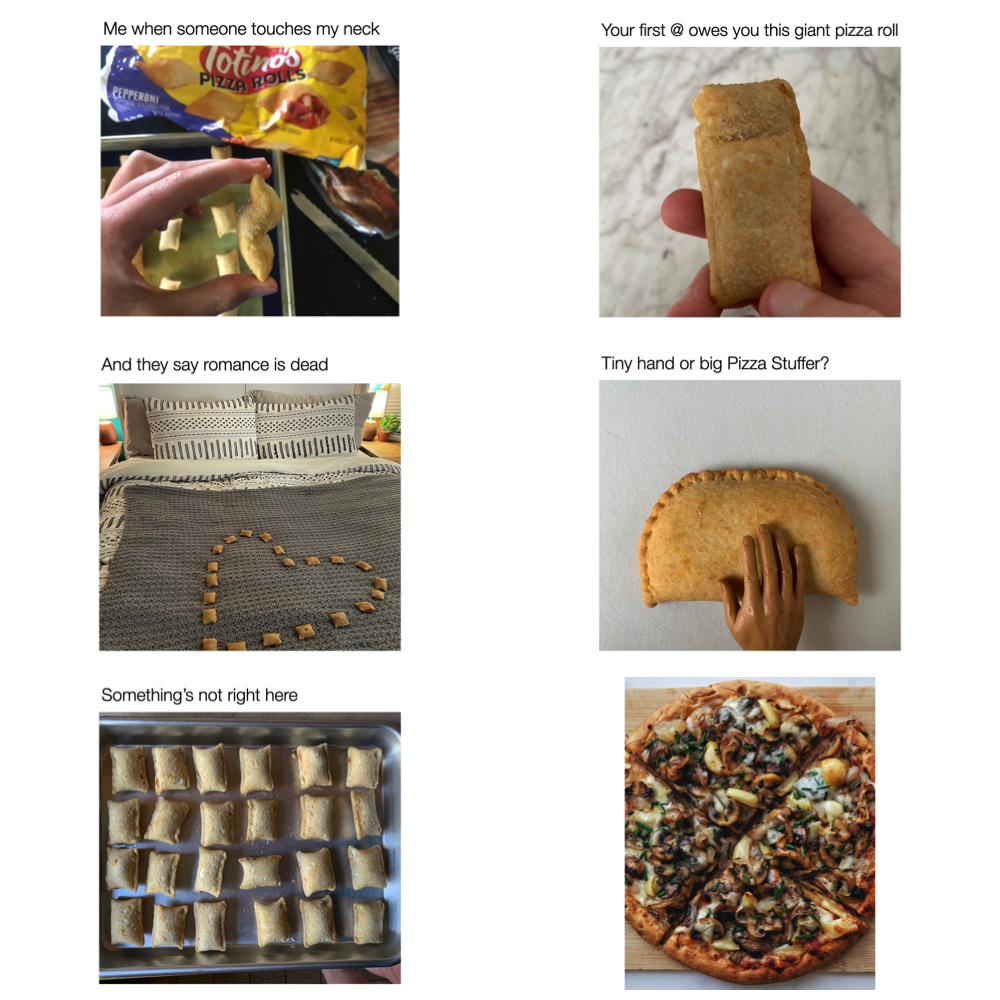}}
\end{figure}
\begin{figure}
    \fbox{\includegraphics[width=0.5\textwidth]{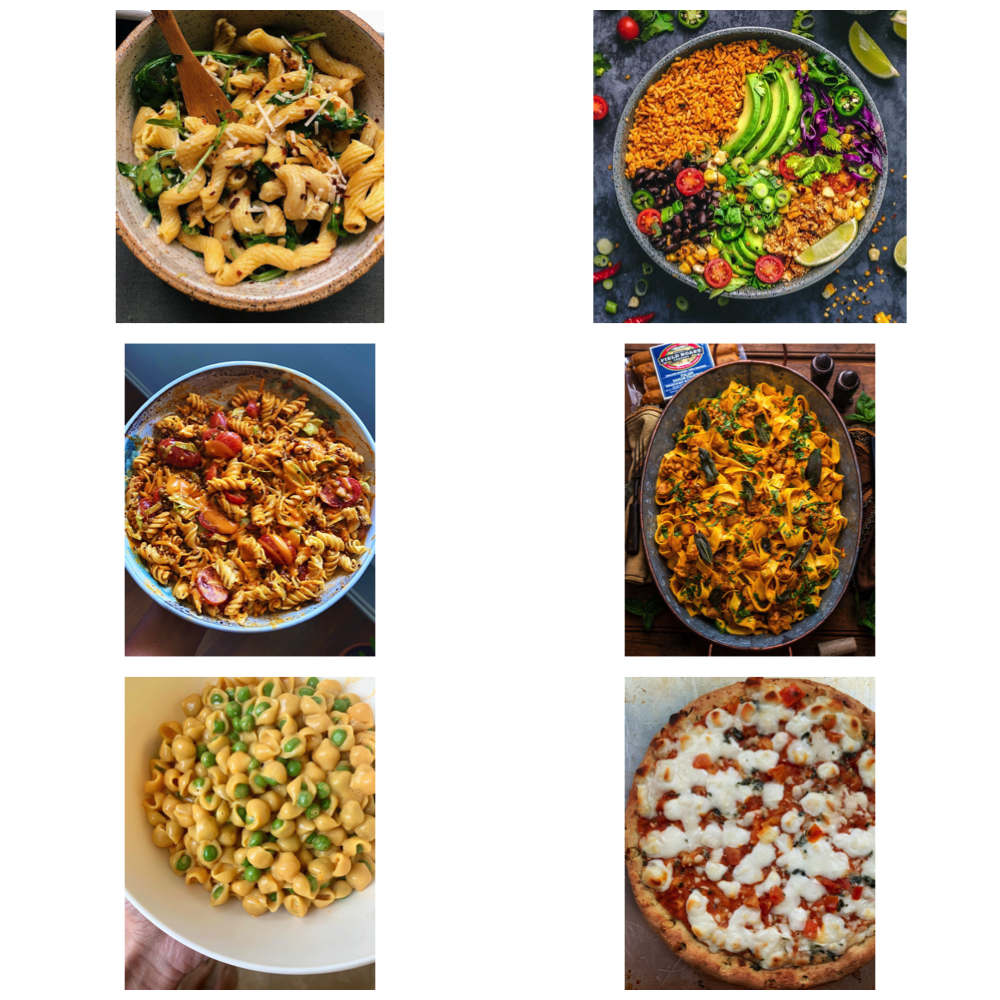}}
    \fbox{\includegraphics[width=0.5\textwidth]{figures/MGs/558.png}}
    \fbox{\includegraphics[width=0.5\textwidth]{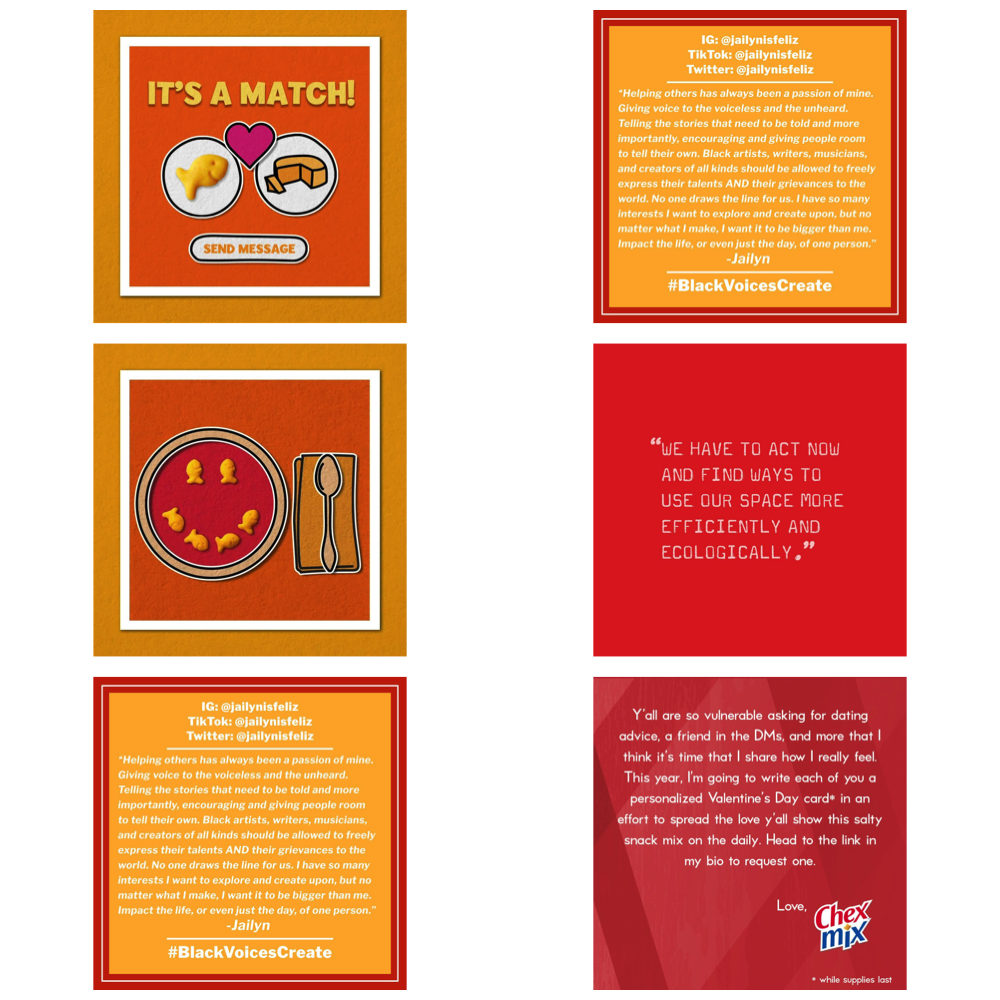}}
    \fbox{\includegraphics[width=0.5\textwidth]{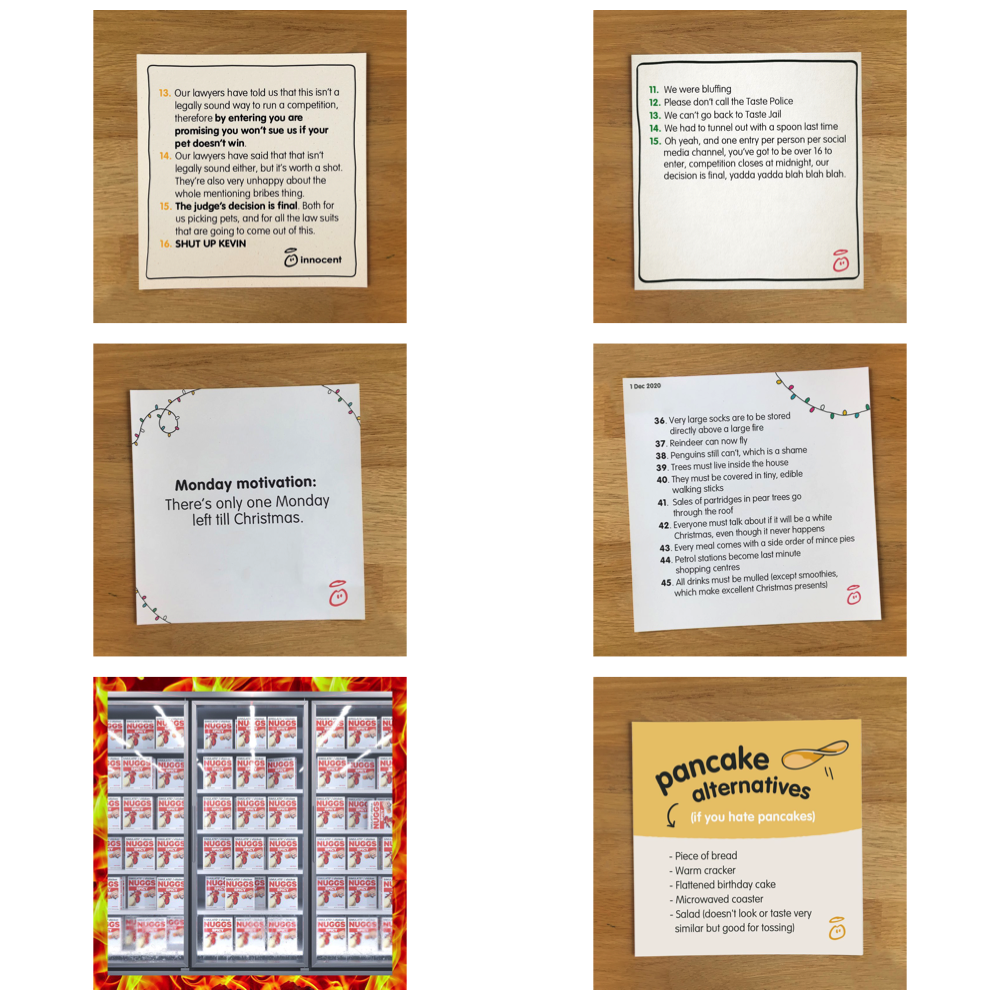}}
\end{figure}
\end{document}